\documentclass[manuscript]{aastex}
\usepackage{emulateapj5}
\usepackage{lscape}
\usepackage{natbib}

\newcommand{\ee}[1]{\mbox{${} \times 10^{#1}$}}
\newcommand{\eten}[1]{\mbox{$10^{#1}$}}

\newcommand{\lteq}{\raisebox{-0.6ex}{$\,\stackrel
{\raisebox{-.2ex}{$\textstyle <$}}{\sim}\,$}}
\newcommand{\tbn}{\tablenotemark}

\newcommand{\kms}{\mbox{km s$^{-1}$}}
\newcommand\cmv{\mbox{cm$^{-3}$}}

\newcommand{\um}{$\mu$m}

\newcommand{\iras}{\mbox{\it IRAS}}
\newcommand{\spitzer}{\mbox{\it Spitzer}}
\newcommand{\irac}{\mbox{IRAC}}
\newcommand{\mips}{\mbox{MIPS}}
\newcommand{\iso}{\mbox{\it ISO}}

\newcommand\fir{far-infrared}
\newcommand\mir{mid-infrared}

\newcommand{\sfr }{\mbox{$\dot M_{\star}$}}

\newcommand{\lsun}{\mbox{L$_\odot$}}
\newcommand{\msun}{\mbox{M$_\odot$}}
\newcommand{\rsun}{\mbox{R$_\odot$}}
\newcommand{\msunyr}{\mbox{M$_\odot$ yr$^{-1}$}}
\newcommand{\msunmyr}{\mbox{M$_\odot$ Myr$^{-1}$}}
\newcommand{\msunmyrpc}{\mbox{\msunmyr pc$^{-2}$}}

\newcommand{\tk}{\mbox{$T_K$}}

\newcommand{\lbol}{\mbox{$L_{bol}$}} 
\newcommand{\lint}{\mbox{$L_{int}$}} 
\newcommand{\tbol}{\mbox{$T_{bol}$}} 
\newcommand{\tbolprime}{\mbox{$T^{\prime}_{bol}$}} 
\newcommand{\lbolprime}{\mbox{$L^{\prime}_{bol}$}} 
\newcommand{\aprime}{\mbox{$\alpha^{\prime}$}} 
\newcommand{\dv}{\mbox{$\Delta v$}}

\newcommand{\mean}[1]{\mbox{$\langle#1\rangle$}} 
\newcommand{\av}{\mbox{$A_V$}} 
\newcommand{\lsmm}{\mbox{$L_{smm}$}} 

\newcommand{\ammonia}{\mbox{{\rm NH}$_3$}}

\newcommand{\coo}{$^{13}$CO}
\newcommand{\cooo}{C$^{18}$O}

\newcommand{\hcop}{HCO$^+$}

\newcommand{\jj}[2]{\mbox{$J = #1\rightarrow#2$}}

\newcommand{\mstar}{\mbox{$M_{\star}$}}
\newcommand{\mcloud}{\mbox{$M(cloud)$}}
\newcommand{\mdense}{\mbox{$M(dense)$}}
\newcommand{\tff}{\mbox{$t_{ff}$}}
\newcommand{\tdep}{\mbox{$t_{dep}$}}
\newcommand{\sfrff}{\mbox{$SFR_{ff}$}}

\newcommand{\tdepdense}{\mbox{$t_{dep}(dense)$}}
\newcommand{\sfrffdense}{\mbox{$SFR_{ff}(dense)$}}
\newcommand{\lto}{\mbox{$\lambda_{turn-off}$}}
\newcommand{\aex}{\mbox{$\alpha_{excess}$}}

\newcommand{\ysotot}{1024}
\newcommand{\itot}{16}
\newcommand{\ftot}{12}
\newcommand{\iitot}{60}
\newcommand{\iiitot}{12}
\newcommand{\tsl}{0.46}
\newcommand{\tsljj}{0.61}
\newcommand{\tzero}{0.16}
\newcommand{\ti}{0.54}
\newcommand{\tf}{0.40}
\newcommand{\tiprime}{0.44}
\newcommand{\tfprime}{0.35}
\newcommand{\tzeroprime}{0.10}
\newcommand{\tslprime}{0.47}

\begin{document}

\title {The Spitzer c2d Legacy Results: Star Formation Rates and Efficiencies; Evolution and Lifetimes}
\author{Neal J. Evans II\altaffilmark{1},
Michael M. Dunham\altaffilmark{1},
Jes K. J{\o}rgensen\altaffilmark{2},
Melissa L. Enoch\altaffilmark{3, 4},
Bruno Mer{\'{\i}}n\altaffilmark{5, 6},
Ewine F. van Dishoeck\altaffilmark{5, 7},
Juan M. Alcal{\'a}\altaffilmark{8},
Philip C. Myers\altaffilmark{9},
Karl R.  Stapelfeldt\altaffilmark{10},
Tracy L. Huard\altaffilmark{9, 11},
Lori E. Allen\altaffilmark{9},
Paul M.  Harvey\altaffilmark{1},
Tim van Kempen\altaffilmark{5},
Geoffrey A.  Blake\altaffilmark{12},
David W. Koerner\altaffilmark{13},
Lee G. Mundy\altaffilmark{11},
Deborah L. Padgett\altaffilmark{14},
Anneila I.  Sargent\altaffilmark{3}}

\altaffiltext{1}{Department of Astronomy, University of Texas at Austin,
  1 University Station C1400, Austin, TX~78712;nje@astro.as.utexas.edu,
  mdunham@astro.as.utexas.edu, pmh@astro.as.utexas.edu}

\altaffiltext{2}{Argelander-Institut f\"{u}r Astronomie, University of Bonn, 
Auf dem H\"{u}gel 71, 53121 Bonn, Germany; jes@astro.uni-bonn.de}

\altaffiltext{3}{Division of Physics, Mathematics, and Astronomy,
  MS~105-24, California Institute of Technology, Pasadena, CA~91125;
afs@astro.caltech.edu}

\altaffiltext{4}{University of California, Berkeley}

\altaffiltext{5}{Leiden Observatory, Leiden University, PO Box 9513, 
NL~2300 RA Leiden, The Netherlands; ewine@strw.LeidenUniv.nl}

\altaffiltext{6}{Research and Scientific Support Department, European
Space Agency (ESTEC), PO Box 299, 2200 AG Noordwijk, The Netherlands;
bmerin@rssd.esa.int}

\altaffiltext{7}{Max-Planck-Institut f\"ur Extraterrestrische Physik (MPE), 
Giessenbachstr.1, 85748 Garching, Germany}

\altaffiltext{8}{INAF-OA Capodimonte, via Moiariello 16, 80131, Naples,
Italy; alcala@oacn.inaf.it}

\altaffiltext{9}{Harvard-Smithsonian Center for Astrophysics, 60
  Garden Street, Cambridge, MA 02138; 
  leallen@cfa.harvard.edu, pmyers@cfa.harvard.edu}

\altaffiltext{10}{Jet Propulsion Laboratory, MS~183-900, California
  Institute of Technology, Pasadena, CA~91109; krs@exoplanet.jpl.nasa.gov}

\altaffiltext{11}{Department of Astronomy, University of Maryland, College
  Park, MD~20742; thuard@astro.umd.edu, lgm@astro.umd.edu}

\altaffiltext{12}{Division of Geological and Planetary Sciences,
  MS~150-21, California Institute of Technology, Pasadena, CA~91125;
gab@gps.caltech.edu}

\altaffiltext{13}{Department of Physics and Astronomy, Northern Arizona
  University, NAU~Box~6010, Flagstaff, AZ~86011-6010;david.koerner@nau.edu}

\altaffiltext{14}{Spitzer Science Center, MC~220-6, California
  Institute of Technology, Pasadena, CA~91125; dlp@ipac.caltech.edu}

\begin{abstract}

The c2d \spitzer\ Legacy project obtained images and photometry with
both \irac\ and \mips\ instruments for five large, nearby molecular
clouds. Three of the clouds were also mapped in dust continuum emission
at 1.1 mm, and optical spectroscopy has been obtained for some clouds. 
This paper combines information drawn from studies of 
individual clouds into a combined and updated statistical analysis of 
star formation rates and efficiencies, numbers and lifetimes for SED classes, 
and clustering properties. Current star formation
efficiencies range from 3\% to 6\%; if star formation continues at current
rates for 10 Myr, efficiencies could reach 15\% to 30\%. 
Star formation rates and rates per unit area vary from cloud to 
cloud; taken together, the five clouds are producing about 260 \msun\ 
of stars per Myr. The star formation surface density is more than an order of
magnitude larger than would be predicted from the Kennicutt relation used
in extragalactic studies, reflecting the fact that those relations apply
to larger scales, where more diffuse matter is included in the gas surface
density. Measured against the dense gas probed by the maps of dust continuum
emission, the efficiencies are much higher, with stellar masses similar
to masses of dense gas, and the current stock of
dense cores would be exhausted in 1.8 Myr on average. Nonetheless, star
formation is still slow compared to that expected in a free fall time, even
in the dense cores. 
The derived lifetime for the Class I phase is
\ti\ Myr, considerably longer than some estimates. Similarly, the lifetime
for the Class 0 SED class, \tzero\ Myr, with the notable exception of 
the Ophiuchus cloud, is longer than early estimates. If photometry is corrected
for estimated extinction before calculating class indicators, the lifetimes
drop to \tiprime\ Myr for Class I and to \tzeroprime\ for Class 0. 
These lifetimes assume a continuous flow through the Class II phase and should
be considered median lifetimes or half-lives.
Star formation is highly concentrated to regions of high extinction, and
the youngest objects are very strongly associated with dense cores.
The great majority (90\%) of young stars lie within loose clusters
with at least 35 members and a stellar density of 1 \msun\ pc$^{-3}$.
Accretion at the sound speed from an isothermal
sphere over the lifetime derived for the Class I phase could build a star
of about 0.25 \msun, given an efficiency of 0.3.
Building larger mass stars by using higher mass accretion rates could be 
problematic, as our data confirm and aggravate the ``luminosity problem"
for protostars. At a given \tbol, the values for \lbol\ are mostly less
than predicted by standard infall models and scatter over several orders
of magnitude. These results strongly suggest that accretion is time variable,
with prolonged periods of very low accretion. Based on a very simple model and
this sample of sources, half the mass of a star would be accreted during
only 7\% of the Class I lifetime, as represented by the eight most 
luminous objects.

\end{abstract}


\section{Introduction}\label{intro}

Star formation in the solar neighborhood has been studied extensively, but
a complete survey for forming stars in nearby molecular clouds has been
lacking. The \iras\ catalogs provided an unbiased survey, but they were limited
in sensitivity, wavelength coverage, and spatial resolution. For objects
in embedded phases of evolution, the luminosity limit was about 
$0.1 (d/140)^2$ \lsun\ with $d$ in pc \citep{myers87}. 
The wavelength coverage from 12 to 100 \micron\ covered the peak of emission 
from typical Class I objects, but deeply embedded Class 0 sources were
often missed, requiring special processing to be detected in \iras\ data
after being found at longer wavelengths \citep{andre00}. In addition,
shorter wavelength data were needed to identify less embedded objects
and to learn more about the embedded ones. Follow-up studies at shorter
wavelengths often found multiple sources, which could be confused within
the large beams of \iras. Detailed studies in the \mir\ from the ground
were hampered by the atmosphere and limited to small regions already
known to have luminous sources or clusters. 

Ground-based near-infrared surveys, enabled by the development 
of large format near-infrared arrays, are a powerful tool for 
finding low luminosity YSOs in nearby clouds with relatively low extinction. 
High spatial resolution and sensitivity in these surveys decrease 
the level of confusion found in older surveys
\citep[e.g.,][]{strom89,Eiroa92}.  Indeed, the 2MASS survey at $J$,
$H$ and $K$ has been used extensively to characterize the young star
population and distribution in a number of clouds
\citep[e.g.,][]{carpenter00}. Many of these studies have focused on
young clusters in more massive star-forming regions, however
\citep[see review by][]{lada03}.

The {\it Infrared Space Observatory} (ISO) improved significantly on
{\it IRAS} in terms of sensitivity and spatial resolution at the
mid-infrared wavelengths. 
The ISOCAM instrument mapped a few square degrees of nearby
star-forming clouds in two bands at 6.7 and 14.3 \micron\ 
(e.g., \citealt{bontemps01}, \citealt{persi03}, \citealt{kaas04}).
The mid-infrared observations revealed the more embedded population compared 
with near-infrared surveys. The number of Class II sources with
$\lbol > 0.03$ \lsun\ in Ophiuchus was doubled \citep{bontemps01}. 
These studies were focused on the densest parts of the clouds, 
but they clearly indicated
the importance of deep mid-infrared observations.

The \spitzer\ mission, combined with ground-based near-infrared
and submillimeter data, can remedy most of the deficiencies of
previous studies. \spitzer\ can provide a survey that is deep, wide, and
relatively unbiased, and that covers the entire wavelength range
needed to characterize the different evolutionary stages, from deeply
embedded YSOs to young stars which have lost most of their disks. Here
we summarize the results
of the \spitzer\ legacy project, ``From Molecular Cores to Planet-forming
Disks", or ``Cores to Disks", further abbreviated to c2d \citep{evans03}.
One of the main approaches of the c2d project 
has been to provide a more complete, less biased
sample of star formation in nearby large clouds and small cores. Toward
that end, we used \spitzer\ to map 15.5 square degrees in five large, nearby
molecular clouds and about 0.6 square degrees in 82 small dense cores. 
For this paper, we focus on the studies of large clouds, where the
mapping efficiency of \spitzer\ enables coverage of large areas, in
contrast to previous surveys, which focused on small areas around
IRAS sources, for example.
The results on the small cores will be summarized by \citet{huard08}.

Based on all of our data and some auxiliary data, we construct a combined
list of young stellar or substellar objects (hereafter YSOs). We discuss
issues of contamination and completeness, and we calculate various
quantities that will be used in the analysis (\S \ref{results}).
We summarize the star formation efficiencies and 
rates (\S \ref{sfesfr}), comparing our results to predictions used
in extragalactic work (\S \ref{exgal}).
These values supercede preliminary values \citep{evans08}.
We compare numbers of sources in
various SED classes and calculate lifetimes for these classes 
(\S \ref{lifetimes}).
We then reexamine the issues of source classification (\S \ref{karlmarx}),
the connection between empirical classes and various stages of star 
formation, and the estimation of lifetimes for these stages 
(\S \ref{stages}). We discuss the spatial distribution of sources
and their clustering properties (\S \ref{clusters}). Finally,
we compare the observational results to the predictions of various
theoretical models (\S \ref{models}), describe future work (\S \ref{future}),
and summarize the main results (\S \ref{summary}).

\subsection{The Sample}\label{sample}

Five large clouds were selected for the c2d project: Serpens \citep{eiroa08}, 
Perseus \citep{bally08}, Ophiuchus \citep{wilking08}, Lupus \citep{comeron08}, 
and Chamaeleon \citep{luhman08}. 
More specifically, we targeted Lupus I, III, and IV  and Chamaeleon II 
(hereafter Cha II).
These were chosen to lie within about 300 pc of the Sun, to span a range of
previously-known star formation activity, and to complement observations of
smaller regions in these clouds obtained by Guaranteed Time Observers (GTOs). 
We have included data from the GTO observations with our data for a complete
picture of the clouds. 

The clouds in this study are listed in Table \ref{tab1}, with the adopted
distances, solid angles and areas with both \irac\ and \mips\ data, 
a measure of turbulence, the total mass in the mapped area, and the
crossing time.  Some of these clouds were targets of \iso\ surveys of
smaller regions: 0.7 sq. deg. in Ophiuchus \citep{bontemps01};
0.13 sq. deg. in Serpens \citep{kaas04}; and 0.2 sq. deg. in Cha~II
\citep{persi03}.
The results presented here are
based on detailed studies of the individual clouds:
Serpens (\citealt{harvey06}; \citealt{harveymips}; \citealt{harveysynth});
Perseus (\citealt{joergensen06}; \citealt{rebull07}; \citealt{laisynth});
Ophiuchus (\citealt{padgett08b}; \citealt{allensynth}); 
Lupus (\citealt{chapman07}; \citealt{merin08a}); 
and Cha II (\citealt{kyoung05}; \citealt{porras07}; \citealt{alcala08}; 
\citealt{spezzi07a}; \citealt{spezzi08}).
Numbers in those papers have been updated based on a consistent analysis
of the combined dataset.

None of these clouds are forming very massive stars; the most massive 
star in the close vicinity of a cloud mapped by c2d is HD147889, 
a B2 (III/IV) star lying within our map of
Ophiuchus. The $\sigma$ Sco group is farther away 
(projected distance  from the cloud of 4 pc), consisting of an O9V star and 
a B2III star and at least two other B stars \citep{pigulski92}. 
The star formation patterns in Ophiuchus may have been affected by
interaction with these stars (\citealt{loren89}, \citealt{nutter06}).
For Perseus, there are hints of an interaction with 40 Per, a B0.5
star in the Per OB Association, which is 26 pc from the L1451 core,
as discussed by \citet{walawender04} and \citet{kirk06}.
Serpens contains a Herbig Ae star, VV Ser, which produces
a large nebulosity (green object toward the south in Figure \ref{fig1})
discussed by \citet{pontoppidan07a} and \citet{pontoppidan07b}.
In addition, Serpens contain 3 other A-type stars and a B8 star
\citep{oliveira08}.
The most massive star in the Cha~II cloud is the F-type
star DK Cha \citep{hughes92}; all the rest have spectral type 
later than K2 \citep{spezzi08}.
The Lupus clouds are in the vicinity of the 
Scorpius-Centaurus association, which flanks the Lupus clouds at a
distance of about 5 degrees, or 17 pc at a distance of 200 pc. 
The sub-groups of the association are called Upper Scorpius (5-6 Myrs) 
and Upper Centaurus-Lupus (14 Myrs) \citep{degeus89}. 
The strong high-energy radiation and supernova remnants from those OB stars
might have had  an important role in the formation and 
evolution of the Lupus clouds \citep{tachihara01}. Specificially,
the highly fragmented cloud structure is thought to be related to the
effects of the nearby OB association.

There are distance uncertainties for all five clouds, which are discussed
in the detailed papers on those clouds. Here we supply our adopted
uncertainties and propagate those into the areas and masses.
Recent parallax observations of radio
emission from young stars in Ophiuchus yield a distance of $120\pm4.5$ pc
\citep{loinard08}. 
By combining extinction maps with parallaxes from Hipparcos and Tycho,
\citet{mamajek08} found a distance of $135\pm8$ pc, while
\citet{lombardi08} have derived a distance of $119\pm6$ pc.
These are all consistent with our assumed distance  and uncertainty 
of $125\pm25$ pc.
Because it is not clear that the whole cloud is at the same distance
(\citealt{loinard08}, \citealt{lombardi08}),
we retain our value and uncertainty in the analysis. A recent astrometric
measurement of water masers in NGC1333, in Perseus, provides a distance
of $235\pm18$ pc \citep{hirota08}. 
Distance estimates for IC348 are larger, so we
retain our standard distance and uncertainty of $250\pm50$ for the 
Perseus cloud as a whole.
\citet{lombardi08} also derived a distance to the Lupus complex of
$155\pm8$ pc, with some evidence that individual Lupus clouds were
at different distances, both consistent with our assumed values of
$150\pm20$ pc for Lupus I and IV and $200\pm20$ pc for Lupus III.
We use a distance to Serpens of $260\pm10$ pc \citep{straizys96}, 
as was used in all
our previous papers on this cloud, but we note that there is some
recent evidence for a distance of $230\pm20$ pc \citep{eiroa08}.

Seeking a common measure of the turbulence of these clouds, we chose
to give the FWHM linewidth of the \coo\ \jj10\ line, ideally averaged over
a full map of the cloud. This was not available in all cases, so
we have resorted to accepting statements like ``a typical
linewidth is ..." \citep{hara99} for Lupus. 
For Perseus, Serpens, and Ophiuchus, J. E. Pineda (pers. comm. 2008) has 
provided the mean and median linewidths for the ensemble of spectra in the
COMPLETE maps \citep{ridge06}, and the average and difference of the
median and mean are given in Table \ref{tab1} as the value and uncertainty.
\citet{tothill08} have provided the same kind of information for the
\jj21\ line of \coo\ in Lupus, based on maps with AST/RO. The linewidths
are based on dividing the integrated intensity by the peak intensity
and correcting for the channel width of 0.91 \kms, assuming Gaussian
lines. The results are substantially larger than the estimates from
the \coo\ \jj10\ lines. For Lupus I, 
$\mean{\dv} = 2.17\pm0.05$;
for Lupus III,
$\mean{\dv} = 2.11\pm0.05$;
and for Lupus IV
$\mean{\dv} = 1.53\pm0.03$.
It would be valuable to have more consistent and sophisticated measures
of the turbulence in these clouds.

The cloud mass (gas and dust)
has been derived from extinction maps made from our c2d data,
together with 2MASS data, toward background stars \citep{evans07}. 
We used the extinction law with $R_V = 5.5$ \citep{weingartner01}, which
reproduces reasonably well the data for molecular clouds 
(\citealt{flaherty07}; \citealt{chapman09}), to determine \av.
The mass was then calculated using the relation 
$N_H/\av = (1.086 C_{ext}(V))^{-1}$ \citep{draine03} and the value
of $C_{ext}(V)$ from the on-line tables\footnote{Available at 
http://www.astro.princeton.edu/~draine/dust/dust.html}
for $R_V = 5.5$.  This grain model results in a conversion from
extinction to hydrogen column density of 1.37\ee{21} cm$^{-2}$ mag$^{-1}$, 
instead of the usual 1.87\ee{21}, established for diffuse ISM gas 
\citep{bohlin78}. In our earlier papers (\citealt{harveysynth}, 
\citealt{alcala08}, \citealt{chapman07}, \citealt{merin08a}), the
value for the diffuse gas was used, resulting in cloud  masses overestimated
by a factor of 1.4 relative to those in Table \ref{tab1}.

The mass includes all cloud mass above an extinction contour with $\av = 2$ mag,
with no assumptions needed about geometry. 
The mass derived from our extinction maps refers to the same area covered in
our surveys, making it ideal for calculations of efficiency, etc.
The two exceptions to this statement are Serpens, for which our survey
was designed to cover completely down to $\av = 6$, and Ophiuchus, for which
we tried to cover completely down to $\av = 3$. The coverages were 
based on extinction maps by \citet{cambresy99}, but some areas down to 
$\av = 2$ were covered incompletely. 
The mass inside the $\av = 6$ contour for Serpens is
1532 \msun, 76\% of the mass within $\av = 2$. The mass inside the $\av = 3$
contour for Ophiuchus is 1914 \msun, 88\% of the mass within $\av = 2$.
We use the value for $\av = 2$ in what follows, but the difference will 
not be large for these clouds.
Table \ref{tab1} also gives the cloud crossing time in Myr, calculated from
the area and the mean speed of turbulent motions, approximated by 
$\mean{v} = 0.68 \Delta v$, with $\Delta v$ the \coo\ \jj10\ linewidth.

With the sensitivity of \spitzer, we can detect sources with luminosity
as low as \eten{-3} \lsun\ at 350 pc, 
but contamination by other kinds of sources ultimately limits
our sensitivity, as discussed below.
A similar study of the Taurus cloud has been done by \citet{padgett08a},
and other clouds in the Gould Belt are being mapped with the same techniques
\citep{allenpasp}. When those studies are completed, the analysis in this
paper can be extended to cover most star formation in large clouds within
about 300 pc of the Sun.

\section{Observations}\label{obs}

The observations here were all obtained by the c2d project or by
GTO observations that we have included in our data. They have
been described in the publications given in \S \ref{sample}.
Here, we give a brief summary of the data used for this paper; a more
complete description can be found in \citet{evans07}.

The \spitzer\ instruments were used to obtain data from 3.6 to 160 \micron,
but the 160 \micron\ data are limited by saturation, incomplete coverage,
and the large beam. Photometry at 160 \micron\ is not available as a standard 
product in the c2d catalogs. We provide and use some limited photometry 
at 160 \micron\ here. 
Data from the four 
\irac\ bands (3.6, 4.5, 5.6, and 8.0 \micron) and data from the 
2 \micron\ All Sky Survey (2MASS) \citep{cutri03} were merged using a
2\arcsec\ matching radius. These sources were merged with photometry
from \mips\ at 24 \micron\ (4\arcsec\ radius). The 70 \micron\ data were
merged with an 8\arcsec\ radius, but human judgment was used when multiple
candidates for merging were available at the shorter wavelengths.
The merging of \irac\ and \mips\ data could be done only for the
area covered by both \irac\ and \mips\ observations, and larger areas were 
usually covered by \mips\ observations. Since we use data from both
instruments to separate YSOs from other sources, we focus here 
on the areas with both \irac\ and \mips\ data.

For one cloud (Cha II), optical photometry has also been published 
\citep{spezzi07a}. Also for Cha II and Serpens, we have reasonably 
complete spectral type information (\citealt{spezzi08} and 
\citealt{oliveira08}). For three clouds (Perseus,
Ophiuchus, and Serpens), complete maps of dust continuum emission at 1.1 mm
were made using Bolocam on the Caltech Submillimeter Observatory. These
have been published (\citealt{enoch06}, \citealt{young06}, \citealt{enoch07})
and the Bolocam data have been combined with \spitzer\ data 
(\citealt{enoch08a}, \citealt{enoch08b}). We will draw on those results here.
We also have less sensitive maps of 1.3 mm emission for Cha II 
\citep{kyoung05} and maps of selected regions at 350 \micron\
\citep{wu07}. Maps of emission from the \jj10\ transitions of CO 
and \coo\ were obtained by the COMPLETE project for Perseus, Ophiuchus, and
Serpens \citep{ridge06}. Maps of the \jj21\ line of \coo\ were obtained
for Lupus and Cha II by \citet{tothill08}.

\section{Results}\label{results}

\begin{figure}
\includegraphics[width=9.5in,angle=90]{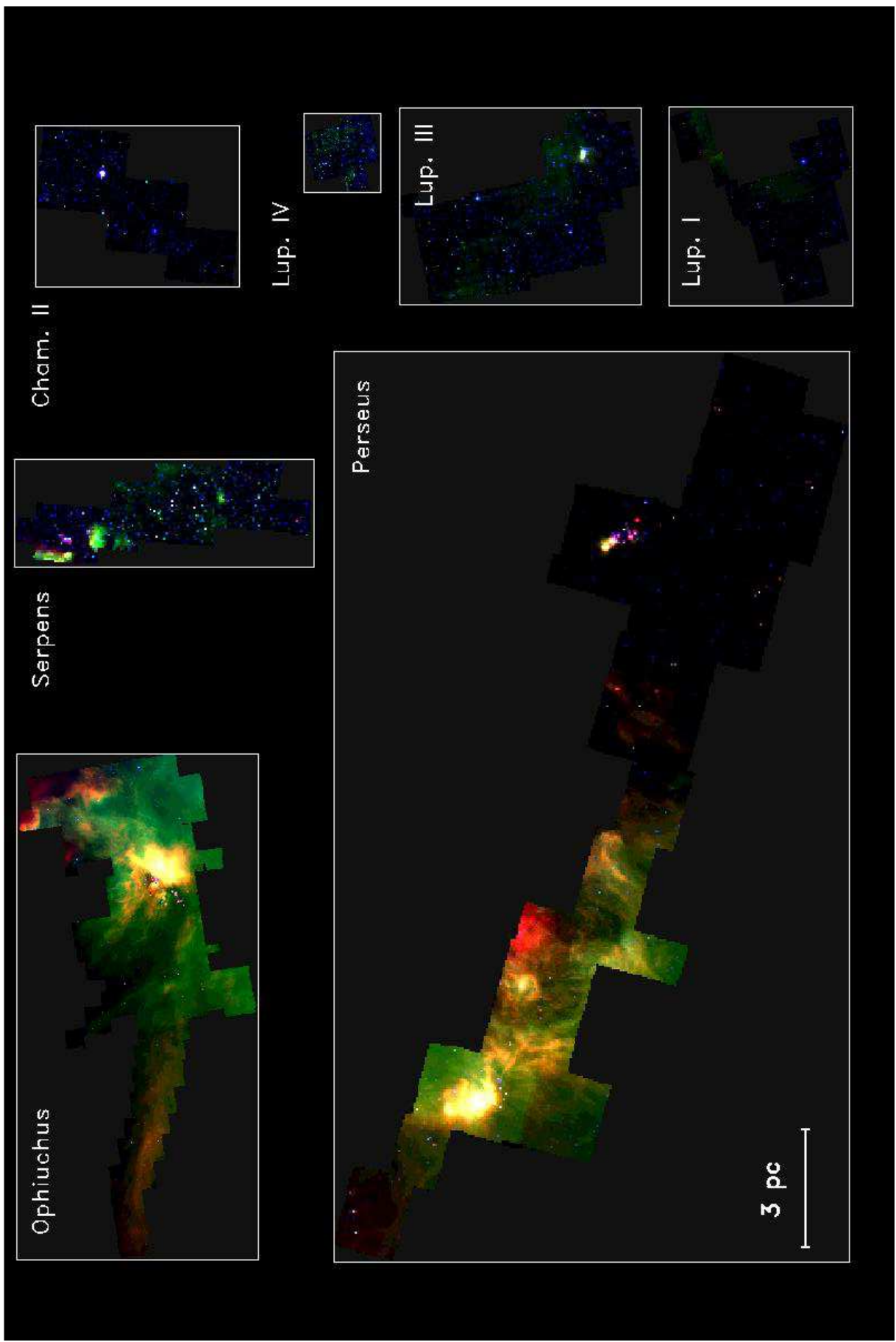}
\figcaption{\label{fig1}
Images of all of the clouds on the same absolute size scale. The color
code is blue (4.5 \micron), green (8.0 \micron), and red (24 \micron).
}
\end{figure}

The fundamental results for the large clouds
are summarized in Figures \ref{fig1} and \ref{fig2} and 
Tables \ref{tab1} to \ref{tabclusters}. 

Figure \ref{fig1} presents three color images of all the clouds on the
same physical scale, with a 3 pc scale bar. Figure \ref{fig2} shows
the extinction in grayscale, with all clouds on the same physical
scale and extinction grayscale. All the identified YSOs are shown,
color-coded by their SED class (\S \ref{intro:classes}). The total area
covered is dominated by Perseus, Ophiuchus, and Serpens.

Tables \ref{tabysolist1} to \ref{tabysoprops} are described in the Appendix.
They are electronic tables providing a full list of YSOs in the five
clouds, based on an updated, uniform analysis of the entire sample,
using the final data release from the c2d project\footnote{Available
at http://ssc.spitzer.caltech.edu/legacy/c2dhistory.html with a thorough
description \citep{evans07}}. The information in the tables in this paper 
is a condensed version of that found in the full c2d catalogs, along with
supplementary information.
Briefly, the c2d catalogs provide flux densities, uncertainties, and
various flags for each wavelength from 3.6 to 70 \micron, similar information
for the 2MASS sources that match our sources, a source type, and a spectral 
index.  Since the source type plays an important role in discussing
contamination, we briefly summarize the nomenclature here.

Only sources with detections in at least three bands could be classified at all.
If an object was consistent with a (possibly reddened) stellar photosphere, 
it was labeled ``star".  Candidates for YSOs required detections in all four IRAC bands
and MIPS-1; those meeting stringent criteria (see \ref{vermin})
are labeled YSOc, with the ``c" emphasizing that they are only 
{\bf candidates}.
They may have one or more appended suffixes, such as ``red", ``PAH\_em", 
or ``star$+$dust". Sources labeled ``red" have flux densities at 24 \micron\
that are at least three times the flux density from the nearest available
\irac\ band. Sources labeled ``PAH\_em" have colors indicative of a peak in the
8 \micron\ band. The ``star$+$dust" designation indicates that
the SED is consistent with that of a stellar photosphere for wavelengths
shorter than a particular band but an excess in that band of at least
3 $\sigma$. This band was appended to make a full name such as
``YSOc\_star$+$dust(MP1)" where MP1 implies the first \mips\ band, 
at 24 \micron. As discussed below, some sources lacked photometry
at enough wavelengths to be classified as YSOc, but were later added
as YSOs based on other information. These may have source types in Table
\ref{tabysoprops} of ``rising", in which the available flux densities 
rise significantly to longer wavelengths (15 cases), or ``red1", in which 
the source is detected only in \irac-4 or \mips-1 (3 cases).

In this paper, we refine and standardize the analysis of the combined
data for the five large clouds, discussing issues of contamination by
background sources (\S \ref{vermin}) and completeness (\S \ref{completeness}).
After describing the standard classification tools provided in the catalogs,
we describe analysis that goes beyond that provided in the catalog.
We discuss extinction corrections (\S \ref{classification})
that can be be made to the flux densities
and the calculation of bolometric luminosity (\lbol) and bolometric
temperature (\tbol). We present in the Appendix the list of YSOs, 
including flux densities from 2MASS ($1.25-2.17$ \um) and \spitzer\ 
($3.6-70$ \um).
Additional flux densities from observations with other 
telescopes, at wavelengths ranging from 0.36 \um\ to 1.3 mm, are 
presented in Tables \ref{tabysolist1}, \ref{tabysolist2}, \ref{tabysolist3}, 
and \ref{tabysolist4}.  
This list of YSOs is improved over the list of YSOc supplied with our
delivery because we have removed some suspect sources, added known sources, 
added data at other wavelengths, and calculated additional quantities, 
provided in Table \ref{tabysoprops}. We describe this process in the
following subsections.

\subsection{Contamination by Other Sources}\label{vermin}

The main challenge is to separate YSOs
from contaminants. By far, most contaminants
are stars, mostly background, without infrared excess. 
These have colors very close to zero in a color-color
diagram using the IRAC bands \citep{allen04} and are easily removed.
The primary remaining contaminants are then
background galaxies with active star formation; these may have colors
very similar to those of embedded young objects, so we must use
magnitude information as well as colors. 

The automated criteria described by \citet{harveysynth} were used for all
five clouds to produce catalogs of YSO {\it candidates} 
(YSOc), which are part of our final data delivery. 
Compared to earlier catalogs, advances in distinguishing 
YSOs from background galaxies \citep{harveysynth} have provided much cleaner 
samples.  These criteria assign
to each source with sufficient information an unnormalized probability
($P(Galc)$) that the source is a galaxy.
The calculation of the probability is complex
and fully described in \citet{harveysynth} and \citet{evans07}, so we summarize it
here. The probability depends on the location of the source in three color-magnitude
diagrams, whether the source is extended, and whether the source is detected above
a threshold at 70 \micron. The assignment of probabilities is entirely empirical, based
on excluding objects in a suitable sample of extragalactic observations.

By comparing to a set of sources from the SWIRE survey of
the ELAIS N1 extragalactic field \citep{surace04}, 
suitably processed to simulate the c2d sensitivity
and the extinction distribution of each cloud, we define a threshold
in $P(Galc)$ below which a source becomes a YSOc. For the catalogs of the
five large clouds, the aggregate plot is shown in Figure \ref{galprob}.
As was true for the individual clouds, most YSOc separate nicely from
galaxy candidates (Galc), but there are some sources in a lower probability
tail from the big peak in the Galc sector that are ambiguous. 
Thus, confusion with star-forming galaxies can be a problem for 
both contamination and completeness.
Sources with log$P(Galc) < -1.47$ were assigned the YSOc moniker in the c2d
catalogs, based on the analysis of Serpens by \citet{harveysynth}.  
From the surface density of objects from the degraded
SWIRE data that would be misclassified as YSOc for each cloud, multiplied
by the surface area of the cloud, we estimate that there could be as
many as 51 contaminating galaxies in our total 15.5 square degrees.
That would still be a small fraction of the total of 1086 YSOc (see below).
We next describe attempts to minimize the number of contaminants.

Certain categories of sources were checked by eye
for all the clouds in an attempt to eliminate residual contaminants. Typically
these were sources that had been ``band-filled" at 24 \micron\ or sources
with peculiar SEDs. Sources that were well detected at some wavelengths
received flux estimates at other wavelengths, based on the known position,
in the process of band-filling.  Some of these flux estimates turned out to be 
emission from the wings of nearby bright
sources and other artifacts. In addition, objects with the ``PAH\_em"
suffix received extra scrutiny because this feature is common in star-forming 
galaxies. Objects that had galactic morphologies were removed from the
final YSO list; however, many star-forming galaxies are point-like to \spitzer. A total of 91 YSOc were eliminated by this process.  
We do not know how many point-like galaxies were removed by 
this process, but the numbers are consistent with removing most of them.

While stars without infrared excess are readily removed from our
sample via color criteria, post-main-sequence stars with circumstellar
shells can masquerade as YSOs. \citet{harveysynth} were able to
identify and remove 4 of these behind the Serpens cloud. Follow-up
optical spectroscopy toward Serpens \citep{oliveira08} have provided 
spectral types for 78 objects, 58 of which had been identified as YSOs.
Based on their analysis, we rejected 11 of those 58 sources (about 20\%) as 
background giants with infrared excesses \citep{oliveira08}.  
All 11 are classified as either Class II or Class III sources, 
with 9 out of 11 (about 80\%) classified as Class III.  
We have removed these 11 sources from the final sample.  We so far
lack the data required to identify and remove background giants in Lupus, 
Perseus, and Ophiuchus.  However, Serpens should be the worst case because it
lies at low galactic latitude and longitude. 
Lupus and Ophiuchus may also be affected, though less so than Serpens.
Additionally, a spectroscopic
study of Cha~II found that 96\% of the YSOs identified in our catalogs
were true cloud members \citep{spezzi08}.  These points, combined with 
the fact that most of the background giants (80\%) are Class III sources, 
to which we are incomplete anyway (\S \ref{completeness}), 
lead us to conclude that contamination 
by background giants will not significantly affect our main results.

Adding these 11 background giants in Serpens to the 
91 sources removed as described above gives a total of 102 sources 
removed from the initial sample of 1086 YSOc (Table \ref{catstats}).  

\subsection{Completeness}\label{completeness}

Since the vast majority of sources in the catalogs are background stars,
we require an infrared excess for a source to be 
considered a candidate YSO. Consequently, our sample is missing 
Pre-Main-Sequence Stars (PMS) that no longer have infrared excess 
but may have H$\alpha$ emission, X-ray emission, etc. 
Complete surveys for such objects are
needed to complete the sample of PMS objects, so we concentrate on YSOs,
{\bf defined here to have an infrared excess}.  More complete samples of PMS objects
exist for some clouds that have the needed complementary data, but not
for all. For example, \citet{alcala08} found 51 certain and 
62 likely PMS stars in Cha II, more than double the number (26) of YSOs.
The ratio of PMS to YSOs in our catalog is 2 for Class II sources and
4.8 for Class III sources.
Cha~II may be a worst case because many of the PMS stars are
not in the region studied by c2d. Restricting attention to that
region, only 4 PMS stars are added to the YSOs found by c2d.
A similar study in Lupus would add 19 PMS stars without infrared excess to
the 94 YSOs \citep{merin08a}.

While objects of very low luminosity may still be lost among the galaxy
background, we estimate that our YSO sample is 90\% complete down to a 
luminosity integrated from 1 to 30 \micron\ of 0.05 \lsun\ and 50\%
complete down to 0.01 \lsun\ \citep{harveysynth} for our most distant cloud 
(Serpens at 260 pc, which also has the highest density of background stars).  
For comparison, the survey of Serpens with \iso\
reached a limit of 0.08 \lsun, but had only two bands and covered
only 0.13 sq. deg. \citep{kaas04}. They found 61 YSOs compared to
our 227, mostly because of their smaller areal coverage.
The 1-30 \micron\ wavelength range covers the bulk of emission 
from YSOs with infrared excesses arising from circumstellar disks; thus these 
completeness limits are a good proxy for our completeness to such objects.  
A 2 Myr old object at the mass boundary between stars
and brown dwarfs has a luminosity of about 0.01 \lsun. Indeed some
objects believed to be brown dwarfs with disks based on more complete analysis
do not make our YSOc list, either because their fluxes are not of high
enough quality or because they have log$P(Galc) \geq -1.47$.  
For example, 2 of the 4 added PMS objects in Cha~II are very low mass brown
dwarfs with excesses \citep{allers07}.

For younger YSOs still embedded within their dense cores, 
this $1-30$ \micron\ wavelength range is less appropriate 
since the bulk of the emission is reprocessed by the surrounding envelope to 
the far-infrared.  A separate search for embedded objects with luminosities 
less than 1 \lsun\ found that we reach a similar completeness limit for 
the internal luminosity of these objects 
(\lint\ of $4 \times 10^{-3}$ \lsun\ at 140 pc, 
or $\sim 0.014$ \lsun\ at 260 pc) \citep{dunham08}, where this limit is 
set by the sensitivity of the 70 \micron\ \mips-2\ observations.
The internal luminosity measures the contribution from the embedded source,
after correction for the effects of heating by the interstellar radiation
field.

We require good quality detections in all 4 \irac\ bands and \mips-1
for a source to be classified as a YSOc. This requirement can result in
our missing two types of sources: deeply embedded sources that were not 
detected in all bands and strong sources that saturated the detectors.  
Note that the latter can include both embedded objects and more evolved 
YSOs no longer embedded within their dense cores.  
A total of 40 sources that are clearly YSOs but which did not make the 
automatically-generated YSOc list for one of the two above reasons were 
added by hand to our final sample, bringing the final sample size to \ysotot.  

Of these 40 sources, 36 were added by comparison to the 
searches for embedded objects presented by \citet{dunham08}, 
\citet{enoch08a}, \citet{joergensen07}, and \citet{joergensen08}.  
Two were added by being well-known YSOs that were saturated in one or more 
of the \spitzer\ bands but not included in the samples of embedded objects 
compiled by the above authors.  For all saturated sources, data from other 
telescopes were substituted for the saturated \spitzer\ data.
Three of the added sources actually have source types of Galc, one each in 
Lupus, Perseus, and Serpens. The source in Lupus was added by
\citet{merin08a} because it lies essentially on the border between 
YSOc and Galc, and it is candidate to be a brown dwarf with a 
disk \citep{allers06}.
The source in Perseus was added because it is contained in the sample 
of embedded objects compiled by \citet{joergensen07}.   
The Serpens source is associated with an outflow.

Turning the luminosity completeness limit into a limit on stellar mass
requires further analysis. For the early, embedded stages, the luminosity
depends on the product of stellar mass and mass accretion rate. As discussed
in \S \ref{lumevol}, accretion at the mean rate expected in a Shu-type model
would predict a luminosity of 1.6 \lsun\ for a central object of 0.08 \msun.
Our luminosity limit of 0.014 \lsun\ would then translate into a mass limit of
7\ee{-4} \msun. However, there is strong evidence for highly variable
accretion rates (\S \ref{lumevol}), so this limit is highly suspect. 
For the later
stages when accretion luminosity is negligible, spectral types and accurate
extinction corrections are necessary to determine masses. For the most
distant cloud (Serpens), a main-sequence luminosity of 0.05 \lsun\
implies a mass of 0.08 \msun, while 0.01 \lsun\ correponds to 0.04 \msun\
\citep{chabrier2000}. At young ages, the mass limits would be lower.
In Serpens, the masses determined by \citet{oliveira08}
from spectral types range from 0.2 to 3.0 \msun. 
In Lupus, masses of YSOs are complete down to 
0.1 \msun\ \citep{merin08a}, and in Cha~II, masses extend well below 
the stellar limit to 0.015 \msun\ \citep{spezzi08}. We do clearly miss some
substellar objects with disks, as discussed above. Until spectroscopy
is available for a larger fraction of the young objects, we can only
estimate our mass completeness to be near the stellar/substellar
boundary.

\subsection{Classification}\label{classification}

For each source with sufficient data, the c2d catalogs provide 
a least squares fit to all photometry between 2 \micron\ 
and 24 \micron\ in the c2d catalog to determine the spectral index $\alpha$: 
\begin{equation}
\alpha = {dlog(\lambda S(\lambda)) \over dlog(\lambda)}
\end{equation}
where $\lambda$ is the wavelength and $S(\lambda$) is the flux density
at that
wavelength. These were used to classify objects into the four classes
defined by \citet{greene94}, as described in detail in \S \ref{intro:classes}:
Class I, Flat, Class II, and Class III.

An alternative classification scheme, used especially for more embedded
objects, employs the bolometric temperature (\tbol).  We will also need the
bolometric luminosity (\lbol) for later analysis. These were computed, using
both \spitzer\ data and auxiliary data at both shorter and longer wavelengths,
by calculating the first two moments of the SED \citep{chen95} using 
integration methods described in more detail by \citet{dunham08}.
Uncertainties in \lbol\ and \tbol\ are dominated for early phases
by incomplete sampling of the SED at wavelengths from 70 to 1000 \micron;
they were estimated to be 20 to 60\% (\citealt{enoch08a}, \citealt{dunham08}).
The poor sampling at \fir\ wavelengths arises in part because of the large
beam of \spitzer, causing confusion in crowded regions, and from the lack
of data at 350 \micron.
A recent analysis of the Serpens B cluster using the fine-scale 70 \micron\
mode and adding 350 \micron\ data \citep{harvey09} was able to separate
sources that were confused in the c2d data.
They find that the values of \lbol\ and \tbol\ change by about the  amount
estimated above. Nonetheless, only one source out of 18 would change 
classification (from Class I to Class 0).

For some of the analysis, it is desirable to correct the flux densities
for extinction. Based only on the c2d data, such corrections would be
highly uncertain.  Optical follow-up studies have provided spectral types 
for the Class II and Class III populations in Chamaeleon II 
(\citealt{spezzi07a}; \citealt{spezzi08}), Lupus 
(F. Comer\'{o}n et al., in preparation; \citealt{merin08a}), and part 
of Serpens (\citealt{oliveira08}).  For Ophiuchus and Perseus, along with 
the remainder of Serpens, we assume a spectral type of K7 for the Class II 
and Class III sources.  We then used the spectral type (known or assumed),
the near-infrared photometry (J, H, and K in order of preference), 
and the \citet{weingartner01} extinction law for $R_V = 5.5$ to correct 
the photometry for extinction.  The extinction law was chosen to match 
that used for determining our extinction maps and cloud masses 
(\S \ref{sample}).  For Class I and Flat objects, we assume the mean 
extinction to all the Class II objects in the same cloud.
The idea is that this correction removes
foreground extinction, but not local extinction from the surrounding
envelope, which will be reradiated in the far-infrared. The mean extinctions
to Class II sources for the five clouds are 
$\av = 3.95$ mag in Cha~II, 
$\av = 2.91$ mag in Lupus, 
$\av = 5.92$ mag in Perseus, 
$\av = 9.57$ mag in Serpens, and
$\av = 9.76$ mag in Ophiuchus. 
Extinction values calculated for Class II and III sources without spectral 
type information should be regarded as highly uncertain.  Extinction 
values for Class I and Flat sources should be regarded as averages for each 
cloud only and not as actual extinctions towards each object.

After these extinction corrections were made, the values of spectral index,
\tbol, and \lbol\ were recomputed and these values are distinguished from
the observed values by primes (\aprime, \tbolprime, and \lbolprime).
Both sets of values are given, along with the value of \av\ used for the
extinction correction, in Table \ref{tabysoprops}. When wavelength
coverage is insufficient to obtain reliable values for \tbolprime\
and \lbolprime, a flag in the table indicates this and appropriate
upper or lower limits are given (see the Appendix for more details).

\subsection{Overall Statistics}\label{sourcestats}

Combining all the clouds, we have the following statistics based on the
final c2d data release. The full catalogs contain a total of 4.26\ee6 entries,
of which 6.14\ee5  are also in the high reliability catalog, which requires
detection in at least one band with S/N $\geq 7$ and a second band, if
available at that position, with S/N $\geq 5$. A total of 6.77\ee5
have detections in at least three bands and can be further classified.
This group provides the parent sample from which our YSOc are drawn.
The great majority are classified as stars (3.32\ee5) or
other (2.55\ee5). The ``other" category contains sources that do not fit
any particular template; the great majority are probably galaxies. Of those
that remain, 2,965 are galaxy candidates and 1,086 are YSO candidates.
The statistics for individual clouds are given in Table \ref{catstats}.

The processes described in \S \ref{vermin} and \S \ref{completeness}
resulted in the list of YSOs used in this paper. There are undoubtedly
still some contaminants in this list, but we believe that they are
a small fraction.  Overall the ratio of YSOs to YSOc is 0.94.
The ratio ranges from 0.85 in the Lupus clouds to 0.99 in
Perseus. In both Perseus and Ophiuchus, the fraction is high because
of the addition of deeply embedded and saturated sources not in the 
original YSOc list.

In the end, there are a total of \ysotot\ YSOs in our sample.
This number represents an order of magnitude increase over previous samples,
and they have been selected in a uniform way from data with very similar
sensitivity.

\section{Star Formation Efficiencies and Rates}\label{sfesfr}

We discuss these topics first because they rely only on the counts of YSOs and
the masses of the clouds obtained from the extinction maps. As such,
the conclusions do not depend on further classification or the choice
of whether or not corrections for extinction are made.

In Table \ref{sfrtab}, we list the number of YSOs in each cloud, 
along with the number per solid angle and the number per area (pc$^{-2}$).
The number of YSOs per area
is clearly highest in Serpens at 14 pc$^{-2}$ with Ophiuchus second and
Perseus, Lupus, and Cha~II somewhat similar. The latter point is a bit
surprising because Perseus has many more YSOs, but it also covers the
largest area (see Table \ref{tab1} and Figure \ref{fig1}). 
While Cha~II is low on YSOs,
it also has the smallest area. The distribution of YSOs in Perseus 
\citep{laisynth} reveals a large, central section of the cloud with almost
no YSOs (see Fig. \ref{fig2}). 
Together with other hints that Perseus may be two overlapping
clouds [see discussion in \citet{enoch06}], 
this distribution suggests that smaller areas for each piece of
Perseus could provide a more relevant comparison.

We also estimate the star formation rate from the number of YSOs by
assuming a mean mass of 0.5 \msun\ and a period of 2 Myr for
star formation. 
As discussed in \S \ref{lifetimes}, the assumption of 2 Myr
is the estimate of the time taken to pass through the Class II SED class,
the latest class to which our study is reasonably complete. We assume a 
mean mass of 0.5 \msun, consistent with studies of the IMF 
(\citealt{chabrier03}, \citealt{kroupa02}, \citealt{ninkovic06}).
There may be variations from cloud to cloud, though small number statistics
are an issue.  In Cha~II, \citet{spezzi08} derive a mean mass of 
$0.52\pm0.11$ \msun\ based on spectroscopic data.
The mean stellar mass may be closer to 0.2 \msun\ in the Lupus clouds
\citep{merin08a}. The mean mass of the YSOs in Serpens is 0.69 to 0.73
\msun, depending on which evolutionary tracks are used \citep{oliveira08}. 

Perseus has the highest star formation rate of 96 \msunmyr, 
slightly higher than the rates for Ophiuchus and Serpens, while Cha~II 
is conspicuously low at 6.5 \msunmyr. If normalized by cloud area, however, the
differences are less striking (see Table \ref{sfrtab}), ranging from 
0.65 \msunmyrpc\ in Cha~II to 3.4 \msunmyrpc\ in Serpens. The rate per
area in Serpens may be higher in part because we covered completely only the
part of the cloud with $\av \geq 6$ mag, rather than 2 mag for the
other clouds.

The number counts of YSOs do not include any corrections for unresolved 
binaries. If
a fraction $f$ of all YSOs are unresolved binaries, the actual number of 
forming stars and the star formation rates should be multiplied by the factor
$(1+f)$, if we ignore even higher multiplicity. Estimates for $f$ range
from 0.3 \citep{lada06a} to $\geq 0.5$ \citep{mathieu94}. Given that
the total stellar mass is always much less than the cloud mass, the star 
formation efficiency would be increased by about the same factor.

\subsection{Comparison to Predictions from Kennicutt Relations}\label{exgal}

Given the mass surface density of the cloud from our extinction maps, 
one can predict the star formation rate surface density from the relations 
employed for other galaxies using the formula from \citet{kennicutt98}. 
\begin{eqnarray}
\Sigma(SF)({\rm \msun\ yr^{-1} kpc^{-2}})  = & & \nonumber\\
 (2.5\pm0.7) \times 10^{-4} (\Sigma(gas)/1 \msun {\rm pc^{-2}})^{1.4\pm0.15} & &
\end{eqnarray}
Taking all clouds together, the surface density is 64 $\msun {\rm pc^{-2}}$,
and the predicted star formation rate would
be $0.08$ \msun\ Myr$^{-1}$ pc$^{-2}$, a factor of 20 below the observed 
value of 1.6. As shown in Figure \ref{kennplot},
all clouds lie well above the prediction of Equation 2, even though
these clouds are forming only low mass stars, which would be largely
invisible to the kinds of tracers, such as H$\alpha$ emission, used to
establish Equation 2.

This difference is not surprising, but it reminds us that
the Kennicutt relation applies to averages over much larger regions than
individual clouds. The coefficient and exponent in such relations 
are clearly functions of the mean density of the region being averaged over.
A census of all star formation within 500 pc should come from
the completion of many \spitzer\ observations; a first crude estimate, along
with information on the surface density of gas, gave rough agreement with
the predictions of Kennicutt's relation \citep{evans08}. 
Note that 85\% of the gas in this region is atomic hydrogen.
\citet{blitz06} give a more detailed discussion of these points. 

Studies of star formation in dense gas have shown a {\bf linear} relation
between star formation rate and the amount of gas traced by HCN emission
(\citealt{gao2004a}; \citealt{gao2004b}; \citealt{wu2005}). As discussed
below, we find that star formation is sharply confined to the dense gas,
so it is two steps removed from the scales on which Equation 2 applies:
the formation of molecular clouds from atomic gas; and the formation of
dense cores from molecular clouds \citep{evans08}. At which of these steps does
the non-linearity (an exponent that is greater than unity) of Equation 2 enter? 

To address that question, studies that resolve individual clouds in other
galaxies are needed. Studies of M51 with 0.5 to 2 kpc resolution continue
to show a non-linear relation with exponent around 1.4 \citep{kennicutt07}.  
Recent observations 
with sub-kpc resolution of both molecular and atomic gas in other nearby 
galaxies \citep{bigiel08} reveal a linear relation between star formation rate
and {\bf molecular} gas density over a range from 3 to 50 $\msun {\rm pc^{-2}}$.

\begin{eqnarray}
\Sigma(SF)({\rm \msun\ yr^{-1} kpc^{-2}})  = & & \nonumber\\
  10^{-2.1\pm 0.2} (\Sigma(H_2)/10 \msun {\rm pc^{-2}})^{1.0\pm0.2} & &
\end{eqnarray}

(Note that this relation is normalized to a different surface density.)
This relation is also shown in Figure \ref{kennplot} over the range
where it was established and extrapolated with a dotted line. 
Equation 3 would predict 0.051 ${\rm \msun\ yr^{-1} kpc^{-2}}$, much
less than our observed value of 1.6. 
\citet{bigiel08} note that they are still measuring the filling
factor of clouds rather than resolving structure within molecular clouds.
This comparison suggests that clouds with the properties we study are
filling only about 3\% of their beams.
As resolution improves further, and as more studies become available 
within our own galaxy, especially of regions forming high mass stars,
further comparisons will be important.

We also plot the relation found for dense gas, as traced both in other
galaxies and in massive dense cores in our Galaxy by HCN, by \citet{wu2005}.
This relation extrapolated to lower surface densities comes closest to
our observed points.

Following \citet{elmegreen02}, a linear relation in molecular gas would
correspond to a threshold core density for star formation of \eten{5} \cmv, 
in good agreement
with typical densities in the dense cores. However, he also predicts that
the fraction of total gas in dense cores would be \eten{-4}, whereas
the fraction in the three clouds with the relevant data is 4.6\%. 
Allowing
for the fact that 85\% of the gas in the local kpc is not molecular gets
the fraction down to 6\ee{-3}, still on the high side.
The non-linear relation that seems to apply on larger scales was interpreted
by \citet{elmegreen02} to mean that the critical criterion is a ratio of
core density to mean ISM density of \eten{5}.
Caution is still warranted as the extragalactic relations need further
analysis, and other factors, such as shear and bars, will affect star 
formation, especially in circum-nuclear regions \citep[e.g.,][]{jogee05}.

\subsection{Star Formation Efficiency}\label{efficiency}

Comparison of the mass in YSOs to the cloud mass gives a measure
of the current efficiency. Since we are only sensitive to YSOs with infrared
excess, this efficiency represents an average over the last 2 Myr. 
We give the star formation efficiency, defined as

\begin{equation}
SFE = {\mstar\over\mstar + \mcloud},
\end{equation}
in Table \ref{tabeff}. The values range from 3\% to
6\% over the various clouds, with a value of 4.8\% taking all clouds
together. Given the star formation rate (\sfr) from 
Table \ref{sfrtab}, we can compute a depletion time for the cloud:
\begin{equation}
\tdep = \mcloud/\sfr.
\end{equation}

As shown in Table \ref{tabeff}, these $\tdep$ are 30 to 66 Myr.
Estimates of cloud lifetimes range from 10 to 40 Myr \citep{mckee07}. 
If clouds produce
stars at the current rates for 10 Myr, the final efficiency when star 
formation has ended in these clouds could be as high as 15\% to 30\%; 
as discussed in \S \ref{ratesdense}, achieving such high efficiences 
would require continued conversion of cloud material into dense cores.

\subsection{The Speed of Star Formation}\label{speed}

\citet{krumholz07} have emphasized that star formation, even in
molecular clouds, is {\it slow} in the sense that the star formation rate
{\it per unit free fall time} is low. \citet{krumholz05} define the star formation rate
per unit free fall time (\sfrff ) to be the fraction of an object's mass that turns into
stars in a free fall time at the object's mean density.
Using our cloud masses from the
extinction maps covering the same areas where we surveyed for YSOs,
we can quantify this, again where we mean star formation over the last
2 Myr since we identify only the sources with infrared excess.
Then we have

\begin{equation}
\sfrff = \sfr \tff / \mcloud = \tff/\tdep,
\end{equation}
where $\tff$ is the free fall time for the mean density of the cloud,
calculated from
\begin{equation}
\tff = 34 {\rm Myr}/\sqrt{\mean{n}}.
\end{equation}

The density \mean{n} in that calculation is the density of all particles,
and we have assumed a mean molecular mass of 2.3 amu. Mean densities
for the clouds, computed from the mass and surface area by assuming
spherical clouds, are given in Table \ref{tab1}. The average over
all clouds is $\mean{n} =  390$ \cmv.
Then, we find $\sfrff$ ranging from 0.028 to 0.064, with an average
over all clouds of 0.040. This value is on the high side of the range of
\sfrff\ inferred from more global considerations by \citet{krumholz07}
and the theory of \citet{krumholz05}.

\subsection{Efficiencies and Rates in Dense Gas}\label{ratesdense}

Figure \ref{fig2} shows a strong correlation of YSOs, especially
the younger ones, with regions of high extinction. For the clouds
with millimeter continuum maps, a strong correlation of the youngest
objects with dense cores identified in the continuum maps 
is apparent \citep{enoch07}. As discussed
by \citet{enoch07}, the cores identifed in this way have minimum
mean densities of about 2\ee4 \cmv\ and more typically more than
\eten5 \cmv, and almost all seem to be gravitationally bound \citep{enoch08b}. 
Their mean densities are 50 to 200 times the mean density of the cloud;
they thus represent quite distinct entities, rather than just modest
peaks in column density.
Thus, it is also interesting to calculate the efficiencies 
and rates using the total mass of gas in dense cores for the three 
clouds with Bolocam data, using the data from \citet{enoch07}. 
The total mass in YSOs in the cloud is quite comparable
to the mass in dense cores (Table \ref{tabeff}); taking all three clouds
together the ratio is 1.3. We compute the time to deplete the 
current stock of
mass in discrete dense cores by dividing the SFR by the mass in
dense cores. We find 0.6 Myr for Ophiuchus, 1.6 Myr for Serpens,
and 2.9 Myr for Perseus.  Taking all clouds together, $\tdepdense =
1.8$ Myr. These times are consistent with the 2 Myr timespan for
detectability of YSOs that we use to calculate lifetimes and with
plausible spreads of formation times in clusters. While star formation
is faster and more efficient in the dense gas probed by Bolocam, it is
still slow compared to a free fall time. Using a mean density of
5\ee4 \cmv\ to calculate the free fall time for the typical core leads to 
$\sfrffdense = 0.05$ for Perseus to 0.25 for Ophiuchus. 

These values for \sfrffdense\ are 
somewhat higher than \sfrff\ for the whole cloud, but they neglect the
fact that the mass of stars is now comparable to the mass of dense gas.
The depletion times also assume that all the gas winds up in stars once
it reaches the density of the cores identified by \citet{enoch07}.
We can make a second calculation in which we assume that star formation
began at some time in the past with a larger reservoir of dense gas and
that a fraction $\epsilon$ of the dense core mass winds up in a star, while
a fraction $1-\epsilon$ is lost to the star forming region by outflows.
By comparing the core mass function to the initial mass function of stars,
\citet{alves07} estimated that $\epsilon = 0.3$ and \citet{enoch08b} finds
$\epsilon > 0.25$. At any given time, $\tdep = \mdense/\sfr$ and the
mass of dense gas decreases exponentially, with a time constant of 
$\epsilon \tdep$. At time $t$ after the start of star formation,
the expression for \tdep\ becomes,

\begin{equation}
\tdep = {t\over\epsilon ln(1+\eta/\epsilon)},
\end{equation}
where $\eta = \mstar/\mdense$ at time $t$. As usual, we take $t = 2$ Myr
as the timescale over which we have complete statistics, assume
$\epsilon = 0.3$, and use the observed value of $\eta$.
This expression then yields values for \tdep\ that are more like 3-6 Myr
and \sfrffdense\ that are 0.03 to 0.06, comparable
to those for the cloud as a whole, as \citet{krumholz07} would predict.

\section{Classes and Lifetimes: The Standard Analysis}\label{lifetimes}

\subsection{Historical Classification Methods}\label{intro:classes}

The current working model for star formation arose in the 1980s with
the merger \citep{als87} of an empirical classification scheme, 
based on the slope
of the spectral energy distribution (SED) between 2 and 20 \micron\ 
\citep{ladawilking84}, with a theoretical picture of star formation involving 
the collapse of an isolated rotating dense core \citep{terebey84} forming
a star and disk \citep{adams85}.
The essential stages of the theoretical model were three: the collapse
of the envelope, forming the protostar and disk; the continued accretion
of disk material onto the forming star; and the dissipation of the
disk by planet formation, evaporation, etc. 
Each of these stages became identified with an empirical SED class as
follows: the envelope collapse as Class I; the accretion disk and star 
as Class II; and the dissipation of the disk during Class III. With
further study, other significant events could be distinguished theoretically,
and further refinements of the classification system were suggested.
We follow the suggestion of \citet{robitaille06} in referring to the
physical arrangement as a ``Stage" and the SED characteristic as a 
``Class".

We focus now on the evolution of the empirical classification system.
Following the recognition by \citet{ladawilking84} that SEDs were falling
into natural groups, \citet{lada87} first codified the tripartite class system
using the spectral index.
The original boundaries were as follows:
\begin{description}
\item[I] $0 < \alpha \lteq 3$
\item[II] $-2 \lteq \alpha \leq 0$
\item[III] $-3 < \alpha \lteq -2$
\end{description}
Note that Lada and coworkers consistently use the symbol $a$, but we have
standardized on $\alpha$.
These were supplemented by descriptions; for example, Class III sources 
could have some ``mid-infrared excess", but ``no or little excess near-infrared
emission." Class II and III objects were all visible, but Class I objects were
all invisible for $\lambda < 1$ \micron. 
Class II sources were identified as T Tauri stars.
It is worth noting that this system and many of the later
elaborations were based on study of the cluster of sources in the L1688
(Ophiuchus) cloud.

By 1989, \citet{wly89} had noted that some SEDs had ``double humps", one
around 1 \micron\ and one in the far-infrared. 
These were put into subclasses, labeled II-D and III-D.
Class II-D sources were T Tauri stars with far-infrared excesses and were
thought to be intermediate between Classes I and II.
Class III-D sources looked like reddened photospheres plus a far-infrared
excess \citep{lada91}. By comparing the numbers of objects in different 
classes, \citet{wly89} also assigned relative lifetimes to the different 
classes (discussed below).

The next major development was the introduction of a class ``before"
Class I by \citet{andreetal93}. Since zero was not part of the original
Roman number system, the Arabic symbol for zero was used. 
Since the Class 0 objects could not then be observed at the wavelengths
originally used for classification, \citet{andre00}
listed three criteria for a Class 0 source:
\begin{enumerate}
\item Indirect evidence for a central YSO, as indicated by, e.g., the detection
of a compact centimeter radio continuum source, a collimated CO outflow,
or an internal heating source;
\item Centrally peaked but extended submillimeter continuum emission
tracing the presence of a spheroidal circumstellar dust envelope
(as opposed to just a disk);
\item High ratio of submillimeter to bolometric luminosity, suggesting that
the envelope mass exceeds the central stellar mass: $\lsmm/\lbol > 0.5$\%,
where \lsmm\ is measured longward of 350 \micron. In practice, this often
means an SED resembling a single-temperature blackbody at $T \sim 15-30$K.
\end{enumerate}
The boundary between Class 0 and Class I was associated with a new
distinction between physical stages \citep{andreetal93}:
the point at which the masses of the protostar and the remaining envelope
were about equal.

The last major development came in 1994, when \citet{greene94}
formalized the 4-class system,\footnote{The class of sources with 
$\alpha = 0.3$ was left undefined in \citet{greene94}, so we have 
arbitrarily assigned such sources to Class I.}, 
again based on the L1688 cluster, as follows:

\begin{description}
\item[I] $0.3 \leq \alpha $
\item[Flat] $-0.3 \leq \alpha <0.3$
\item[II] $-1.6 \leq \alpha < -0.3$
\item[III] $\alpha < -1.6$
\end{description}
\citet{greene94} described the evolutionary status of sources in the
new Flat Class as ``uncertain", but \citet{calvet94} showed that these
could be interpreted as infalling envelopes.
\citet{greene94} noted that they found no sources with $-1.9 < \alpha < -1.6$ or
$ -0.3 < \alpha < 0$. These classifications were based on the slope between
2 and 10 \micron, though they claimed that comparison to classes based
on slopes between 2 and 20 \micron\ showed that this did not matter.
\citet{greene94} did not list Class 0 in their list, but they did acknowledge
\citet{andremontmerle94}, who pointed out that millimeter wavelength
emission  became much weaker for $\alpha < -1.5$, as a rationale for the 
revised boundary between Classes II and III.

Improving sensitivity at millimeter wavelengths led to the detection of
some starless cores \citep{ward-thompson94}, and these were 
called ``pre-protostellar cores" (PPCs) or later ``prestellar cores". 
To fit them into the class 
system, one might use Class $-1$, but \citet{boss95} argued that they should
instead be Class $-2$ to make room for a theoretically important event,
the formation of the first (molecular) hydrostatic core, which would then
be called Class $-1$. These terms have not caught on, and the recent
definitive review by \citet{difrancesco07} simply distinguishes ``prestellar"
cores among the larger set of starless cores as being gravitationally bound.

With the increasingly baroque nature of the class system, it was natural 
to seek a continuous variable to capture the transitions. By analogy with
the effective temperature of stars and spectral classes, \citet{myersladd93}
suggested the use of a bolometric temperature, defined as the temperature
of a blackbody with the same flux-weighted mean frequency as the actual SED,
considering all wavelengths with available data. \citet{chen95}
showed that the traditional classes, including 0, could be associated with
certain ranges of \tbol, as follows. 
\begin{description}
\item[0] $\tbol < 70 $
\item[I] $70 \leq \tbol \leq 650$
\item[II] $650 < \tbol \leq 2800$
\end{description}

For very early stages, data at short wavelengths
were lacking, and \tbol\ could be quite sensitive to whether or not such
data existed. Detections at shorter wavelengths could increase \tbol\ enough
to move a source from Class I to II. Orientation of aspherical envelopes
and disks can also affect \tbol\ substantially.
At the other wavelength extreme, improved availability of submillimeter data
considerably increased the number of Class 0 sources (\citealt{visser02},
\citealt{young03}).

For distinguishing Class 0 from Class I, the ratio of luminosities
($\lbol / \lsmm$) may be most useful. In this scheme, a ratio exceeding 200, 
the inverse of criterion 3 of \citet{andre00}, marks the transition 
from 0 to I. This form of the ratio has the virtue of increasing with
the evolutionary progression, as does \tbol, and \citet{youngevans05}
found that this ratio was a more robust indicator than \tbol\  of the ratio of
mass in the star to mass in the envelope, based on models of collapsing
cores. Both \citet{visser02} and \citet{young03} found that quite a few
objects would be classified as 0 by the ratio, but as I by \tbol.
A drawback to using $\lbol / \lsmm$ is the lack of complete data
at 350 \micron, the shortest wavelength included in \lsmm; we lack such
data for many of our the sources in this study.
Clearly, conclusions about evolution will depend on establishing a
clear connection of such parameters to physically relevant stages of
evolution.

It remains difficult to capture the
increasing amount of information in any single parameter.
With the greatly increased  and more uniform data set available from
the c2d project, we will reevaluate the different tracers to learn
what, if any, well defined criteria can trace evolution through
all stages. The distinction between physical stages of evolution and
SED classes, however defined, has been usefully emphasized recently by
\citet{robitaille06} and \citet{robitaille07}, who presented large grids
of models SEDs from 2-D radiative transfer calculations, following earlier
work by \citet{whitney03}, which demonstrated the importance of inclination
effects. Similarly,
\citet{crapsi08} have discussed the possibility of confusion between stages
and classes in the context of a grid of 2-D models.

\subsection{Lifetimes: Previous Estimates} \label{times}

One of the goals of research into star formation is to constrain the
lifetimes associated with different stages of evolution. As noted above,
\citet{wly89} used the numbers of objects in various classes to 
establish relative lifetimes. This method assumes that the census is
complete and that star formation in the sample has been a continuous 
process at a steady rate for longer than the dwell time in the slowest
phase considered. 
When extrapolated from a single region, it also assumes that other 
variables are irrelevant, such as the total mass available, 
the conditions in the region, such as turbulence, etc.

With these caveats in mind, \citet{wly89} used their data on the Ophiuchus
cluster to suggest lifetimes. 
They estimated the ages of the Class II sources, finding an average of
$0.39 \pm 0.17$ Myr, with the oldest at about 1.5 Myr.
They found roughly equal numbers of Class I and Class II sources, 
suggesting equal lifetimes.  With some other considerations, 
they suggested a duration for the Class I phase of 0.2 to 0.4 Myr.
Later, \citet{greene94} found that Class I plus Flat SED lifetimes
in Ophiuchus were 75\% of the Class II lifetime, which they took to be 0.4 Myr.
A somewhat older population in Ophiuchus, with ages around 2 Myr,
was later established by \citet{wilking05}.
 
\citet{kh95} analyzed data from the Taurus cloud and found about
ten times as many Class II plus Class III sources as Class I sources, 
very different from the situation in Ophiuchus. They found
a similar number of Flat SED sources and suggested lifetimes of 0.1 to 0.2
Myr for each of the Class I and Flat classes, based on a duration for
Class II plus Class III
of 1 to 2 Myr, which they found from comparison to evolutionary tracks. 

More recent analyses of the Perseus cloud find a lifetime for the Class
0/I SED of 0.25 to 0.67 Myr within 95\% confidence level \citep{hatchell07}.
These estimates include \spitzer\ data on embedded sources and the
analysis of the age spread  of the IC348 cluster from \citet{muench07}.
Early estimates of 0.01 Myr for the lifetime of the Class 0 phase were based 
on the small number of Class 0 sources in Ophiuchus \citep{andreetal93}. 
Because the first examples of Class 0 sources found were quite luminous with
powerful, collimated outflows, the short lifetime suggested a phase of 
rapid accretion in which at least half of the final stellar mass was 
accreted \citep{andre00}. 
 
Most of these estimates were based on samples of 50 to 100 sources and in
particular clouds. They are subject to small number statistics and
to possible differences from cloud to cloud, especially for the earlier
classes.  An exception was a compilation of 95 Class 0/I objects 
\citep{froebrich05}. The positions of these sources in \tbol-\lbol\ plane
was compared to grids of models, and \citet{froebrich06} favored a Class 0
lifetime of 0.02 to 0.06 Myr.

\subsection{Numbers and Lifetimes: The c2d Large Cloud Sample} \label{c2dtimes}

Table \ref{tabclasses} presents the numbers of YSOs in each \citet{greene94} 
class as determined by 2MASS - \spitzer\ 24 \micron\ photometry.
They are also shown graphically in Figure \ref{piecharts}.
We remind the reader that \citet{greene94} used photometry only out to 
10 \micron, but found no significant differences in $\alpha$ for sources 
with data out to 20 \micron.
We have not yet separated Class 0 sources from Class I sources in 
Table \ref{tabclasses}; this separation will be discussed in \S \ref{early}.
In addition, Class III is certainly missing sources, as discussed
in \S \ref{results}.  So at this point,
we restrict comparison to Class I, Flat, and Class II sources. 

With \ysotot\ YSOs and more than 100 in each
class, the uncertainties from small number statistics are decreased to less
than a 10\% effect in any class, almost certainly smaller than other
sources of uncertainty.
We find that \iitot\% of our YSOs are in Class II, with \itot\% in 
Class I, \ftot\% in Flat, and \iiitot\% in Class III. 
The results with classes defined by \aprime\ (after extinction corrections)
are also shown in Table \ref{tabclasses} for each cloud and summarized
in Table \ref{tabdefs}. 
The main effect of extinction corrections is to decrease the number of
Class I (to 14\%) and Flat (to 9\%), while increasing the number of
Class II (to 64\%) and Class III sources (but still at 13\%). 

With or without extinction corrections, our result is quite unlike that of 
\citet{wly89} who found roughly equal numbers of Class I and Class II
objects in Ophiuchus. We find many more Class II objects, as also
found by \citet{wilking05}. The variations from cloud to cloud are
substantial, as can be judged from Figure \ref{piecharts}, though
the effects of small number statistics are substantial for Cha~II and
Lupus (see Table \ref{tabclasses}). Among the three clouds with substantial
numbers of sources in early classes, Perseus stands out as particularly
rich in Class I sources.

If star formation has been continuous over a period longer than the
age of Class II sources and if we can average all our clouds, we can
obtain relative lifetime estimates for each phase by taking the ratios of
number counts in each class, and multiplying by the lifetime for
Class II. Recall that this method assumes that
all objects of all masses behave identically and that no other variables
enter importantly. Since these assumptions are not very realistic, caution
is needed in interpreting these, or any previous, estimates for lifetimes.

The lifetime of the Class II phase is still uncertain. 
In the classic study of 83 cTTS in Taurus, \citet{strom89} found
that 60\% of those with ages less than 3 Myr had K-band excesses.
The sample of cTTS would probably be biased towards those with long-lived
disks. More recently, \citet{haisch01} studied young clusters of different ages
and showed that half the stars had lost
their disks, as indicated by an L-band excess, in $\leq 3$ Myr.
A study of disk frequency in wTTs \citep{cieza07} shows that half have
lost their disks within about 1 Myr; since the wTTS sample is undoubtedly
biased towards stars that lose disks early, this is probably a lower
limit for the disk half-life.  The median age of stars
in IC348, assuming constant formation rate and the distance we
have adopted, is  3.0 Myr \citep{muench07}. The ratio
of wTTS to cTTS in IC348 is 1.5, and somewhere between 30\% \citep{lada06b}
and 22\% \citep{cieza06} of the wTTs have disks. However many of these
would not be Class II objects. These numbers suggest an upper limit
of 3 Myr for the Class II phase.
For Cha~II, \citet{spezzi08} find a mean age for the YSOs of about 2 Myr.
Both \citet{bontemps01} and \citet{kaas04} also assumed 2 Myr in their
analysis of ISO data.
Based on all these pieces of information, it seems that the best estimate
for the lifetime of the Class II phase is $2\pm1$ Myr. The uncertainty
is dominated by uncertainties in stellar ages, which are uncertain
by factors of 2 at these early times [e.g., \citet{haisch01}, \citet{hillenbrand08}].
Since the marker used to define the transition
was that half the stars lacked infrared excess, this lifetime may be best
thought of as a half-life rather than a lifetime which is the same for
all objects [see also \citet{hillenbrand08}]. 
In particular, there is some evidence for longer lifetimes
for infrared excesses around lower mass stars or brown
dwarfs \citep{allers07}.

If we take 2 Myr for the duration of the Class II phase, then the lifetime
of the Class I phase, $t(I) = $\ti\ Myr, with an additional $t(F) = $\tf\ Myr 
for the Flat SED phase. Using the numbers after extinction corrections,
we obtain $t^{\prime}(I) = $\tiprime\ Myr and $t^{\prime}(F) = $\tfprime\ Myr.
With or without extinction corrections, these estimates are 
substantially longer than some recent estimates 
for the Class I lifetime of 0.1-0.2 Myr \citep{greene94},
but this estimate depends on the assumption of an age for the Class II phase.
Interestingly, our estimate agrees better with the original estimate of 0.39 Myr
for the Class I stage by \citet{wly89} because they assumed an age of
0.4 Myr for the Class II stage in the L1688 cluster and found equal
numbers. So we obtain the same answer but with totally different data
and assumptions. 

The uncertainties in these numbers are clearly not dominated by small 
number statistics any longer. While the original definition of classes did
not include extinction corrections, it appears that an approximate correction
leads to 10\% and 20\% shorter lifetimes for the Flat and Class I phases.
The limitations of the underlying assumptions dominate the uncertainties. 
In particular, the continuous formation assumption
is dubious. If star formation in a particular region decreased or stopped
within the last 2 Myr, we would underestimate the lifetime for the Class I
and Flat classes. Conversely, if star formation began or increased within
the last 2 Myr, we would overestimate those lifetimes. Evidence for these
effects can be seen in the lifetimes calculated cloud by cloud in Table
\ref{tabclasses}. For Class I, these vary from $t(I) = 0.2$ Myr (Cha~II and Lupus)
to $t(I) = 0.77$ Myr in Perseus, which supplies over half the Class I
objects and nearly 40\% of all YSOs in our sample. Since Perseus clearly
had significant star formation in the IC348 region 3 Myr ago, it does not
obviously violate the continuous formation assumption, but there may have
been a recent burst. If we take an extreme position and leave Perseus out
of the average, $\mean{t(I)} = 0.40$ Myr.

Our only defense against these issues is to hope that
averaging over all the clouds provides some cancellation of biases.
Since we selected our clouds to have a range of known
star formation activity, the sample should not be too heavily biased, but it
does contain three of the most active nearby regions. Once the data from 
the Taurus cloud and the Gould Belt Survey are available in the same 
form as the c2d data, these numbers should be recomputed.

One possible systematic bias towards overestimating Class I lifetimes 
would arise if low mass objects are found in our sample when they are
young and accreting but fall into the galaxy confusion region during
the Class II stage. We start to lose Class II sources around the 
stellar/substellar
boundary, based on a few examples. If we assume that we miss all
substellar objects, which is too extreme, we would miss 10\% to 30\%
of Class II sources (e.g., \citealt{andersen06}, \citealt{luhman07}), 
which would introduce a comparable percentage error in the Class I lifetime.

\section{Alternative Classifications}\label{karlmarx}

In this section, we raise some questions about the use of the class system
defined by \citet{greene94} to categorize YSOs.
First we retain $\alpha$ as the discriminant, and discuss uncertainties
attendant on the choice of wavelengths and method of calculating $\alpha$. 
We then discuss alternatives for very early and very late stages.

\subsection{Definition of Classes with $\alpha$}\label{definitions}

The original definition \citep{ladawilking84} used wavelengths between
2 and 20 \micron\ to define $\alpha$. We use a least-squared fit to any
data in Table \ref{tabysolist1} between 2 and 24 \micron. 
More recently, \citet{lada06b}
have used a fit to only the 4 \irac\ bands (3.6 to 8.0 \micron) in a study
of disks in IC348. How much difference does the choice of wavelength
range make? We classified sources both ways (Table \ref{tabdefs}).
Using only \irac\ bands moves sources from earlier to later classes in 
such a way
that we get more Class III (by 47\%), fewer Class I (by 8\%), fewer 
Flat SEDs (by 15\%), 
and fewer Class II (by 8\%). Reddening will have a larger effect at shorter
wavelengths, which biases $\alpha$ upward, producing
``Flat" SEDs that are clearly reddened Class II sources when the full SED is
examined. Thus, different definitions of $\alpha$ can result in 
10\% to 15\% difference in inferred lifetimes for the earlier SED phases.
We did another experiment, using sources in Cha~II, for which we have
spectral types \citep{spezzi08}, allowing photometry to be corrected for
extinction.  Indeed, the values of $\alpha$ calculated after extinction 
corrections were smaller,
with a mean difference of 0.2. This was not enough to affect the
classification of sources in Cha~II. With the approximate extinction
corrections used for the full data set, we find somewhat
larger effects, but still 10\% to 20\%. 
These effects are probably smaller than other sources of uncertainty, 
inherent in
assumptions such as the continuous formation assumption (\S \ref{times}).

\subsection{Definition of Classes with \tbol}\label{tbolclasses}

Values for \lbol\ and \tbol\ were calculated as described in 
\S \ref{classification} and are given in Table \ref{tabysoprops}.
The {\bf observed} values of \lbol\ and \tbol\ are easily
compared to most other data on embedded sources in the literature,
but
in the paper connecting \tbol\ to the classes defined by $\alpha$, 
\citet{chen95} corrected the observed flux densities for extinction 
before computing
\tbol. For less embedded objects with \av\ determinations from optical data,
they used those values. If a source had no measured \av\ but had a known
spectral type, they used the spectral type and $V-R$ to calculate \av.
If no spectral type was known, they assumed an M0 spectral type.
For sources with no optical data, they assumed an average $\av = 1$ for 
Taurus and off the core region in Ophiuchus, and $\av = 10$ for the core
sample in Ophiuchus. Our extinction corrections are similar to what they
assumed in Ophiuchus and substantially larger in other clouds (see \S
\ref{classification}).
They analyzed the effects of extinction; these could
be substantial in both \tbol\ and \lbol, primarily for the sources
with higher intrinsic \tbol, where most of the energy emerged at shorter 
wavelengths, and these sources generally had reasonably well determined
values for \av. They considered the uncertainties introduced by extinction
uncertainties and found a factor of two for high \tbol\ sources and 10\% for
low \tbol\ sources \citep{chen95}. 
Most subsequent application of \tbol\ 
has been to deeply embedded sources, so no corrections for extinction are
typically made; the energy absorbed at short wavelengths is assumed to be
re-radiated at longer wavelengths. 

Our method of extinction corrections attempts to capture the essence
of the \citet{chen95} approach. If we compare the numbers in classes
defined by \tbolprime, the effects are more dramatic than was the case for
classes defined by $\alpha$.
The observed fluxes produce only one Class III source with the \tbol\
dividing line from \citet{chen95}, while using \tbolprime\ give 485
Class III sources! Clearly, the boundary in \tbol\ between Class II and III
is a sensitive function of extinction.

Class 0 objects cannot be separated from Class I objects by using $\alpha$.
Previously this statement was true because Class 0 objects could not be
detected in the mid-infrared \citep{andre00}. With \irac\ on \spitzer, we
now detect many, though not all \citep{joergensen05}, Class 0 sources.
However, the resulting values of $\alpha$ from the c2d catalog
scatter widely and do not
correlate with other discriminants such as \tbol\ for the early
classes \citep{enoch08a}. We show the full sample of sources with
enough data to constrain both $\alpha$ and \tbol\ in Figure \ref{avstbol}.
While there is some tendency for Class 0 sources, as identified by \tbol,
to have larger values for $\alpha$ than Class I sources, 
the scatter precludes any Class 0
criterion based only on $\alpha$. These conclusions do not change
using the photometry after extinction corrections to calculate \tbol\ 
(right panel of figure).
The range of $\alpha$ may be due in part to geometry of the 
envelope-disk interface, the inner radius of the envelope \citep{joergensen05},
different amounts of scattered light, emission
from jets in the 4.5 \micron\ band, and deep ice features (see, e.g.,
\citealt{huard06}; \citealt{dunham06}; \citealt{bourke06};
\citealt{boogert08}). The fluxes do not 
increase monotonically from 3.6 to 24 \micron\ in some Class 0 sources. 

Conversely, there was no definition of the Flat SED class in terms of 
\tbol. Figure \ref{avstbol} shows a strong concentration around 650 K,
the Class I/II boundary defined by \citet{chen95}, with some outliers.
Histograms of the number of sources in linear bins of \tbol\ and \tbolprime\
are shown in Figure \ref{tbolhist}.
The different classes according to $\alpha$ and \aprime\ are color-coded.
The mean value of \tbol\ for Flat SED sources is 649 K.
We suggest a range of \tbol\ from 350 to 950 K for Flat SED sources;
this range includes 79\% of the Flat SED sources and contains only
23\% of the Class I sources (by $\alpha$) and 22\% of the Class II/III
sources. \citet{enoch08a} have also suggested a boundary between
Stage I and Stage II sources in the 400 to 500 K range.

The distribution of Flat SED sources in extinction-corrected 
\tbolprime\ is somewhat broader
and shifted to higher \tbolprime, with a mean value of 844 K
(right panel of Fig. \ref{avstbol}).
A range of 500 to 1450 K contains 77\% of the Flat SED sources,
while only 14\% of Class I and 18\% of Class III sources lie in this
range.  Figure \ref{tbolhist} suggests that
Flat SED sources do form a distinct class of objects.
However, this seeming homogeneity can in part be caused by observing
a flat SED over a restricted wavelength range; \tbol\ has little freedom
to vary without data at shorter and longer wavelengths. 

Class II and III sources have barely distinct peaks in Figure \ref{tbolhist}, 
but substantial overlap. This figure illustrates the sensitivity of the
Class II/III boundary to extinction corrections that was noted above.
It also suggests that \tbol\ is not particularly good at distinguishing
Class II from Class III objects and $\alpha$ is preferred.

\subsection{Source Types from c2d Catalogs}\label{sourcetypes}

We can also compare the source types in the c2d catalogs, as described
in \S \ref{results}, to classifications based on $\alpha$ or \tbol.
Figure \ref{map2vstbol} shows the source types versus \tbol, both with
and without extinction corrections, with the
color code showing the $\alpha$ SED class. The categories such as
YSOc\_red, red, and rising are almost exclusively Class 0/I or Flat.
The source type YSOc\_star+dust(IR1) contains a broad range of
classes; an excess starting at 3.6 \micron\ can continue to increase
(Class I), stay constant (Flat), or fall (Class II or Class III). 
When the excess begins at longer wavelengths, such as YSOc\_star+dust(MP1), 
almost all are Class III, with a few exceptions. The exceptions may
include the ``cold disks," such as those studied by \citet{brown07}.
The generic YSOc class contains a heterogeneous sample of classes.
Similar patterns are seen in Figure \ref{map2vsa}, which shows the
source type versus $\alpha$.

\subsection{Source Types by Color-Color Diagrams}\label{ccdiagrams}

Color-color diagrams constructed from \irac\ photometry \citep{allen04}
or from combined photometry from \irac\ and \mips\ \citep{muzerolle04}
have also proved useful in separating sources in different phases of
evolution. These are shown for our YSO data in Figures \ref{iraccc}
and \ref{iracmipscc}, both with data after extinction corrections in the
right panel. The color codes correspond to the classification by
$\alpha$ except that Class 0 and I sources are separated by \tbol.
The overall correspondence is good, with Class II sources heavily
concentrated in the expected locations in the two diagrams, as indicated
by the boxes, taken from \citet{allen04} and \citet{muzerolle04}.
Flat SED sources were not separated by \citet{allen04} or
\citet{muzerolle04}, but they tend to lie between the Class II and 
Class 0/I regions [see also \citet{allen07}]. 

The Class 0/I sources scatter more widely but 
generally have redder colors than the Flat SED sources. The Class 0 and I
sources are not clearly separated in these diagrams, similar to what we
found in Figure \ref{avstbol}, but on average, Class 0 sources
have redder colors. There are exceptions, such as Class 0 sources with 
[5.8]$-$[8.0] colors near zero. Also, there are some Class II sources
rather far from their home box, especially in [3.6]$-$[4.5], many of
which find their way home in the figure corrected for extinction.

Our large sample suggests that some adjustments to the box locations
and perhaps shapes would be advisable for better segregation of classes in 
the [3.6]$-$[5.8] versus [8.0]$-$[24] diagram. In particular, the divisions
between Class II, Flat, and Class 0/I may be captured better by diagonal
lines in Figure \ref{iracmipscc}. The bottom two panels of Figure
\ref{iracmipscc} show the divisions between the simulated SEDs for
sources in different stages \citep{robitaille06}, which do capture the
divisions seen in the data well, especially for the dereddened data.
This general congruence indicates that class definitions generally
agree with the physical stages, despite exceptions. We consider some
of the remaining issues and extend the analysis to classes not separated
in Figure \ref{iracmipscc} (prestellar, 0, and Flat) in the next
section.

\section{Connection to Physical Stages}\label{stages}

In this section, we discuss the relations between SED classes,
defined above and physically meaningful stages in the evolution.
We will estimate lifetimes for Class 0 and prestellar phases.

\subsection{The Early Years: Class 0 and Before}\label{early}

As discussed previously, distinguishing Class 0 from Class I sources
can be ambiguous without very well sampled SEDs through the submillimeter
region. Lacking data at 350 \micron\ for many of our sources, we will
use $\tbol = 70$ K to define the Class boundary \citep{chen95}.

To include still earlier phases, prestellar cores, we rely on 
surveys in the millimeter continuum. We have complete surveys at
1.1 mm with reasonably uniform sensitivity for Perseus, Serpens, and
Ophiuchus and a less sensitive survey of Cha II. Focusing on the first
3 clouds, \citet{enoch08b} report on a total of 201 cores (defined
as sources of millimeter continuum emission above some level). By
comparing these cores to the \spitzer\ data, \citet{enoch08a} separate
these into cores with and without central luminosity sources.
The latter are assigned to prestellar cores.  Cores with central
luminosity sources are divided into Class 0, I, and II using the boundaries
in \tbol\ listed in \S \ref{intro:classes}, based on \citet{chen95}.
There are only a handful of Class II sources associated with dense cores,
and these are likely due to projection effects.
The results are given in Table \ref{tabdefs}. 

To get lifetimes for these classes, we must bootstrap, using the 
lifetime derived above for Class I (now reinterpreted as the lifetime
for Class 0 and Class I combined). Using the same continuous formation
assumptions and a lifetime for Class 0/I of  \ti\ Myr, we would infer
a mean lifetime for Class 0 of \tzero\ Myr for all three clouds combined. 
If we use the same sample, but use  \tbolprime, calculated
after extinction corrections, the resulting $t^{\prime}(0) = $ \tzeroprime\ Myr.
The 60\% decrease in lifetime results partly from the shorter lifetime for
Class I, but primarily from the smaller number of Class 0 sources
after extinction corrections (Table \ref{tabdefs}).
Doing the calculation cloud by cloud, with the values of
$t(I)$ for that cloud,  yields a wide range,
from Ophiuchus with $t(0) = 0.043$ Myr to Perseus with 0.32 Myr. The
Perseus result is consistent with the results of \citet{hatchell07}.
While the Ophiuchus lifetime is 3 times higher than the original estimate 
of \citet{andreetal93}, it is much shorter than for the other clouds. 
In addition, it could be overestimated, as it rests on three Class 0 sources,
two of which are borderline Class 0/I sources.
Either Class 0 sources in Ophiuchus do evolve very quickly into Class I
sources, or the continuous formation approximation breaks down there.
A decrease in star formation rate beginning \tzero\ Myr ago would result
in very few Class 0 sources now. The variation from cloud to cloud
is less if the same value of $t(I)$ of 0.54 Myr is used for all clouds,
as in \citet{enoch08a}. Note that these results use the ratio of Class 0
to Class I sources with both required to have millimeter emission.

To get a lifetime for the prestellar ($t(pre)$) cores that are seen in the Bolocam
surveys, \citet{enoch08b} compared the number of cores that remain starless
to the number of embedded protostars. The latter number is slightly less
than the number of Class 0/I sources because a small number of
Class I objects do not have distinct Bolocam cores. Since this number
is small, we still use the lifetime
for the combined Class 0 and I phase to represent the lifetime of such
protostellar cores. The result, using all cores in all clouds, is 
$t(pre) =$ \tsl\ Myr. In this case, extinction corrections remove enough
Class 0/I sources to balance the decrease in Class I lifetime, so the
extinction-corrected result is $t^{\prime}(pre) = $\tslprime\ Myr. 
These numbers differ
very slightly from that (0.45 Myr) given by \citet{enoch08b} because
of small differences in the source list.
One must immediately caution that this is the lifetime for
cores remaining starless {\it after} they become dense enough to be
detected with Bolocam. Based on careful analysis of the instrumental
limitations, \citet{enoch07} concluded that the minimum mean density is
about 2\ee4 \cmv. The variation in $t(pre)$ from cloud to cloud is 
0.23 (Serpens) to 0.78 (Perseus) Myr if we use the $t(I)$ for each cloud; 
Ophiuchus is not unusual in this category with a 
value (0.37 Myr) near the mean. However the total {\it mass} of the cores in 
Ophiuchus is small; it may be a cloud that has just finished a burst of star 
formation that has exhausted much of its dense gas. Again, the range is
less if the same $t(I)$ is used for all clouds, as in \citet{enoch08b}.

Comparison to a separate study using SCUBA data can provide some idea
of the uncertainties in these estimates. \citet{joergensen07} compared
SCUBA data in Perseus to c2d infrared data, and \citet{joergensen08} have
extended this study to Ophiuchus. Compared to the Bolocam data, the 
SCUBA data are taken at a slightly
shorter wavelength (0.85 mm versus 1.1 mm), have a smaller beam (15\arcsec\
versus 30 \arcsec), and are more sensitive to low mass cores ($\sim 0.5$ \msun\
versus 0.8 \msun\ for Perseus), but they do not cover the clouds as completely.
The sensitivity to density is similar, but they may be probing objects
of somewhat lower mean density. Other differences exist in detail in the
methods of selecting sources, so comparison to the Bolocam
results will give a good idea of the current systematic uncertainties.
\citet{joergensen08} find more starless cores in Ophiuchus and fewer in 
Perseus than does \citet{enoch08b}. Calculating the lifetimes in the
same way, we find a prestellar core lifetime of 0.55 Myr for Perseus and
0.75 Myr for Ophiuchus. Taking the two clouds together yields a value
of \tsljj\ Myr, somewhat larger than the value of \tsl\ Myr,
obtained by \citet{enoch08a}.

Considering the two studies, a reasonable estimate for the lifetime of
prestellar cores above a mean density of about 2\ee4 \cmv\ would be
about $0.5\pm0.1 (\pm0.3)$ Myr, where the uncertainty in parentheses
includes the range over different clouds. These results are in general
agreement with a third analysis of Perseus by \citet{hatchell07}.

In comparing our results to previous results, summarized in Fig. 2 of
\citet{wardthompson07}, our lifetime would appear to be lower than
what they found if plotted at our minimum density of $n = 2\ee4$ \cmv.
However, most cores are denser than the minimum value, and 
\citet{enoch08b} have done a more careful analysis, dividing
the prestellar cores into two bins of mean density greater than or less
than $n = 2\ee5$ \cmv. The inferred lifetimes of 0.32 Myr and 
0.11 Myr are quite consistent with values found by \citet{kirk05} and
with the overall relation found by \citet{wardthompson07} indicating
lifetimes decreasing toward, but not yet reaching, the free-fall
time (\tff) as the
density increases, as originally suggested by \cite{jessop00}.
At this point, we should recall the assumptions required for these
estimates. In particular, we assume no dependence on core mass in
deriving a lifetime. \citet{clark07} suggested that lower mass cores
collapse more quickly, while \citet{hatchell08} note that, to the
contrary, a higher
fraction of the more massive cores in Perseus contain embedded objects,
suggesting faster evolution for more massive cores. The core mass distribution
for Bolocam cores with embedded sources is broader and flatter than the
prestellar core distribution as well \citep{enoch08b}, also suggesting faster
evolution for the more massive cores, though other explanations are
possible. The issues connecting the core mass function and the initial
mass function of stars are explored in a number of papers (\citealt{enoch08b},
\citealt{hatchell08}, \citealt{clark07}, \citealt{goodwin08},
\citealt{andre07}, and \citealt{swift08}).

Note that we have referred to these starless cores as prestellar cores
because
almost all with suitable molecular line data were found to be gravitationally
bound \citep{enoch08b} and hence worthy of the ``prestellar" moniker, 
as defined by \citet{difrancesco07} and \citet{wardthompson07}.

With the large sample of embedded objects now available, \citet{enoch08a}
have subdivided the Class 0 and Class I SEDs into early and late versions.
While the SED evolution appears to be continuous, suitable choices of
\tbol\ can produce distinctive SEDs in the mean. By counting the number
in each bin, they found a variation in \tbol\ that increases with time:
$\tbol = 25 {\rm K} + C(t/\eten5 yr)^{1.8}$, with $C = 51$ K,
though the evolution must slow down
to match the Class II lifetime.

\subsection{Mid-Life Crisis: Class I, Flat, and Class II}\label{midlife}

The boundary between Class I and II is also problematic.
As discussed in \S \ref{intro:classes},
we will use the terminology of ``stage" to describe the physical
configuration and ``class" to describe the SED. Models with 2-D 
radiative transfer by \citet{robitaille06} and \citet{crapsi08} show that
Stage II objects with large inclination angles can have Class I SEDs. 
Stage I and II objects occupy largely distinct regions of the various
color-color diagrams but there is also significant overlap in some parts
of these diagrams \citep{robitaille07}. Some Stage I objects viewed 
relatively face-on will be classified as Class II objects using popular
color-color diagrams such as Figure \ref{iraccc}.
Longer wavelengths are helpful in distinguishing these
situations; in particular, extended millimeter continuum emission or
tracers of dense gas, peaked on the source in either case,
unambiguously indicate the presence of an envelope.

Based on his grid of models, \citet{crapsi08} argues that 34\% of 
Stage II (star+disk)
sources are misclassified as earlier types because of inclination effects.
Potentially, all Flat spectrum sources could be nearly edge-on Stage II
objects \citep{crapsi08}. If correct, the effect on lifetimes of 
{\bf physical stages} could be
substantial; further comparison of the properties of Flat spectrum sources
to known edge-on disks is needed to evaluate this suggestion.
In addition, \citet{crapsi08} argues that half of the objects
with Class I SEDs could be part of the edge-on Stage II population. If
correct, the lifetime for envelope dissipation would be less than half that of
the combined lifetime for Class 0 and Class I derived above. 
\citet{crapsi08} considered a number of methods for determining the SED
class, including $\alpha$, \tbol, and the absorption or emission in the
silicate feature; all were subject to confusion by inclination effects.

The most reliable indicator of how much envelope remains is the ratio
of single dish to interferometer flux density at millimeter wavelengths;
these probe the envelope and disk, respectively \citep{crapsi08}. 
When such observations become available for the large number of 
Class I sources in our sample, Stage I lifetimes can be refined. 
For the three clouds with complete Bolocam data, we can already 
determine what fraction of objects we classify as Class I based on
$\alpha$ can be associated with millimeter continuum emission.
Overall, 65\% can be associated, larger than predicted by \citet{crapsi08}. 
If we assign all Flat SED class and 33\% of the Class I objects to Stage II,
the lifetime of Stage I would be 0.28 Myr.
Broken down by cloud, the percentage that lie
within one FWHM of a dense core is 69\% for Perseus, 75\% for Serpens, and
49\% for Ophiuchus. Again, these numbers differ slightly from those
in \citet{enoch08b}, owing to slight differences in Class I source
definition.  A random distribution of infrared sources would
produce about 5 more-evolved objects projected by chance on a millimeter
core for all three clouds taken together \citep{enoch08b}. 
A similar comparison for the SCUBA data finds
that 66\% and 53\% of Class I sources in Perseus and Ophiuchus lie within
30\arcsec\ of the center of a SCUBA core, and one false identification
is predicted among the 66 SCUBA cores in Ophiuchus \citep{joergensen08}.

The large fraction of the Class I sources without 1-mm emission 
in Ophiuchus may result from the 
generally high extinction in front of the YSOs in that cloud. As discussed
earlier, high extinction will increase the value of $\alpha$.
This problem has been explored by \citet{kempen08}, who mapped
objects in Ophiuchus that had been previously classified as Class I using
\hcop\ \jj43\ and \cooo\ \jj32\ emission. They found that half (11 of 22)
of the objects assigned to Class I by their c2d value of $\alpha$ 
had no \hcop\ emission peaked on the source. They found a similar
result (6 of 13) for the Flat SED sources.
They reclassified those objects as disks. The very high extinction toward
many sources in Ophiuchus had increased the value of $\alpha$ derived
from observations, by amounts as high as 1.1. 
As can be seen from Table \ref{tabclasses}, extinction corrections
decrease the number of Class I sources in Ophiuchus from 35 to 27, 
using $\alpha$.  There is a smaller decrease for Flat sources, from 47 to 44.
The core of Ophiuchus has even higher extinction than our assumed value
of $\av = 9.76$ mag. Calculating separately the extinction toward Class II
sources inside the contour of $\av = 15$ in the extinction maps, the mean
\av\ is 15.7 mag inside the contour and 5.7 mag outside. 
Using extinction corrections based on those numbers for
Class I and Flat SED sources inside or outside that contour,
the number of Class I sources is unchanged, but the number of Flat SED
sources drops to 28. Thus, very high extinction may substantially
inflate the number of Flat (by 68\%) SED and Class I (by 30\%) sources
as classified by observed $\alpha$.
Ophiuchus seems to be exceptional in this regard, as can be seen in
Figure \ref{extalpha}. There are many Flat SED sources in regions of
very high extinction, as measured by the cloud extinction maps. These
would deredden to Class II sources.

The objects classified as Class I by $\alpha$ but lacking dense gas tracers
tended to have characteristic SEDs which peaked at 24 \micron\ and fell
toward longer wavelengths. Similarly Class I sources without millimeter
continuum emission tended to have such SEDs (\citealt{enoch08a},
\citealt{dunham08}). This characteristic SED may be a way to identify
suspicious Class I sources. Another approach to sorting out these borderline
sources may be studies of veiling of absorption lines \citep{casali92};
a few sources in Ophiuchus without evidence of dense gas do have veilings
that are generally associated with Class I objects \citep{doppmann05}.

\subsection{Senior Citizens: A 2D Disk Classification System}\label{senior}

Finally, the simple division between Class II and Class III sources 
may be obsolete.
Note first that there is no minimum value for $\alpha$ in the 
\citet{greene94} scheme, as stars with no infrared excess, but other
evidence that they are pre-main-sequence, can be included in Class III.
Values as low as $\alpha = -3$ are possible for the Rayleigh-Jeans tail
of a hot star.
For most of our clouds, we lack a complete sample of such objects, because
we require a significant infrared excess to identify a source as a YSO, 
thus effectively setting a lower limit on $\alpha$. 
Thus our counts of Class III sources, on this definition, are incomplete.

A second point is that the classical idea that disk evolution leads
to a more negative value of $\alpha$ is based on all the excess fluxes
decreasing roughly together. In fact, we see a wide variety of SED
shapes around the Class II/III boundary, leading us to suggest a new,
two-dimensional classification system.

As discussed in \citet{cieza07} and in more detail by \citet{merin08b}, 
the diversity of SEDs for star and disk
systems can be captured better in a two-dimensional system.
Many stars show no excess below a certain wavelength. We
define $\lambda_{turn-off}$ as the last wavelength with no detectable
excess over a photosphere, as fitted to the photometry at shorter
wavelengths. Then we define $\alpha_{excess}$ as the slope of the SED
beyond $\lambda_{turn-off}$. Thus $\alpha_{excess}$ characterizes 
the properties of the excess emission, rather than mixing the stellar
and disk properties.

As shown by \citet{cieza07} and \citet{merin08b},
classical T Tauri stars have excesses that start short ($\lto \leq 2$ \micron)
and a tight distribution of \aex, around $-1$. In contrast weak-line
T Tauri stars with excesses usually have $\lto \geq 2$ \micron, with 
a wider range of \aex. The spread in \aex values grows with increasing
\lto\ until it covers the range from $-3$ to 1, similar to the
range seen for debris disks \citep{chen05}.

The increase in \lto\ values may be caused by several things, including
unresolved binaries, either stellar or substellar in mass, and 
inside-out photo-evaporation. The c2d sample includes a substantial 
number of ``cold" disks, with large \lto\ and positive \aex, but
they remain a small fraction of the sample, suggesting that the duration
of this phase is short. These issues will be explored by
\citet{merin08b}.

\section{Clustering}\label{clusters}

\begin{figure}
\includegraphics[width=9.5in,angle=90]{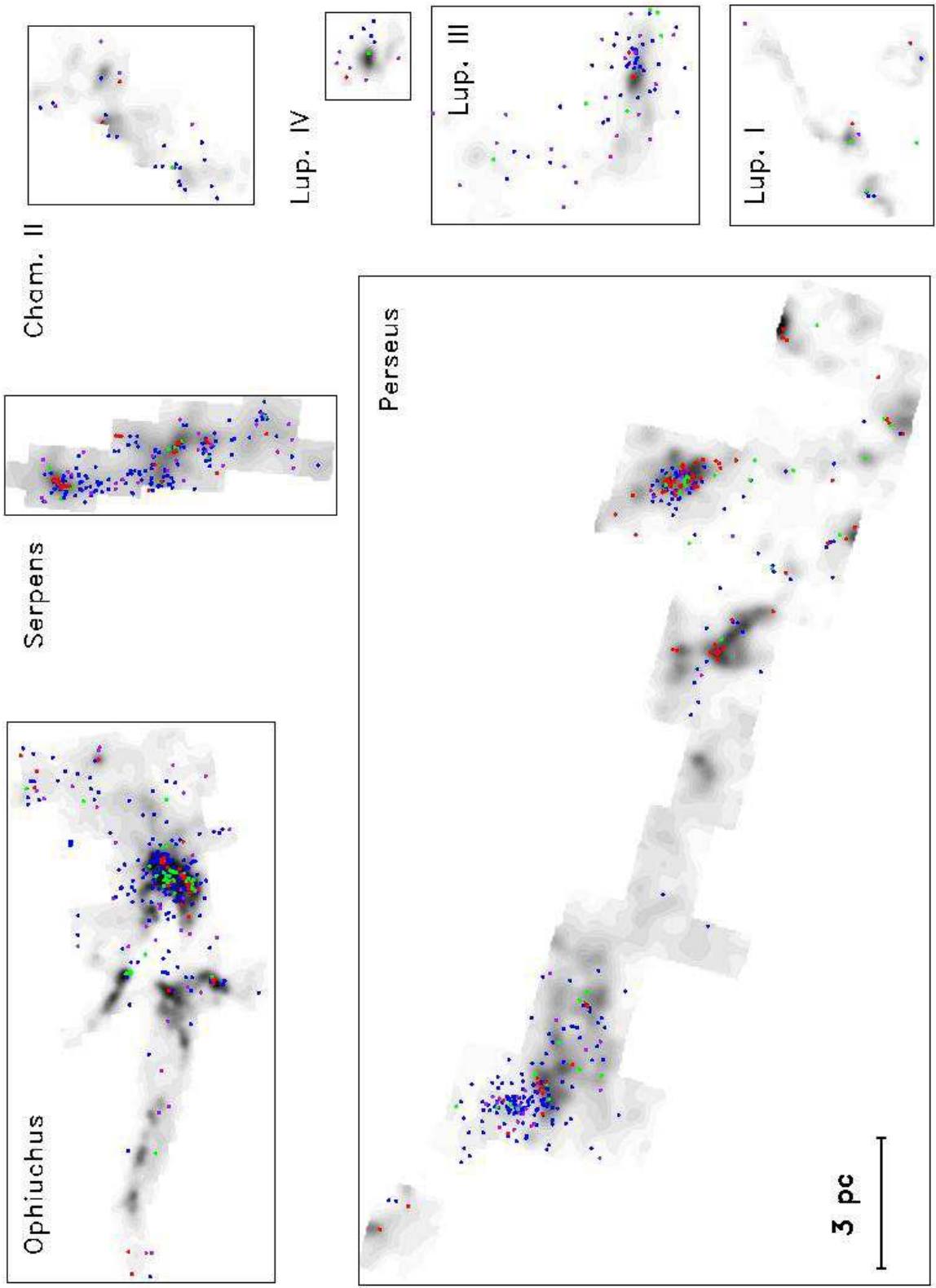}
\figcaption{\label{fig2}
Gray scale images of the extinction of all of the clouds on the same absolute 
size scale as shown by the 3 pc scale bar. All YSOs are plotted with the
following colors red (I), green (flat), blue (II), and purple (III).
The gray scale is linear from $\av = 1$ to 25 mag.
}
\end{figure}

\begin{figure}
\includegraphics[width=5.5in,angle=90]{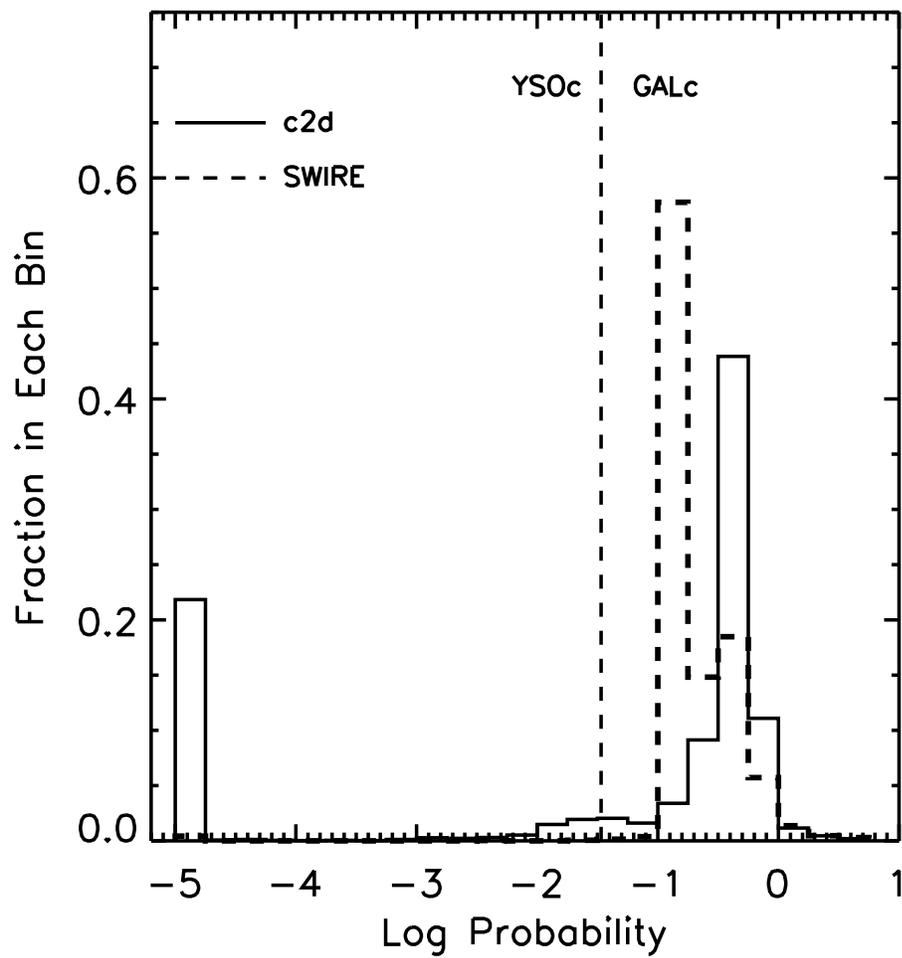}
\figcaption{\label{galprob}
A histogram of the number of objects in bins of the log of the
probability that the source is a galaxy. The combined data for the five 
large clouds are represented by the solid line and the results for the
SWIRE data (degraded to simulate our data) are shown with dotted lines.
The vertical dashed line at log$P(gal) = -1.47$ is the threshold used
to separate YSOc from Galc (galaxy candidates).
}
\end{figure}

\begin{figure}
\includegraphics[width=5in,angle=-0]{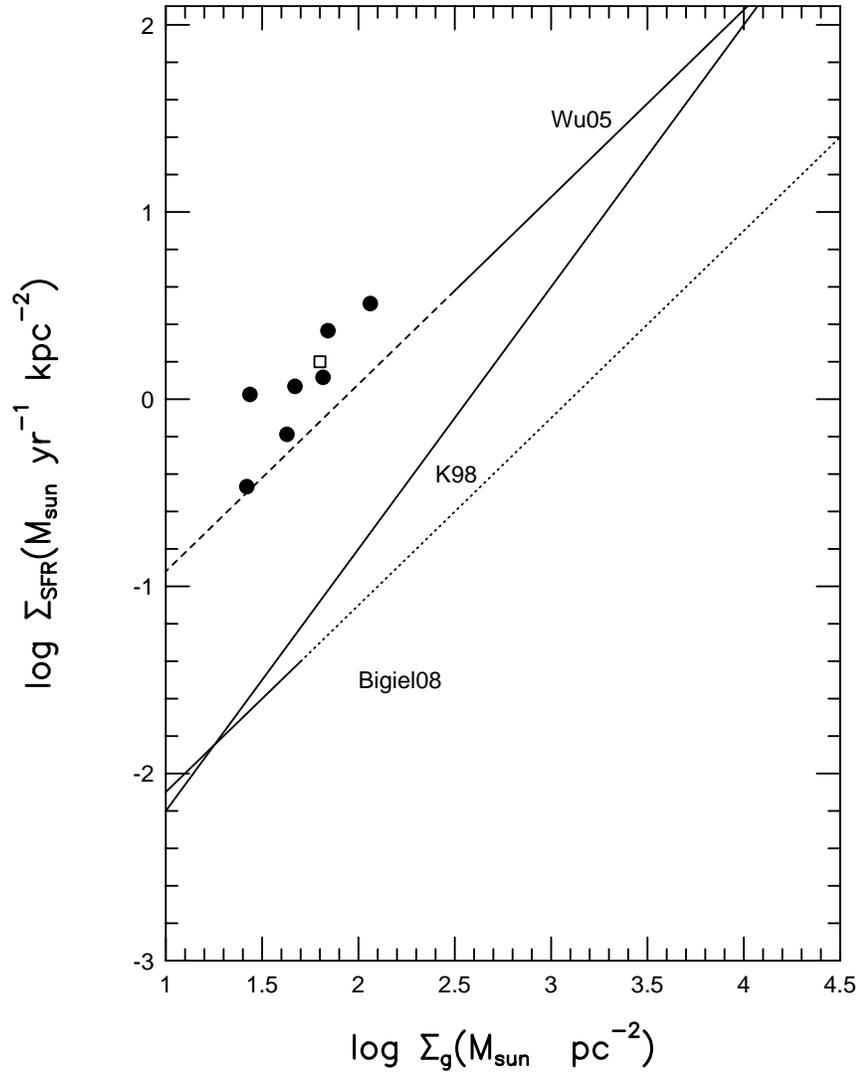}
\figcaption{\label{kennplot} 
The surface density of star formation is plotted versus the mass surface 
density.  The solid line marked K98 shows the relation from \citet{kennicutt98},
and the filled circles show our data.  
The three Lupus clouds are plotted separately.
The open square shows the average for all our clouds.
The line labeled Bigiel08 shows the linear relation from \citet{bigiel08} 
(solid over the range of their study and dotted as extrapolated to
higher surface densities).
The line labeled Wu05 shows the relation for dense gas traced by
HCN \citep{wu2005} (solid over the range studied in dense cores
in our Galaxy and dashed as extrapolated to lower surface densities). 
}
\end{figure}

\begin{figure}
\includegraphics[width=4.5in,angle=0]{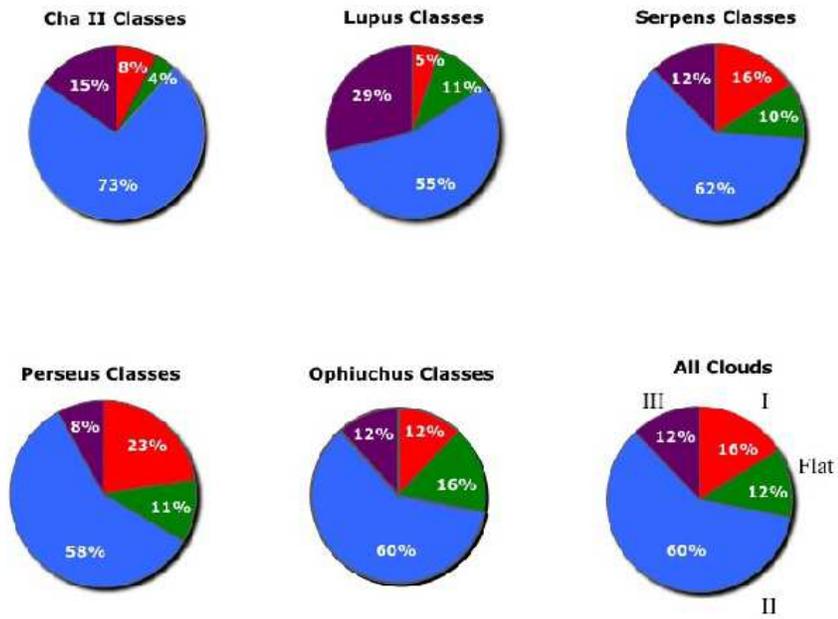}
\figcaption{\label{piecharts}
Pie charts for each cloud and for the total, showing the percentage
of sources in each SED class. Red is Class I, green is Flat, blue
is Class II, and purple is Class III.
}
\end{figure}

\begin{figure}
\includegraphics[height=7.5in,angle=90]{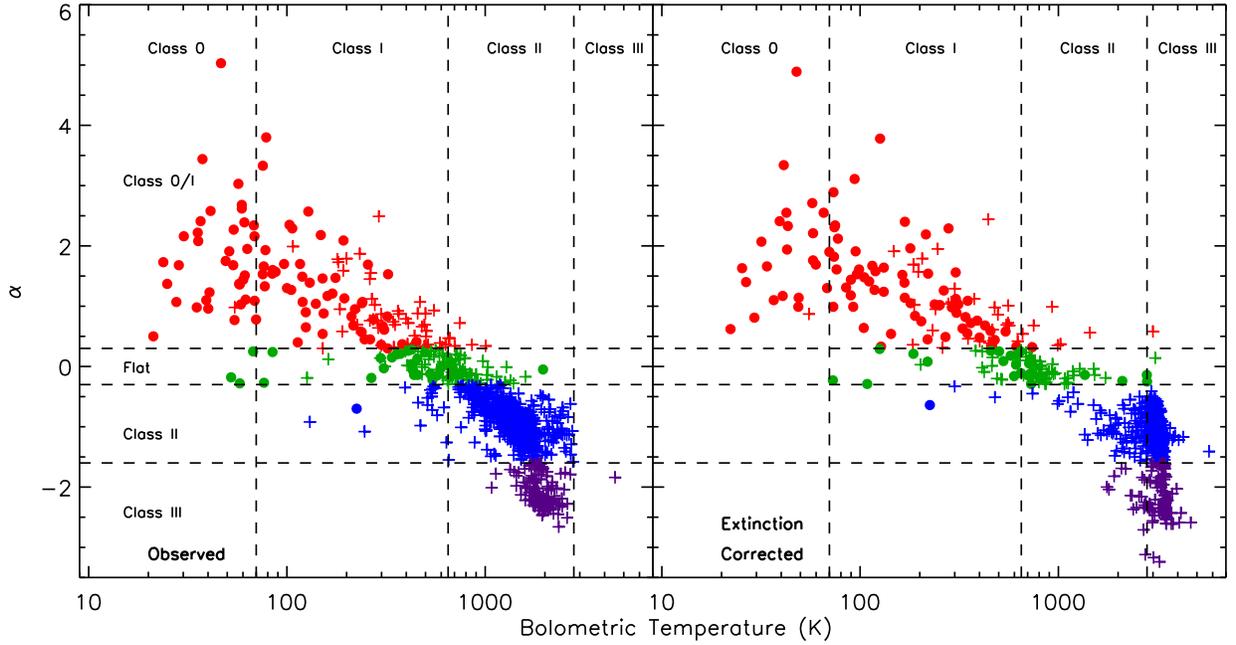}
\figcaption{\label{avstbol}
The value of $\alpha$ is plotted versus \tbol\ for each source with
well-determined values. The color code is based on the Lada class 
of each source, as defined by $\alpha$, with Class I 
plotted as red, Flat as green, Class II as blue, and Class III as purple.  
Filled circles indicate sources associated with envelopes as traced by 
millimeter continuum emission, while plus signs indicate sources with 
no such associations.  The right panel shows \aprime\ and \tbolprime\ 
calculated after corrections for extinction were applied to the observations, 
as described in \S \ref{classification}. 
The vertical dashed lines show the boundaries between classes, 
as defined by \citet{chen95}.
}
\end{figure}

\begin{figure}
\includegraphics[height=7.5in,angle=90]{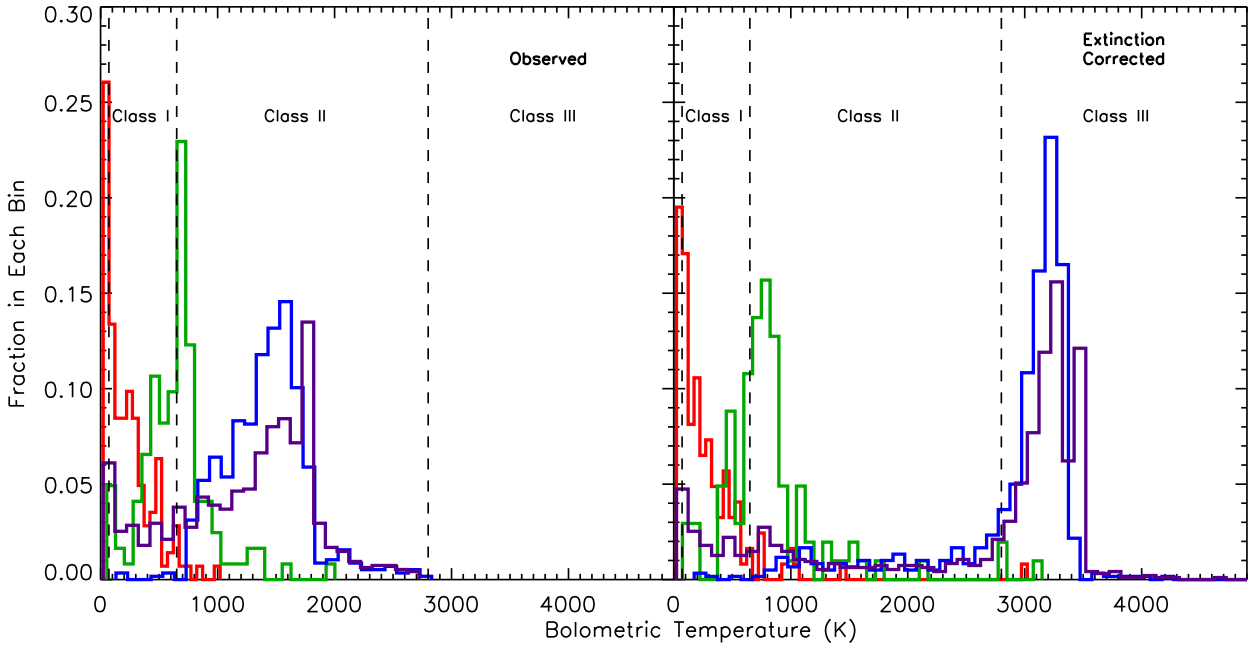}
\figcaption{\label{tbolhist}
The distribution of \tbol\ for sources in each of the Lada classes, as 
defined by $\alpha$, in linear \tbol\ bins.  
The color code is the same as in Figure \ref{avstbol} (Class I plotted as 
red, Flat as green, Class II as blue, and Class III as purple).  
The right panel shows the same distributions after corrections for 
extinction were applied to the observations, as described in 
\S \ref{classification}.
The vertical dashed lines show the boundaries between classes, 
as defined by \citet{chen95}.
}
\end{figure}

\begin{figure}
\includegraphics[height=7.5in,angle=90]{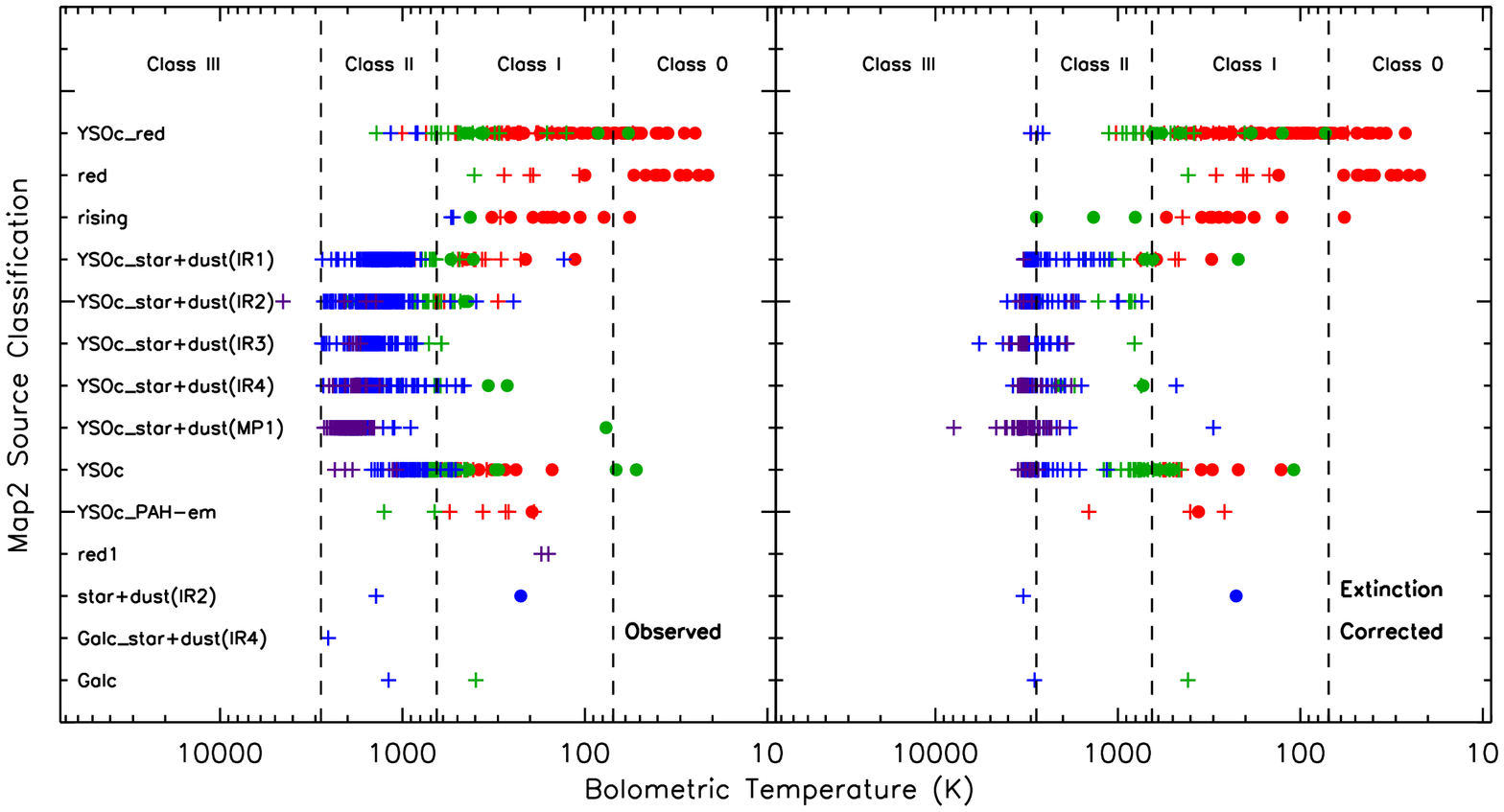}
\figcaption{\label{map2vstbol}
The c2d source type is shown versus \tbol\ for each source with
well-determined values. The color code is the same as in Figure 
\ref{avstbol} (Class I plotted as red, Flat as green, 
Class II as blue, and Class III as purple).  Also as in Figure \ref{avstbol}, 
filled circles indicate sources associated with envelopes as traced by 
millimeter continuum emission, while plus signs indicate sources with 
no such associations.  The right panel shows the c2d source type versus 
\tbolprime, calculated after corrections for extinction were applied to 
the observations, as described in \S \ref{classification}.
The vertical dashed lines show the boundaries between classes, 
as defined by \citet{chen95}.
}
\end{figure}

\begin{figure}
\includegraphics[height=7.5in,angle=90]{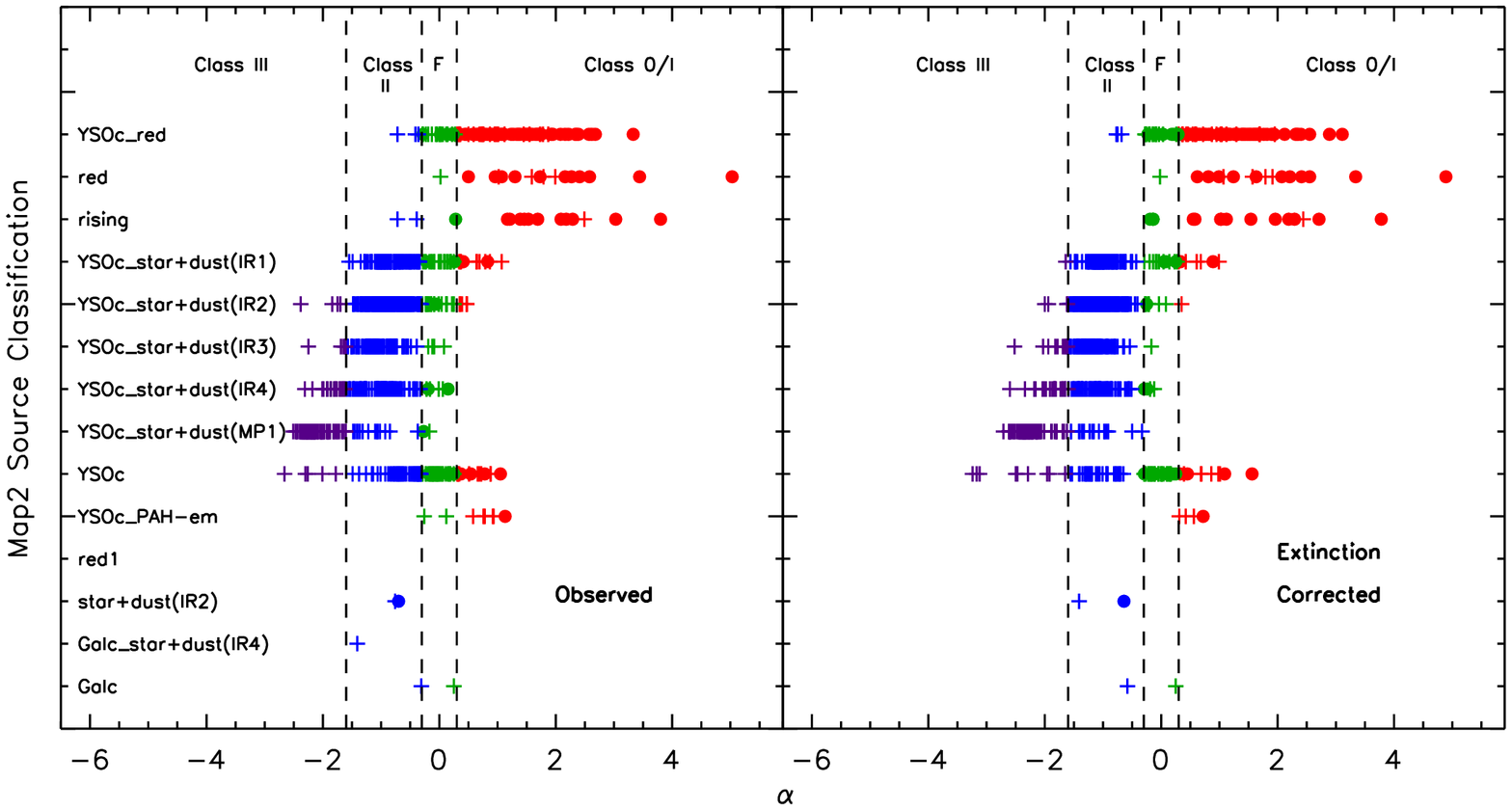}
\figcaption{\label{map2vsa}
The c2d source type is shown versus $\alpha$ for each source with
well-determined values. The color code is the same as in Figure 
\ref{avstbol} (Class I plotted as red, Flat as green, 
Class II as blue, and Class III as purple).  Also as in Figure \ref{avstbol}, 
filled circles indicate sources associated with envelopes as traced by 
millimeter continuum emission, while plus signs indicate sources with 
no such associations.  The right panel shows the c2d source type versus 
\aprime, calculated after corrections for extinction were applied to 
the observations, as described in \S \ref{classification}.
The vertical dashed lines show the boundaries between classes, 
as defined by \citet{greene94}.
}
\end{figure}

\begin{figure}
\includegraphics[height=7.5in,angle=90]{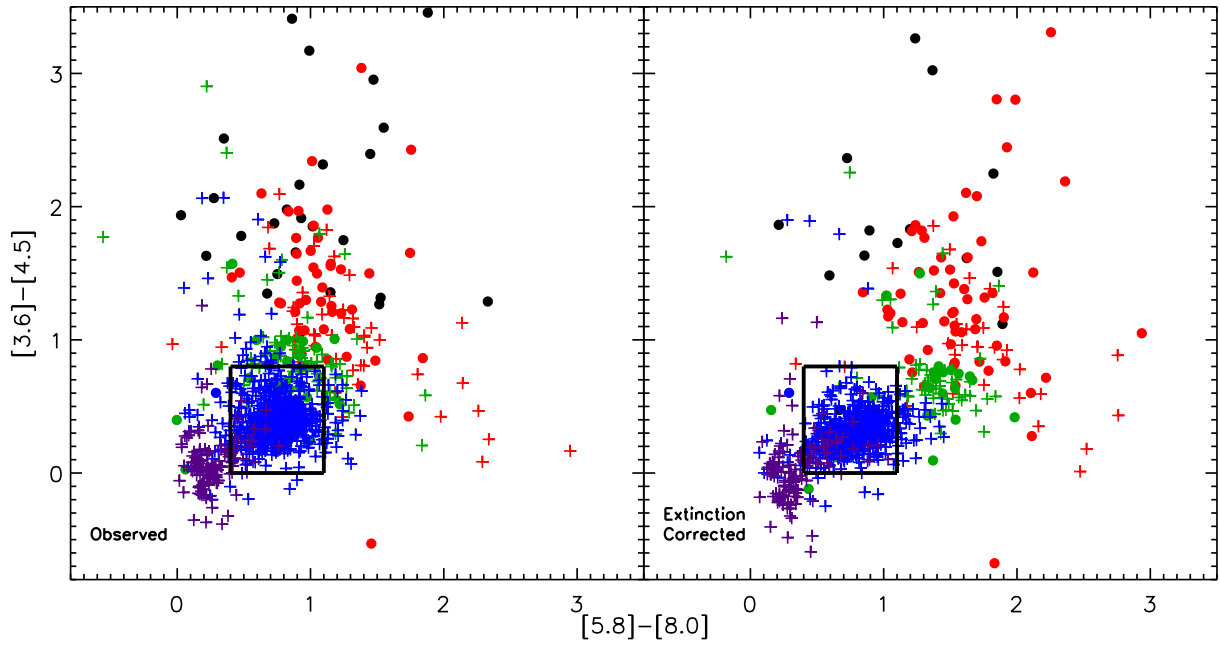}
\figcaption{\label{iraccc}
The color-color diagram for the 4 \irac\ bands. The diagram on the left
is constructed from the observed photometry in Table \ref{tabysolist1}, while
the one on the right is constructed after approximate corrections for
extinction. The color code is the same as in Figure \ref{avstbol} 
(Class I plotted as red, Flat as green, Class II as blue, and Class III as 
purple) with the addition that Class 0 sources are plotted in black.  
Also as in Figure \ref{avstbol}, 
filled circles indicate sources associated with envelopes as traced by 
millimeter continuum emission, while plus signs indicate sources with 
no such associations.  The box indicates the area identified with Class
II sources by \citet{allen04}.
}
\end{figure}

\begin{figure}
\includegraphics[height=6.5in,angle=90]{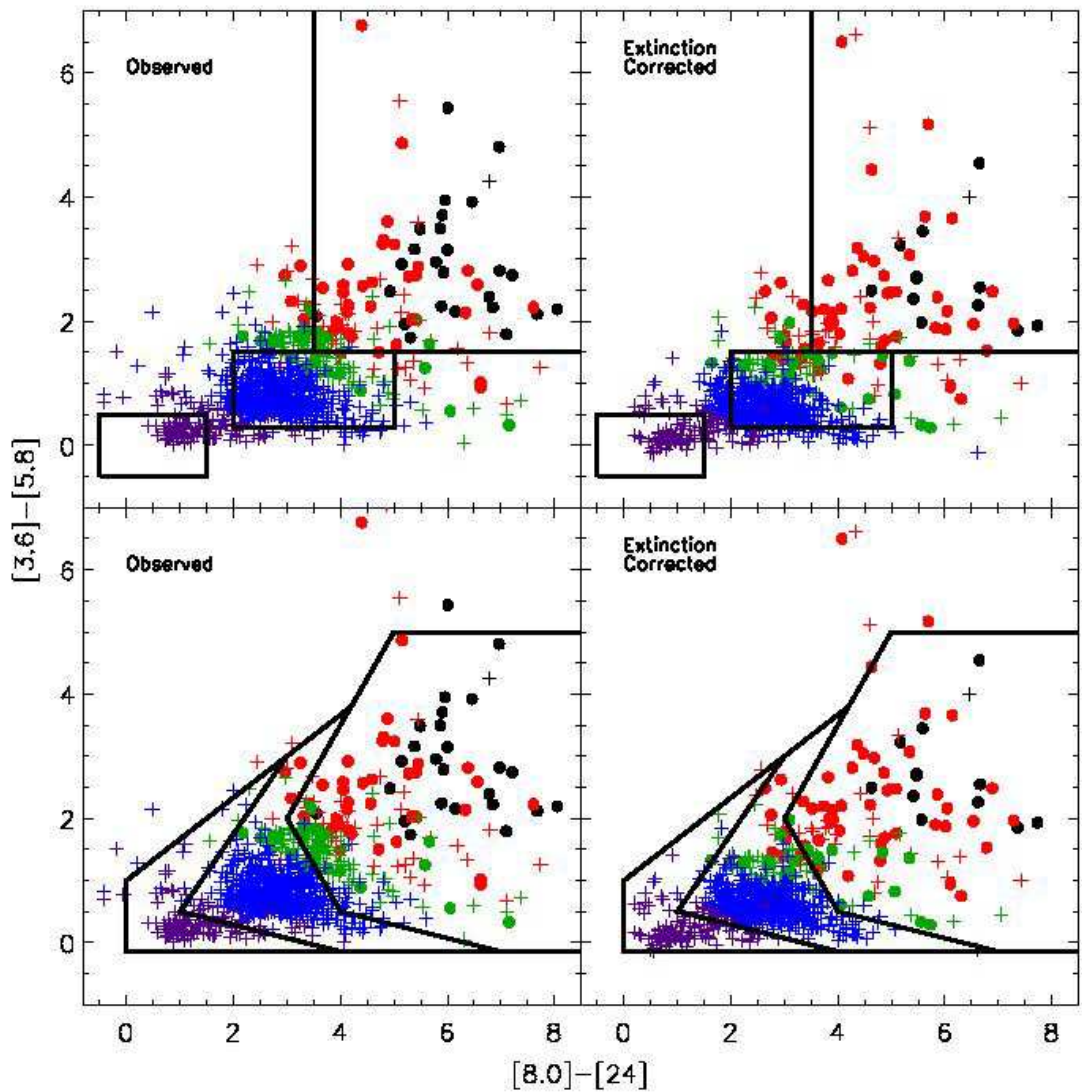}
\figcaption{\label{iracmipscc}
The color-color diagram using 3 of the 4 \irac\ bands and the first \mips\
band. The diagrams on the left
are constructed from the observed photometry in Table \ref{tabysolist1}, while
the ones on the right are constructed after approximate corrections for
extinction. The color code is the same as in Figure \ref{iraccc} 
(Class 0  plotted as black, Class I plotted as red, Flat as green, 
Class II as blue, and Class III as purple). 
Filled circles indicate sources associated with envelopes as traced by 
millimeter continuum emission, while plus signs indicate sources with 
no such associations.  The boxes in the upper two panels indicate the 
areas identified with Class III/stellar, Class II, and Class 0/I 
sources by \citet{muzerolle04}, going from lower left to upper right.
The bottom two panels overlay the areas filled by Stage III, II, and I
sources moving from lower left to upper right, 
taken from \citet{robitaille06}.
}
\end{figure}

\begin{figure}
\includegraphics[width=6in,angle=0]{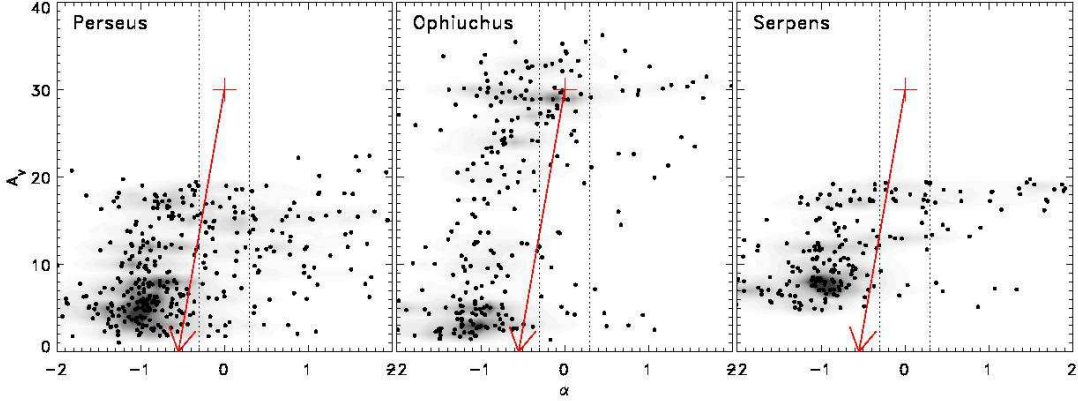}
\figcaption{\label{extalpha} 
The extinction for each YSO in Perseus, Ophiuchus and Serpens versus its 
value of $\alpha$. The extinction for each source is extracted at its
position in the map of extinction based on background stars. 
The extinction vector (red arrow) shows the effect on $\alpha$ assuming 
that half the extinction lies in front of the source. The opacity law of
\citet{weingartner01} with $R_V = 5.5$ is used. 
The gray-scale shows the surface density of
points.  Ophiuchus has a distinct population of
flat spectrum sources at an $A_V$ of around 30, which could be reddened
from the main locus of the Class II sources at low extinctions.
The vertical dotted lines show the boundaries between Class II, Flat,
and Class I sources as defined by \citet{greene94}.
}
\end{figure}

Inspection of Figure \ref{fig2} shows clearly that the YSOs are
not spread uniformly over the cloud. The earlier classes are particularly
tightly clustered in regions of high extinction, as is even more
apparent in the figures in the papers on individual clouds. Furthermore,
\citet{enoch07} have shown that the dense cores are also restricted to
regions of high extinction and strongly clustered. We provide some
simple quantification of these statements in this section. We draw on
previous studies of clustering by \citet{carpenter00},
\citet{lada03} and \citet{allen07}.

We have categorized the environment of the YSOs using simple criteria
of volume density and number of YSOs. We distinguish loose 
(density above 1 \msun\ pc$^{-3}$) and tight (density above 25 \msun\
pc$^{-3}$) groups ($5 \leq  N < 35$) and clusters ($N \geq 35$).
The number of 35 is taken from \citet{lada03}.
If the density falls below the loose criterion, YSOs are classified
as distributed. 
The criterion for loose clustering is also taken from the review of
\citet{lada03} as the minimum density for a cluster to avoid disruption
by passing interstellar clouds \citep{spitzer58}.
Since this criterion results in most sources within our clouds belonging
to a loose group or cluster, we experimented with higher density contours
to find one that captured our intuitive notion of a tight group or
cluster. More sophisticated analysis of the clustering behavior of 
YSOs is beyond the scope of this paper.

The densities are computed from the surface density of YSOs using a
nearest neighbor method, assuming local spherical symmetry to convert to
volume density and assuming a mean stellar mass of 0.5 \msun.
All the YSOs, color-coded by SED class,  are shown on gray-scale images 
of the extinction in Figure \ref{fig2}.
The clustering properties  are discussed in more
detail in other papers [e.g., Perseus and Ophiuchus in \citet{joergensen08};
Lupus in \citet{merin08a}; Cha~II in \citet{alcala08}].

Here we aggregate all the clouds to discuss the total numbers in these
different environments (Table \ref{tabclusters}). 
Note that a source may appear more than once. For example, a source
may lie within a tight group that itself is part of a loose cluster.
Only 9\% of the YSOs in our five clouds are truly distributed, while
91\% are found in loose clusters and 54\% are in tight clusters.
Tight groups account for 13\% and loose groups for 7\%, illustrating
the fact that most tight groups are themselves part of loose clusters.
The low fraction of distributed YSOs is consistent with previous studies
(\citealt{elada91}, \citealt{carpenter00}), with the caveat that an older
distributed population without infrared excess could be present 
\citep{carpenter00}.

The age of the population, measured by the ratio of Class I and Flat
SED sources to Class II and III (Table \ref{tabclusters}), 
depends on the environment as well.
The highest ratios, hence youngest populations, are in groups, while
the lowest ratios are found in the distributed populations, with clusters
intermediate. The statistics for groups are heavily influenced by Perseus, 
which has a number of groups dominated by early SED classes. For clusters,
the young cluster around NGC1333 is balanced by the older cluster around
IC348. The preponderance of more evolved sources in the distributed
population is consistent with the idea that nearly all stars form
in clustered environments and then disperse.

The cloud crossing times, given in Table \ref{tab1}, range from about
3 to 8 Myr. At the high end, these exceed the average 2 Myr time over which
we expect to be at least 50\% complete to YSOs. However, Figure \ref{fig2}
clearly shows that the distribution over Perseus and Ophiucus is far from
uniform. \citet{joergensen08} identifies two loose clusters in Perseus, labeled
e-NGC1333 and e-IC348 in their Table 6. The crossing times for these
loose clusters are 2.4 and 3.4 Myr, respectively, using the same mean speed
as used for Table \ref{tab1}. Only one loose cluster can be identified in
Ophiuchus (e-Oph) and it would have a crossing time of 7.2 Myr. As might
be expected, the YSOs fill this region very unevenly (see Figure \ref{fig2} and
Figure 13 in \citet{joergensen08} for a figure showing the boundaries of the
cluster). For the L1688 cluster itself, the crossing time is 2.2 Myr.
Overall, the distributions of sources are broadly consistent with a picture
in which most stars form in clusters or groups and disperse at a rate
reflecting the cloud turbulence. More detailed analysis will be needed before
more precise statements can be made.

The definition of loose clusters also allows computation of a star
formation efficiency for regions intermediate between the cloud as a whole 
and the dense cores \citep{joergensen08}. With the revised cloud masses,
we infer efficiencies of 9.1\% and 7.7\% for the e-NGC1333 and e-IC348 
loose clusters in Perseus, substantially higher than the 3.8\% we find 
for the cloud as a whole.
For Cha~II, the loose cluster is nearly as large as the cloud, so the
star formation efficiency only increases to 3.9\% for the YSOs 
\citep{alcala08}.

The highly clustered nature of star formation in these clouds may have
implications for discussions of the Sun's birthplace.
Both the abrupt truncation of the Kuiper Belt near 50 AU
and the orbits of detached Kuiper Belt objects such as Sedna are
likely the result of stellar encounters within the first few million
years of solar system history (\citealt{brown04}; Kenyon \& Bromley 2004).
The presence of short-lived nuclei such as $^{60}$Fe, $^{41}$Ca, $^{53}$Mn,
and $^{26}$Al in planetesimals must, in part, arise from the timely
nucleosynthetic input from nearby AGB stars and/or supernovae 
\citep{wasserburg06}.  The abundance and temporal duration of the injection of
$^{60}$Fe in particular has been used to argue for the formation of the
Sun in a massive molecular cloud and dense stellar cluster complete with
the forerunners of Wolf-Rayet stars and Type 1b/c supernovae 
(\citealt{looney06}, \citealt{bizzarro07}).
The iron yield is a strong function of metallicity, however,
with rather lower mass stars capable of producing the amount of $^{60}$Fe
required by the meteorite data \citep{tachibana06}. Thus, the highly
clustered nature of star formation revealed by the c2d survey and the
nearby presence of early B stars to the Ophiuchus and Perseus clouds
argues that such locations should also be considered as possible sites of
the formation of the solar system.

\section{Comparison to Models of Star Formation}\label{models}

\subsection{Does Turbulence Matter?}\label{turb}

The five clouds in this study span a small range in level of turbulence,
as crudely probed by the mean linewidth of the \coo\ \jj10\ line.
We have searched for correlations between linewidth and 
cloud mass, star formation rate, depletion time,
and the lifetime for Class I sources. None are apparent, probably
because the range of turbulence over our cloud sample is small.

\citet{schmeja08} have used an earlier c2d delivery to count and
classify YSOs, and they compared the nature of the clusters in
Perseus, Serpens, and Ophiuchus. They concluded that the most centrally
condensed cluster (Serpens) was associated with the lowest turbulence,
as measured by the \ammonia\ emission lines from the region of the cluster.
In contrast, Table \ref{tab1} indicates that Serpens has the {\bf largest}
linewidth averaged over the cloud, as measured in \coo\ \jj10. Clearly,
the characterization of the turbulence can change
conclusions. In addition, the earlier delivery was considerably more
contaminated with background galaxies. Inclusion of background galaxies
would tend to make earlier classes look less tightly clustered.

Some pictures of star formation invoke a very dynamic picture with
protostars or dense cores competing for gas (e.g., \citealt{bonnell04}).
We can assess the relevance of such pictures to the regions we study
by computing the timescale for collisions between dense cores from
$t_{coll} = (n_{core} \pi r^2 v)^{-1}$. For the
Serpens cluster A \citep{harveysynth}, the data from \citet{enoch07}
on $n_{core}$ and $r$, combined with a velocity dispersion among cores
of 1 \kms\ \citep{williams00} yields $t_{coll} = 3.1$ Myr, 6 times the
Class I lifetime and 3 times the combined Class I and prestellar core lifetimes
spent above a density of about 2\ee4 \cmv. For this cluster, we conclude
that collisions of dense cores {\it at the stage that we observe them}
are not dominant. Rather the cores we observe are more likely to evolve
individually to form stars.  \citet{andre07} reached a similar conclusion
for the main cluster in Ophiuchus, and even for very compact subclusters
within the main cluster. Also, \citet{walsh04} found very small 
($\leq 0.1$ \kms) motions of dense cores relative to surrounding 
less dense gas in 
a larger survey of dense cores.  We caution, however, that the answer 
may be very different in regions forming still richer clusters, such as Orion.

\subsection{Luminosity Evolution}\label{lumevol}

We can make some comparisons of the lifetimes derived from
the relative numbers of objects in different classes or stages
in \S \ref{c2dtimes} to models of star formation. As noted by 
\citet{enoch08b}, the timescale for prestellar cores to begin forming
stars, once their mean densities are sufficient for detection by
Bolocam surveys, is a few times the free-fall time at those densities,
consistent with earlier work summarized by \citet{wardthompson07}.
Such lifetimes are broadly consistent with the evolution of magnetically
controlled models (\citealt{tassis05}, \citealt{adams07}) 
once the density exceeds the
threshold for detection by Bolocam. The figure from \citet{wardthompson07}
that summarizes lifetimes as a function of mean density suggests that
they get shorter, relative to a free fall time, as the density increases,
also broadly consistent with magnetic models.

The lifetime for the Class 0 stage averaged over all clouds is
about 0.16 Myr, but only 0.043 Myr in Ophiuchus. The very short lifetime 
often quoted for the Class 0 phase (0.01 Myr) was originally derived from
observations in Ophiuchus \citep{andremontmerle94}, but this now
seems anomalous. 
If star formation in Ophiuchus has declined from rates
that were higher about 0.1 Myr ago, the number of Class 0 sources
would be much lower compared to Class I sources. Such a scenario is
broadly consistent with the fact that Ophiuchus also has the highest
ratio of mass in YSOs to mass in dense gas of the three clouds with
Bolocam data (Table \ref{tabeff}). Another factor that may have 
contributed somewhat to the short Class 0 lifetime in Ophiuchus
is an overestimate of the number of Class I objects (\S \ref{midlife})
because of very high extinction in the Ophiuchus core.

\begin{figure}
\includegraphics[height=7.5in,angle=90]{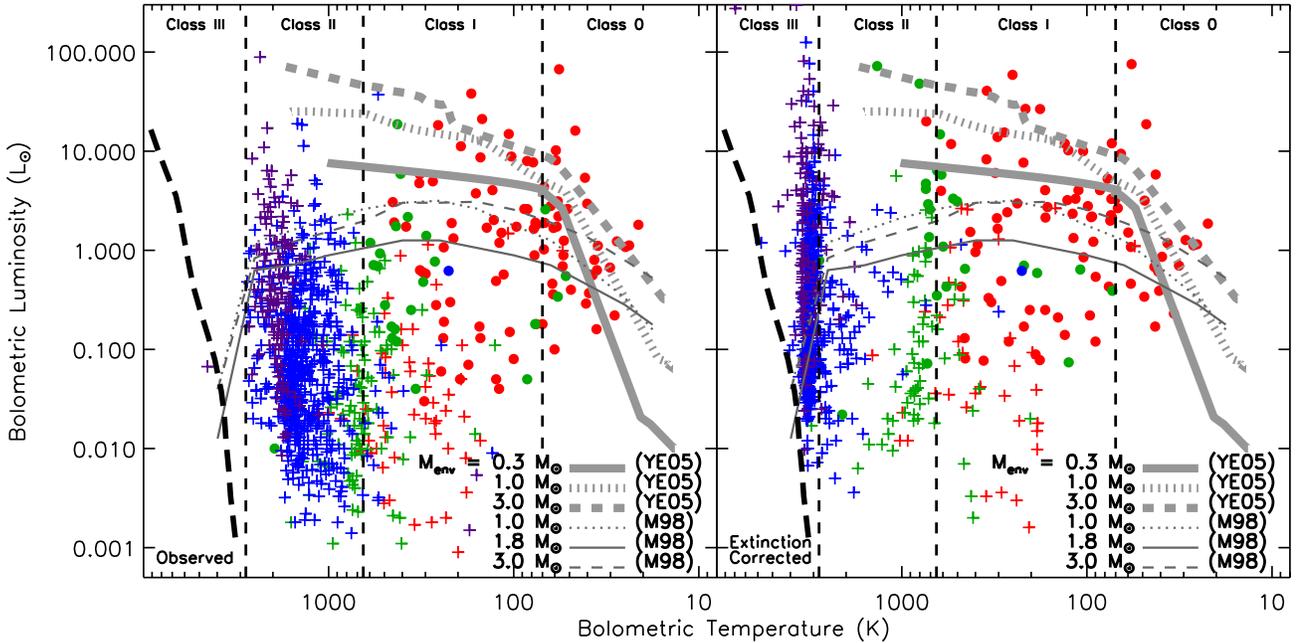}
\figcaption{\label{bltplot} 
The bolometric luminosity (\lbol) is plotted versus the bolometric
temperature (\tbol) for all sources with sufficient data to properly
constrain the values.  Similar plots for the embedded sources only and
for the individual clouds are in \citet{enoch08a}. 
The color code is the same as in Figure 
\ref{avstbol} (Class I plotted as red, Flat as green, 
Class II as blue, and Class III as purple).  Also as in Figure \ref{avstbol}, 
filled circles indicate sources associated with envelopes as traced by 
millimeter continuum emission, while plus signs indicate sources with 
no such associations.
The three thick lines are model tracks for
initial core masses of 0.3, 1.0, and 3.0 \msun, collapsing according to 
a Shu inside-out collapse model in which all mass becomes part of the
star \citep{youngevans05}; they end when all infall stops, but the later
stages are highly uncertain.  The three thin lines are model tracks with 
initial core masses of 1.0, 1.8, and 3.0 \msun, collapsing with an 
accretion rate that decreases exponentially with time and efficiencies such 
that the final stellar masses are 0.5, 0.3, and 0.5 \msun, 
respectively \citep{myers98}.
The right panel shows \lbolprime\ and \tbolprime\ calculated after
corrections for extinction were applied to the observations, as described in
\S \ref{classification}.
The vertical dashed lines show the boundaries between classes, 
as defined by \citet{chen95}.
The heavy dashed line on the left is the ZAMS \citep{dantona94} from
0.1 to 2 \msun.
}
\end{figure}

The longer lifetime for the Class 0 phase in most clouds is more 
consistent with the luminosity distribution shown in Figure \ref{bltplot}
and \citet{enoch08a}. In particular, there is no clear evidence
in our data for an early phase of very rapid mass accretion, which 
would produce luminosities higher than the mean values we observe.
In fact, the luminosities seen in Class 0 sources are mostly similar to
or less than predictions \citep{youngevans05} of the evolution
in \lbol-\tbol\ space  based on simple inside-out collapse 
models \citep{shu77}, with constant accretion rates and unit efficiency.
Models with {\bf mean} accretion rates onto the star faster than 
the \citet{shu77} rate are not supported by these observations. 

For Class I sources, most of the observed \lbol\ fall well below
the predictions by \citet{youngevans05}, suggesting that accretion
occurs at a rate even lower than that of the Shu picture, which is the
smallest among star formation models. The mass in the envelope, on the other
hand, decreases about as predicted by this model \citep{enoch08a}. 
This result suggests that infall occurs at about the rate expected in the
Shu picture, but that accretion onto the star occurs at a much lower
rate most of the time. The lower rate may be partially explained if
only a fraction $\epsilon$ of the available mass in a dense core winds
up on the star. As discussed earlier, comparison of the core mass function
and the IMF suggest $\epsilon \approx 0.3$. This helps explain the low
luminosities, but it is insufficient to explain the very low values
that we see. In addition, the spread in \lbol\ of two to three orders of
magnitude at a given \tbol\ range practically demand a picture in which
accretion onto the star is not steady. While no Class 0 sources have 
$\lbol < 0.1$ \lsun\ in Figure \ref{bltplot}, this can be misleading;
heating of a substantial envelope just by the interstellar radiation field
can produce an apparent luminosity at this level \citep{evans01}. The effect
decreases as the envelope dissipates, but is still present during the Class I
phase.

\begin{figure}
\includegraphics[width=5in,angle=90]{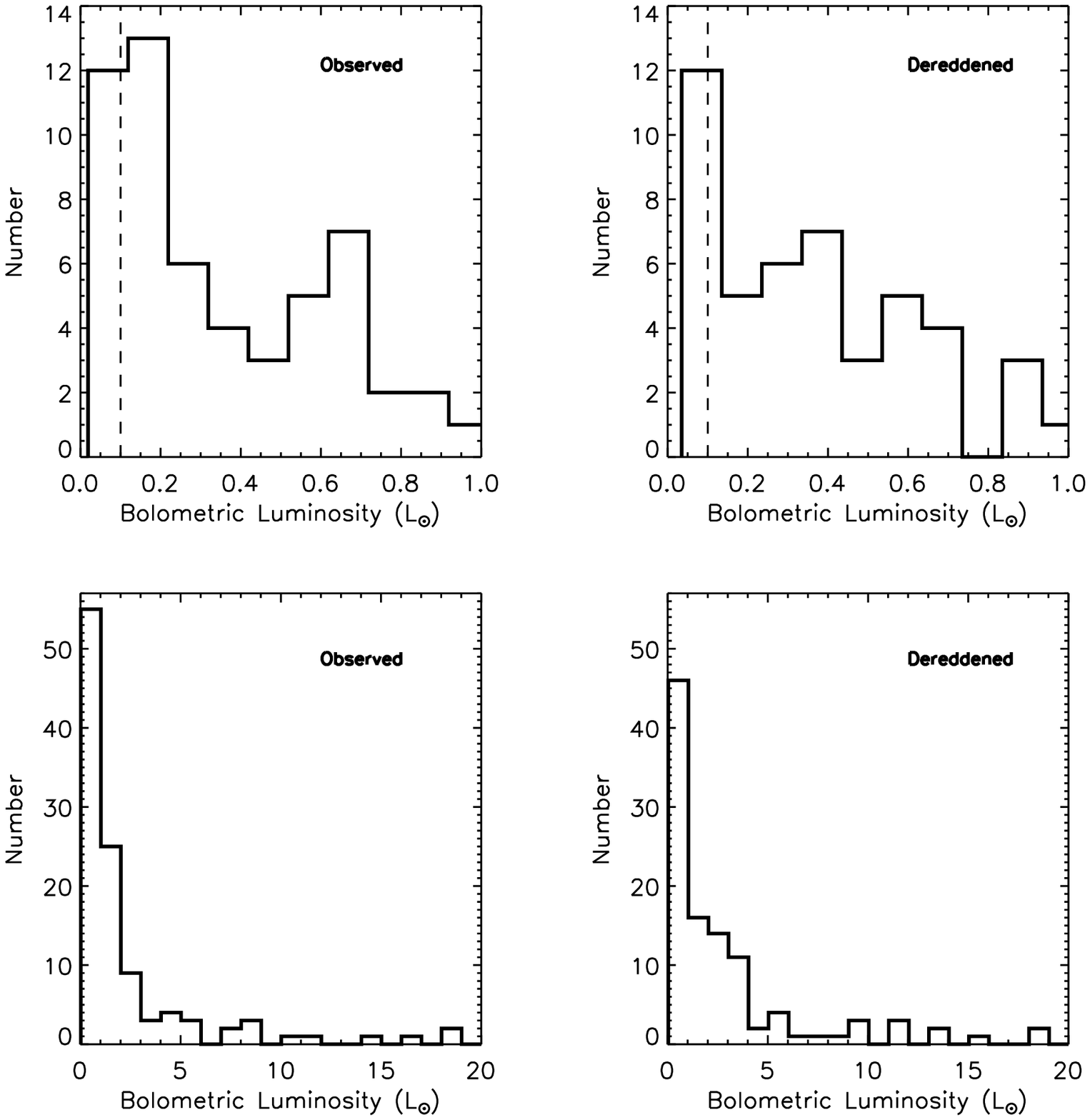}
\figcaption{\label{lumhist} 
The number of embedded sources (associated with envelopes) in luminosity bins.
The lower plots show all sources up to 20 \lsun. There are three sources
with higher \lbol, with $\lbol \sim 70 $ \lsun\ as the largest. The top	
panels show a blow up of the range below 1 \lsun. The panels on the
right show the histograms obtained after correcting for extinction.
The vertical dashed line
indicates the level of \lbol\ that can easily result just from heating from
the interstellar radiation field. This value can vary by factors a 
few, depending on the mass of the envelope and the strength of the
radiation field. There is a selection bias against lower
luminosities. This low luminosity regime is studied in more detail by
\citet{dunham08}. 
}
\end{figure}

The distribution of luminosities for all Class 0 or I sources with
millimeter emission is shown in Figure \ref{lumhist}. The panels on the
bottom cover 1 to 20 \lsun; there are three sources more luminous, with
the most extreme at 70 \lsun. However the vast majority of sources 
have $\lbol \le 3$ \lsun. As shown in the top panels, the trend of
increasing numbers at lower luminosity continues down to 0.1 \lsun, 
where selection effects come into play. A more thorough study \citep{dunham08}
of the low luminosity end that dealt with these effects found 15 objects with
{\bf internal} luminosity, $\lint \le 0.1$ \lsun. \citet{dunham08} was able
to rule out a continued rise in the luminosity distribution to lower
luminosities, which could have called into question whether any cores
were truly lacking in internal sources. In agreement with an earlier
analysis \citep{kirk07}, \citet{dunham08} found that 75\% to 85\%
of cores thought to be starless before \spitzer\ remain so down to 
a limit of $\lint > 4\ee{-3} (d/140 pc)^2$ \lsun.  
As discussed by \citet{dunham08},
the luminosity distribution in Figure \ref{lumhist} is quite inconsistent
with constant mass accretion, which would dominate the luminosity
in the early stages. For an object at the stellar/brown dwarf boundary
of 0.08 \msun\ and a radius of 3 \rsun, the standard accretion rate
from a Shu model ($\sfr = 2\ee{-6}$ \msunyr) would produce
$\lint = 1.6$ \lsun. For the sources in Figure \ref{lumhist}, 59\%  have
$\lbol < 1.6$ \lsun; \lbol\ generally provides an upper limit to \lint\ 
for deeply embedded sources.
Our results reinforce and aggravate the ``luminosity problem" first noted 
by \citet{kenyon90}. As suggested by \citet{kenyon90} and further developed
by \citet{kenyon94} and \citet{kh95}, non-steady accretion is
the obvious solution, and it is nearly required by our results. Furthermore,
it appears to begin in the very early, Class 0, phase. Such a scenario
also invalidates any attempt to infer stellar mass from luminosity in phases
where accretion luminosity dominates stellar luminosity.

Non-steady accretion could be
caused by a magnetic wall that forms in some models, resulting in
``spasmodic" accretion as the wall periodically fails \citep{tassis05}.
Presumably, turbulence in the infalling envelope
could also result in variations in infall rates
as well. Simulations of star formation in turbulent clumps show
variations in mass accretion rates of about one order of magnitude
(e.g., \citealt{offner08}).
Because almost all the luminosity of accretion is released when 
material falls onto the star, relatively constant infall can store 
mass in a disk, producing very little luminosity until a disk instability
dumps a lot of mass onto the star, as in FU Orionis events \citep{kh95}. 
Statistical evidence for episodicity
extends into the Class 0 stage \citep{dunham08} and there is direct
evidence in one case for a current accretion rate much lower than a
time average rate \citep{dunham06}. 

Episodic accretion at peak rates higher than predicted by a Shu model
would explain the few sources that lie above the tracks from 
\citet{youngevans05} in Figure \ref{bltplot}. Short periods of rapid
accretion would presumably also drive strong outflows. If such
episodes are more frequent in the Class 0 period, that could explain the
higher ratio of CO outflow luminosity to \lbol\ seen in Class 0 objects
\citep{andre00}. Some evidence for that picture might be seen in
Figure \ref{bltplot}, if one accounts for the lower luminosity
predicted in the Class 0 phase for constant accretion models.

\subsection{Comparison to a Specific Model}\label{shucomp}

We can also apply our derived lifetimes to simple models in a consistency
check. Considering a Shu model \citep{shu77}, 
the wave of infall propagates outward at
the sound speed until it reaches the outer bound of the region that will
collapse. This should occupy roughly half the time spent in the Class I 
phase, as the other half is occupied by the last shell falling inward.
In this oversimplified picture, the feeding zone for a protostar would
be given by 

\begin{equation}
r_f = 0.5 a t(I),
\end{equation}
where $a$ is the sound speed and $t(I)$ is the lifetime for the Class I
objects. For a purely thermal sound speed at $\tk = 10$ K, $r_f = 0.096 t(I)$ pc with
$t(I)$ in Myr, or 0.052 pc if $t(I) = $ \ti\ Myr, or 0.042 pc using
$t^{\prime}(I) = $ \tiprime\ Myr. 
These radii are comparable to the typical size of cores \citep{enoch08b}, 
and the modal spacing between YSOs in a study of a large number
of clusters is also about 0.05 pc \citep{gutermuth08}.
The age we derive for the Class I phase, together with a Shu model,
is consistent with the data at this level.
However, core sizes are defined more
by observational techniques than by evidence of a sharp boundary.
In clustered regions, the core size is set by the overlap with nearby
cores, as seen in projection. Sizes of cores taken from CLUMPFIND
operating on the SCUBA data give somewhat smaller sizes (0.036 pc in
Perseus and 0.018 pc in Ophiuchus), suggesting that resolution
effects are important \citep{joergensen08}. In much more crowded
environments, such as the Orion cluster, the feeding zone could
be even smaller. 

How much mass would accumulate in this time?  For the inside-out
collapse model \citep{shu77}, the mass infall rate is given by
\begin{equation}
\dot M = m_0 a^3/G,
\end{equation}
with $m_0 = 0.975$ and $a$ is the sound speed. As discussed by 
\citet{shu87}, the form of this expression is general, but 
$\dot M$ is constant only for certain initial density distributions.
For a temperature of 10 K, we have $\dot M = 1.6\ee{-6}$ \msunyr.
With an efficiency of $\epsilon$ for matter to wind up in
the star, $\sfr = \epsilon 1.6\ee{-6}$ \msunyr, and the final
stellar mass would be $\mstar = 0.85 \epsilon$ \msun\ if $t(I)$ is
\ti\ Myr. For $\epsilon = 0.3$, suggested by \citet{alves07}, this
produces a star of 0.25 \msun\ (0.20 \msun\ using $t^{\prime}(I) = $ \tiprime\
Myr), near the modal mass of the IMF. If the
Flat SED objects correspond to the infall of the remnant envelope, one
could add another  0.20 \msun\ in the time attributed to that class,
but we suspect that many Flat sources do not have any substantial 
envelope, as discussed above. Previous, shorter lifetimes for the 
Class I phase would have predicted lower final masses, so it is
not trivial to understand how one builds any stars of mass exceeding
0.5 \msun. One could
make more massive stars by increasing $\epsilon$ or the sound speed,
but the latter solution would predict even 
larger \lbol, and we already have a luminosity problem, illustrated
in Figure \ref{bltplot}. Models with faster mean infall rates will have to 
explain the low mean values and broad spread of luminosities in this
figure, probably with recourse to spasmodic or episodic accretion.

To test whether or not a picture of episodic mass accretion could result
in a typical mass star, we consider a very simple model.  The observed \lbol\
for each source classified as Class 0 or Class I according to \tbol\ and
associated with an envelope (112 sources) is turned into a mass accretion rate,
\sfr\, by

\begin{equation}
\sfr = \frac{\lbol R}{G \mstar}.
\end{equation}
$R$ is the stellar radius, assumed to be constant at 3 \rsun, and \mstar\
is the protostellar mass, assumed to be 0.25 \msun\ for this calculation.
This value of \mstar\ is chosen because it is half of the mean mass of
0.5 \msun\ assumed in \S \ref{sfesfr}.
In reality, \mstar\ will grow as the protostar accumulates mass.  However,
if we assume that the \emph{average} mass accretion rate stays constant
throughout the embedded phase, the under- and over-estimates in \sfr\
introduced by including sources with masses above and below 0.25 \msun\
should roughly cancel out.  The final stellar mass accumulated,
$\mstar (t_f)$, is then
\begin{equation}
\mstar (t_f) = \displaystyle\sum_{\sfr} \sfr t(\sfr),
\end{equation}
where $t(\sfr)$ is the amount of time spent in each accretion state 
and is given
by the number of sources at each \sfr\ divided by the total number
of sources, multiplied by the Class I lifetime of \ti\ years.
With a bin size in
\sfr\ of $1.0 \times 10^{-8}$ \msun\ yr$^{-1}$, we calculate
$\mstar (t_f) = 0.7$ \msun.  Smaller bin sizes result in identical results, 
while
larger bin sizes overestimate $\mstar (t_f)$ since very low mass accretion rates
are lumped together in one bin.  Even though 92 of the 112 sources (82\%)
have \lbol\ implying $\sfr \leq 1.6 \times 10^{-6}$ \msun\ yr$^{-1}$,
the end result is still a typical mass star.  In this picture,
much of the mass of stars accumulates during brief periods of high
accretion. For this specific model and data sample, half the final 
mass is contributed during the 0.039 Myr (7\% of the Class I lifetime) 
represented by the eight highest luminosity sources.
Since the duration of the events with the highest accretion rates
is the shortest, even larger samples may be needed to find the rarest
accretion events. 

\citet{myers08} has pointed out that the dense cores do not have sharp
boundaries, but instead inhabit ``clumps" of lower, but still substantial,
density. In this case, the final stellar mass may be determined by a 
competition between free fall onto the protostar and dispersal of the
clump gas. More massive stars could be formed in this way. If the infalling
material from the clump has sufficient opacity to make the SED fall into
Class I, the time limit on this process would still be the lifetimes
derived above, and those would limit the final stellar mass to values
similar to those derived above.

To compare more specifically to theoretical models, it is necessary to
translate the models into simulated observations. \citet{youngevans05} did that
for the standard Shu model \citep{shu77}, assuming unit efficiency and
spherical radiative transfer.
They predicted that the {\bf observed} Class 0
phase would be short, about 0.05 Myr, terminated by the collapse of the
first hydrostatic core as the
central source shrank to 3 \rsun\ and accretion luminosity became
important, and significant emission in the \mir\ began to be produced.
That lifetime was very uncertain because it depended on assumptions about
the first hydrostatic core, so the time may be longer; alternatively,
episodic accretion could move sources back and forth across the Class 0/I
boundary. The Class 0 sources with high \lbol\ would need to have thicker
envelopes or be viewed edge-on through a non-spherical envelope.
The transition from Class I to Class II in the models of \citet{youngevans05}
depended strongly on the mass of the initial envelope, ranging from
0.1 to 0.4 Myr for masses of 0.3 to 3 \msun. This transition occurred
near, but not at, the end of infall. The time spent in the range of
\tbol\ we might attribute to the Flat stage was very short, reaching
0.1 Myr only for the model with a mass of 3 \msun. Clearly, the effects
of non-spherical models and non-steady accretion will have to be considered
to bring observations and models into better congruence.

\section{Future Work}\label{future}

While the c2d study has provided answers to many questions, much work
remains to be done. 
It will be important to incorporate results from the Taurus
and Gould Belt surveys. Spectral types for the remaining visible sources are
necessary, both to weed out remaining contaminants and to estimate masses
of the forming stars. Searches for binaries would be very valuable, 
especially for the cold disks. More complete lists of PMS stars without
infrared excess are needed in most clouds. Near-infrared surveys that
go much deeper than 2MASS would better match the sensitivity of the
Spitzer data, aiding greatly in discriminating against background galaxies. 
Low-noise millimeter continuum maps of Lupus and Cham~II
are needed to put those clouds on the same footing as the northern clouds.
Episodic accretion could leave signatures in the envelope chemistry or
in outflows, so studies of multiple molecular lines are needed for the
embedded sources.

\section{Summary}\label{summary}

The data from five large clouds studied by the \spitzer\ c2d legacy
project have been combined and reanalyzed in a consistent way.
The five clouds combined contain \ysotot\ objects we classify as YSOs.
We estimate that there are no more than about 50 (5\%) contaminating background
galaxies in the sample and probably many fewer because of human examination.
We estimate that the sample is complete down to $\lbol \approx 0.05$ \lsun\
over a wide range of SED types. This limit arises from increasing contamination
by background, star-forming galaxies at low flux densities rather than 
sensitivity limits. With better discrimination against galaxies, the fundamental
luminosity limit could be lowered to about \eten{-3} \lsun.

The five clouds are forming about 260 \msun\ of stars per Myr. Using cloud 
masses determined from c2d extinction maps, the star formation efficiency
over the last 2 Myr
is 3\% to 6\%, depending on the cloud. The time to deplete the current cloud
material at the current rate exceeds most estimated cloud lifetimes,
but final efficiencies could be 15\% to 30\% if the clouds form stars
at the current rate for 10 Myr. That would require continued creation of
dense cores from less dense cloud material at a rate that replenishes
the current stock in about 2 Myr.

The star formation surface density is about 20 times larger
than would be predicted by relations used for extragalactic
star formation, whether non-linear \citep{kennicutt98} or 
linear \citep{bigiel08} in gas surface density.
The results are more consistent with relations found for dense gas 
\citep{wu2005}.

For three clouds, comparisons to dense gas, as mapped by continuum emission
from dust, are possible. The efficiencies are much higher when measured against
the available mass of dense gas. Depletion times for the dense gas are about
1.8 Myr, similar to timescales for cluster formation. However, star formation
is still slow compared to a free-fall time, even in the dense gas.

We use the spectral index in the near-infrared to mid-infrared to place
YSOs into traditional SED classes. Combining the number counts with an
estimate of 2 Myr for the lifetime of Class II YSOs,  and assuming a continuous
flow through the classes, we derive lifetimes for prestellar, Class 0,
Class I, and Flat SED phases of \tsl, \tzero, \ti, and \tf\ Myr, respectively.
If corrections to flux densities are made for measured or estimated
extinctions before computing evolutionary indicators, the lifetimes become
\tslprime, \tzeroprime, \tiprime, and \tfprime\ Myr for prestellar, Class 0, 
Class I, and Flat SED classes. These lifetimes should be viewed as median
lifetimes or half-lives, rather than a lifetime that applies to all objects.
The prestellar lifetimes refer to cores whose mean density is at
least 2\ee4 \cmv\ and more typically above \eten5 \cmv. 
Lifetimes of cores with lower mean densities are clearly longer \citep{wardthompson07},
and may vary from cloud to cloud [cf. the Pipe nebula \citep{lada08}].

The connection between the traditional classes and actual physical stages
is uncertain, particularly in regions of very high extinction, such as
the core of Ophiuchus. Class II objects behind heavy extinction can be
classified as Flat or even Class I objects. Our statistics after extinction
corrections illustrate the uncertainties (typically 10\% to 20\%), but
the effects can be larger. As discussed in more detail by \citet{merin08b},
the traditional evolution from Class II to Class III does not capture the
diversity in SEDs found in our survey. We suggest a two-parameter
classification as an intermediary to more physical interpretations.

Star formation is far from uniformly distributed over the clouds.
YSOs are highly concentrated in regions of high extinction and the
younger YSOs are even more concentrated in regions where millimeter
continuum emission indicates the presence of dense cores.
The majority (54\%) of YSOs are in tight clusters (at least 35 YSOs
and a density above 25 \msun\ pc$^{-3}$), and 91\% lie within loose
clusters (at least 35 YSOs and a density above 1 \msun\ pc$^{-3}$). 
Only 9\% are truly distributed.

We find no obvious trends with the degree of turbulence in the clouds as
a whole, but the range is very small. Estimates of collision times between
dense cores suggest that they are unlikely to collide during their 
remaining lifetime, suggesting that individual cores may retain their
identity and form a small number of stars. Thus our data are consistent
with a mapping of core mass function onto stellar mass function, as
discussed by \citet{enoch08b}. However, denser star-forming clusters, such
as Orion, may be quite different. 

Comparison to a simple inside-out collapse model suggests that a star of
mass 0.25 \msun\ could form during the lifetime of the Class I stage, assuming
an efficiency of 0.3 for infalling mass to wind up in the star. However,
we see a spread of at least three orders of magnitude in \lbol\ in the
embedded phases, with most YSOs below the luminosity expected in an
inside-out collapse model. The data strongly indicate that accretion onto
the forming star must be transient, with very large fluctuations. This
result confirms the conclusion of \citet{kh95} and extends it to earlier
stages. A very simple model of episodic mass accretion applied to our
sample suggests that a typical star could assemble half its mass in
a few episodes of high accretion, corresponding to the high luminosity
sources in our sample. These episodes could occupy only about 7\% of
the full Class I lifetime.

\acknowledgments

We thank the referee, Charles Lada, for a very thought-provoking report,
which helped us to improve the clarity of the paper.
We thank the Lorentz Center in Leiden for hosting several meetings
that contributed to this paper. Part of the work was done while in
residence at the Kavli Institute for Theoretical Physics in Santa
Barbara, California. We thank M. Krumholz for enlightening discussions
and derivations of the speed of star formation in dense gas and
F. Comer{\'o}n for drawing our attention to biases caused by low
mass objects dropping out of the sample as they age.
We are also grateful to D. Ward-Thompson for helpful suggestions.
Support for this work, part of the \spitzer\ Legacy Science Program, 
was provided by NASA through contracts 1224608, 1230779, 1288658,
1288664, and 1230782 issued by the Jet Propulsion Laboratory,
California Institute of Technology, under NASA contract 1407. 
Additional support came from NASA Origins grants NNG04GG24G and
NNX07AJ72G to NJE.  While several coauthors were at the Kavli Institute, 
the research was supported in part by the NSF under Grant No. NSF PHY05-51164.
MLE acknowledges support of an NSF Graduate Research Fellowship and 
a Spitzer Space Telescope Postdoctoral Fellowship.
Astrochemistry in Leiden is supported by a NWO Spinoza grant and
a NOVA grant.

{\it Facilities:} \facility{Spitzer}, \facility{CSO}





\begin{table}
\caption{Facts about Clouds \label{tab1}}
\begin{tabular}[htb]{lrrrrrrrl}\hline
\hline
Cloud  &  Solid angle & Distance & Area & $\Delta v$  & Mass\tbn{a} & \mean{n}\tbn{b} & t(cross)  & Refs \\
       &  (deg$^{2}$)  &  (pc)    & (pc$^2$) & (\kms) & (\msun) & (\cmv) & (Myr) &  \\
\hline
Cha~II	& 1.038	& $178\pm18$ 	& $10.0\pm2.0$	& 1.2 & $426\pm86$ & 345 & 3.7 & 1, 2 \\
Lupus   & 3.101	& $150\pm20$\tbn{c} & $28.4\pm6.5$ &  1.2 & $816\pm188$ & 381 & 4.7\tbn{d} & 3, 4 \\
Perseus	& 3.864	& $250\pm50$	& $73.6\pm29.4$	& $1.54\pm0.11$ & $4814\pm1925$& 196 & 7.8 &5, 6 \\
Serpens	& 0.850 & $260\pm10$	& $17.5\pm1.4$ & $2.16\pm0.01$ & $2016\pm155$ & 707 & 2.7 & 7, 6 \\
Ophiuchus & 6.604 & $125\pm25$	& $31.4\pm12.6$	& $0.94\pm0.11$ & $2182\pm873$\tbn{e} & 
318 &  8.4 & 8, 6 \\
\hline
Total	& 15.457 &\nodata	 & $160.9\pm51.9$ & \nodata & $10254\pm3228$ & 389 & 
\nodata  \\
\hline
\end{tabular}
\tablenotetext{a}{The masses are computed from extinction maps with 240\arcsec\
resolution based on c2d and 2MASS data, except for Ophiuchus, which uses the 270\arcsec\
resolution map; the mass refers to the area with $\av \geq 2$ mag, and the
uncertainty reflects only the distance uncertainty.}
\tablenotetext{b}{The mean density of the cloud, calculated from the mass and the
surface area, assuming a spherical cloud. For Lupus, the value is an average over the
three clouds.}
\tablenotetext{c}{The Lupus III cloud is at $200\pm20$ pc.
This is accounted for in the total area and mass.}
\tablenotetext{d}{This is the crossing time for Lupus III, the largest subcloud.
The time for Lupus I is 3.6 Myr and for Lupus IV, it is 1.9 Myr.}
\tablenotetext{e}{This mass excludes Ophiuchus-North, a disconnected piece
of the northern streamer.}
\tablerefs{
(1) \citet{whittet97}; 
(2) \citet{vilas94}; 
(3) \citet{comeron08}; 
(4) \citet{hara99}; 
(5) \citet{enoch06}; 
(6) \citet{ridge06}; 
(7) \citet{straizys96};  
(8) \citet{degeus89};
}
\end{table}

\begin{table}
\caption{Number of Objects of Various Types in Catalogs \label{catstats}}
\begin{tabular}[htb]{lrrrrrrrr}\hline
\hline
Category         & Cha~II & Lupus I & Lupus III & Lupus IV & Perseus & Serpens & Ophiuchus \\
\hline
Full Archive	 & 265607 & 410067  & 480766    & 177727   & 777484  & 377456  & 1629665 \\
High Reliability &  41787 &  47597  & 101507    &  33340   &  58312  & 104098  &  211638 \\
Three Bands      &  47952 &  67494  & 113515    &  47367   &  70071  &  91555  &  207496 \\
Stars            &  22958 &  30371  &  56813    &  21494   &  25753  &  57784  &  106964 \\
YSOc             &     29 &     20  &     79    &     12   &    387  &    262  &     297 \\
GALc             &    298 &    337  &    299    &     79   &    826  &    208  &     842 \\
Other            &   5036 &   6539  &   9859    &   3026   &  11529  &   9807  &  207496 \\
\hline
\end{tabular}
\end{table}

\begin{table}
\caption{Numbers, Densities, Star Formation Rates \label{sfrtab}}
\begin{tabular}[htb]{lrrrrrl}\hline
\hline
Cloud  &  N(YSOs) & N/$\Omega$  & N/Area & SFR  & SFR/Area & Notes \\
 &  & (deg$^{-2}$)  & (pc$^{-2}$) & (\msunmyr) & (\msunmyr pc$^{-2}$) &   \\
\hline
Cha~II	& 26		& 25	&  2.6	& 6.5 & 0.65  & \citet{alcala08} \\
Lupus\tbn{a}   & 94	& 30	&  3.3 & 24    & 0.83  & \citet{merin08a} \\
Perseus	& 385		& 100	&  5.2 	& 96  & 1.3  & \citet{laisynth} \\
Serpens	& 227	 	& 267	& 13.0	& 57  & 3.2  & \citet{harveysynth} \\
Ophiuchus & 292	 	& 44	&  9.3	& 73  & 2.3  & \citet{allensynth}  \\
\hline
Total   & 1024		& 66	&  6.4	& 256 & 1.6  & \nodata  \\
\hline
\end{tabular}
\tablenotetext{a}{A sum over the three Lupus 
clouds of the values for each cloud.
}
\end{table}

\begin{table}
\caption{Efficiencies and Depletion Timescales \label{tabeff}}
\begin{tabular}[htb]{lrrrrrl}\hline
\hline
Cloud  &  ${M_*\over(M(cloud) + M_*)}$ & $M_*/M(dense)$  & $t_{dep}(cloud)$ & 
$t_{dep}(dense)$  & \sfrff\ & Notes \\
 &  &   & (Myr) & (Myr) &  &   \\
\hline
Cha~II	& 0.030		&       & 66 	&\nodata & 0.028 & \citet{alcala08} \\
Lupus\tbn{a} & 0.054	& 	& 35    &\nodata & 0.050 & \citet{merin08a} \\
Perseus	& 0.038		& 0.69	& 50 	& 2.9  & 0.049   & \citet{laisynth} \\
Serpens	& 0.053	 	& 1.2	& 35    & 1.6  & 0.036   & \citet{harveysynth} \\
Ophiuchus & 0.063	& 3.3   & 30	& 0.6  & 0.064   & \citet{allensynth} \\
\hline
All Clouds\tbn{b} & 0.048 & 1.2	&  40	& 1.8  & 0.040    & \nodata  \\
\hline
\end{tabular}
\tablenotetext{a}{A sum over the three Lupus 
clouds of the values for each cloud.
}
\tablenotetext{b}{For all but \sfrff, this number is calculated by adding all
clouds with the relevant data together; for \sfrff, it is the average over
all clouds of the individual values.
}
\end{table}

\begin{table}
\caption{Number of YSOs by Cloud and Class \label{tabclasses}}
\begin{tabular}[htb]{lrrrrrrl}\hline
\hline
Cloud  &   I & Flat    &  II   & III    &$t(I)$ & $t(F)$ &Reference \\
       &	&	&	&	& (Myr)	& (Myr)	&	     \\
\hline
Cha~II	& 2	&  1	&  19	&  4	& 0.21	& 0.11	& \citet{alcala08} \\
Lupus\tbn{a} &5 & 10	&  52	& 27	& 0.19	& 0.38	& \citet{merin08a} \\
Perseus	& 87	& 42	& 225	& 31 	& 0.77	& 0.37	& \citet{laisynth}   \\
Serpens	& 36	& 23	& 140	& 28 	& 0.51	& 0.33	& \citet{harveysynth}\\
Ophiuchus & 35	& 47	& 176	& 34	& 0.40	& 0.53	& \citet{allensynth} \\
\hline
Total	&165	&123	& 612	&124	& 0.54	& 0.40	& \nodata  \\
\hline
After E.C.\tbn{b} &  &  &  &	&$t^{\prime}(I)$ & $t^{\prime}(F)$ & \av\ Used  \\
\hline
Cha~II	& 1	&  2	&  18	&  5	& 0.11	& 0.22	& 3.95  \\
Lupus\tbn{a} &5 &  7	&  54	& 28	& 0.19	& 0.26	& 2.91 \\
Perseus	& 76	& 35	& 244	& 30 	& 0.62	& 0.29	& 5.92  \\
Serpens	& 32	& 22	& 140	& 33 	& 0.46	& 0.31	& 9.57  \\
Ophiuchus & 27	& 44	& 179	& 42	& 0.30	& 0.49	& 9.76  \\
\hline
Total	&141	& 110	& 635	&138	& 0.44	& 0.35	& \nodata  \\
\hline
\end{tabular}
\tablenotetext{a}{A sum over the three Lupus
clouds of the values for each cloud.
}
\tablenotetext{b}{After correction for extinction, as described in \S 
\ref{classification}.
}
\end{table}

\begin{table}
\caption{Dependence of Class on Definition \label{tabdefs}}
\begin{tabular}[htb]{lrrrrrl}\hline
\hline
Method  &  Class 0 & Class I & Flat & Class II & Class III & Notes \\
\hline
Catalog & \nodata	& 165	& 123	 & 612	& 124 	& fit to 2-24 \\
IRAC   & \nodata	& 145	& 104	 & 580	& 187 	& fit to 3.6-8 \\
De-reddened & \nodata	& 141	& 110	 & 635	& 138	& fit to 2-24 \\
\tbol(obs)	& 40	& 156	&\nodata & 753	& 0	& \nodata   \\
\tbol(dered)	& 24	& 125	&\nodata & 223	& 577	& \nodata   \\
\tbol(obs)	& 44	&  88	&\nodata &   0  &   0   & with mm core   \\
\tbol(dered)	& 26	&  95	&\nodata &  11  &   0   & with mm core \\
\hline
\end{tabular}
\tablenotetext{a}{The Catalog values of $\alpha$ are from the c2d catalog,
based on a least-squares fit to any available data from 2 to 24 \micron.
IRAC values are based only on IRAC wavelengths. De-reddened values of
$\alpha$ or \tbol are based on extinction estimates from spectral types
where available, assumption of a K7 spectral type otherwise for Class II
and III sources. For Flat and Class 0/I sources, the mean extinction toward
Class II sources in that cloud was assumed. The last two lines are based on
\citet{enoch08a}, who require a detection with Bolocam and thus refer only
to the three clouds with such data.
}
\end{table}

\begin{table}
\caption{Sources by Class and Environment \label{tabclusters}}
\begin{tabular}[htb]{lrrrrrl}\hline
\hline
Environment  &  Class I & Flat & Class II & Class III & Total & I+F/II+III \\
\hline
Distributed & 11	& 7   & 43  & 32  & 93   & 0.24       \\
Loose Group & 28	& 5   & 30  & 7   & 70   & 0.89       \\
Tight Group & 34	& 25  & 64  & 8  & 131   & 0.82       \\
Loose Cluster & 127	& 112 & 559 & 131 & 929  & 0.35       \\
Tight Cluster & 90	& 79  & 322 & 63  & 554  & 0.44       \\
\hline
\end{tabular}
\end{table}

\clearpage

\appendix

\section{Tables of YSOs}\label{appendix}

Table \ref{tabysolist1} contains the list of YSOs analyzed in this paper.
The columns include a running index number, which is consistent throughout
these tables, the JHK flux densities from
2MASS, flux densities from the four IRAC bands, and flux densities from 
the two shortest wavelength bands of MIPS.  
Additional flux densities from observations with other telescopes, 
at wavelengths ranging from 0.36 \um\ to 1.3 mm, are presented in 
Tables \ref{tabysolist2}, \ref{tabysolist3}, and \ref{tabysolist4}.
The data are taken from 
\citet{alcala08},
\citet{brede08},
\citet{davis99},
\citet{djupvik06},
\citet{enoch06},
\citet{enoch07},
\citet{hatchell05},
\citet{hatchell07},
\citet{henning93},
\citet{johnstone00},
\citet{kirk06},
\citet{merin08a},
\citet{stanke06},
\citet{tachihara07},
\citet{wu07},
and
\citet{young06}.

Table \ref{tabysoprops} contains the derived properties for each source.
The columns list, in order, the running index number, the name of the
 cloud containing the source, the official c2d source name, 
which includes the position information in J2000 coordinates, the c2d
source classification, and the values for $\alpha$, \tbol, and \lbol, 
calculated from the observed flux densities. Following are the same
three quantities (\aprime, \tbolprime, and \lbolprime),
calculated after correcting the flux densities for
extinction, as described in \S \ref{classification}. 
The next column indicates whether there is evidence for an envelope,
as traced by extended millimeter continuum emission (e.g., \citealt{enoch06},
\citealt{young06}, \citealt{enoch07}). The last column has
information on the reliability of the \tbol\ and \lbol\ values.  
``G'' indicates existing photometry provides sufficient spectral coverage for 
reliable calculations; the uncertainties in both quantities are dominated 
by errors introduced by incomplete, finite sampling of the source SED and 
are typically $20-60$\% (\citealt{enoch08a}; \citealt{dunham08}). 
``L\_I'' indicates the source is associated with an envelope and thus likely
 an embedded source, but lacks spectral coverage between 24 \um\ and 1 mm.  
The calculated values of \lbol\ and \tbol\ are thus considered lower and 
upper limits, respectively.
``L\_II'' indicates the source is not associated with an envelope and has an 
observed $\alpha \leq$ $-$0.3, thus is not likely an embedded source, but no 
extinction correction could be derived based on available data, usually 
because of a lack of sufficient coverage in the near-infrared.  The calculated 
values of \lbol\ and \tbol\ are thus considered lower limits.

The full tables are quite long and available only electronically. We provide
samples of the tables in the print version.

\clearpage

\rotate
\tabletypesize{\scriptsize}
\begin{table}
\caption{\label{tabysolist1}2MASS and \emph{Spitzer} Flux Densities, in milli-Jy, of YSOs in the c2d Clouds}
\begin{tabular}[htb]{rrrrrrrrrr}\hline
\hline
   &  \emph{J} & \emph{H} & \emph{K$_{s}$} & IRAC  & IRAC & IRAC & IRAC  & MIPS & MIPS \\
 Index  & 1.25 \um   &  1.65 \um   & 2.17 \um & 3.6 \um & 4.5 \um & 5.8 \um & 5.8 \um & 24 \um & 
70 \um \\
\hline
   1 &     3100 $\pm$      29 &     5400 $\pm$      99 &     5900 $\pm$      54 &     2600 $\pm$     180 &     1500 $\pm$      92 &     1400 $\pm$      68 &      870 $\pm$      42 &      300 $\pm$      28 &       75 $\pm$      8.2 \\
   2 &      390 $\pm$      11 &      770 $\pm$      25 &      880 $\pm$      31 &      460 $\pm$      24 &      210 $\pm$      12 &      210 $\pm$      10 &      130 $\pm$      6.1 &       31 $\pm$      2.8 &             ... \\
   3 &      300 $\pm$      8.1 &     1600 $\pm$      75 &     5800 $\pm$     170 &             ... &
          ... &    19000 $\pm$    3600 &    11000 $\pm$    3500 &             ... &    37000 $\pm$    3400 \\
   4 &      0.89 $\pm$    0.070 &       3.5 $\pm$     0.15 &       9.5 $\pm$     0.29 &       18 $\pm$
    0.87 &       25 $\pm$      1.2 &       34 $\pm$      1.6 &       51 $\pm$      2.4 &      610 $\pm$      57 &     1500 $\pm$     150 \\
   5 &       51 $\pm$      1.5 &       78 $\pm$      2.4 &       86 $\pm$      2.5 &       73 $\pm$
  4.0 &       67 $\pm$      3.3 &       61 $\pm$      2.9 &       57 $\pm$      2.7 &       51 $\pm$
   4.7 &       40 $\pm$      5.7 \\
   6 &       12 $\pm$     0.35 &       1.0 $\pm$     0.29 &       8.7 $\pm$     0.25 &       5.9 $\pm$
    0.29 &       4.9 $\pm$     0.24 &       3.9 $\pm$     0.20 &       3.1 $\pm$     0.15 &      0.82 $\pm$     0.21 &             ... \\
   7 &      160 $\pm$      4.4 &      270 $\pm$      7.5 &      300 $\pm$      8.2 &      290 $\pm$
  15 &      250 $\pm$      13 &      260 $\pm$      13 &      330 $\pm$      16 &      380 $\pm$      36 &      170 $\pm$      19 \\
   8 &       1.5 $\pm$    0.060 &       4.2 $\pm$     0.12 &       6.3 $\pm$     0.21 &       7.1 $\pm$     0.35 &       6.8 $\pm$     0.34 &       6.7 $\pm$     0.33 &       7.0 $\pm$     0.33 &       8.1
$\pm$     0.76 &             ... \\
   9 &             ... &       2.7 $\pm$     0.12 &       38 $\pm$      1.1 &      210 $\pm$      10 &
     440 $\pm$      26 &      640 $\pm$      31 &      700 $\pm$      34 &     3600 $\pm$     340 &
 6200 $\pm$     600 \\
  10 &     1200 $\pm$      29 &     1800 $\pm$      55 &     1700 $\pm$      46 &      800 $\pm$      41 &      450 $\pm$      24 &      390 $\pm$      19 &      320 $\pm$      15 &      260 $\pm$      24 &             ... \\
\hline
\end{tabular}
\end{table}

\rotate
\tabletypesize{\scriptsize}
\begin{table}
\caption{\label{tabysolist2}$0.36-0.96$ \um\ Flux Densities, in milli-Jy, of YSOs in the c2d Clouds}
\begin{tabular}[htb]{rrrrrrrrr}\hline
\hline
   &  \emph{U} & \emph{B} & \emph{V} & \emph{R}  & \emph{H$\alpha$12} & \emph{I} & \emph{m}914  
& \emph{z}  \\
 Index  & 0.36 \um   &  0.44 \um   & 0.55 \um & 0.64 \um & 0.665 \um & 0.79 \um & 0.915 \um & 
0.96 \um  \\
\hline
   1 &             ... &      0.48 $\pm$     0.12 &       4.2 $\pm$      1.0 &       12 $\pm$      2.8
&       16 $\pm$      1.9 &      510 $\pm$      19 &             ... &             ... \\
   2 &             ... &      0.16 $\pm$    0.040 &      0.65 $\pm$     0.16 &       3.6 $\pm$    0.030 &       2.4 $\pm$     0.29 &       33 $\pm$     0.60 &       90 $\pm$      6.6 &       49 $\pm$      2.3 \\
   3 &             ... &   0.00100 $\pm$   0.0020 &      0.14 $\pm$   0.0040 &      0.83 $\pm$  0.00100 &      0.78 $\pm$    0.100 &       6.7 $\pm$    0.060 &       17 $\pm$      1.2 &       19 $\pm$     0.36 \\
   4 &             ... &             ... &             ... &   0.00100 $\pm$  0.00100 &   0.00100 $\pm$  0.00100 &     0.040 $\pm$   0.0020 &     0.090 $\pm$  0.00100 &      0.11 $\pm$  0.00100 \\
   5 &             ... &      0.40 $\pm$  0.00100 &       1.3 $\pm$    0.040 &       4.4 $\pm$    0.040 &       4.0 $\pm$     0.48 &       14 $\pm$     0.12 &       22 $\pm$      1.6 &       22 $\pm$     0.20 \\
   6 &       2.8 $\pm$     0.18 &       5.3 $\pm$     0.15 &       7.5 $\pm$     0.21 &       8.8 $\pm$      1.2 &       1.0 $\pm$      1.2 &       11 $\pm$    0.100 &       11 $\pm$     0.81 &       11 $\pm$    0.100 \\
   7 &             ... &      0.95 $\pm$     0.23 &       8.2 $\pm$      2.0 &       11 $\pm$    0.100
&       14 $\pm$      1.6 &       35 $\pm$     0.64 &       59 $\pm$      4.3 &       63 $\pm$      1.2 \\
   8 &             ... &             ... &             ... &    0.0030 $\pm$  0.00100 &             ... &     0.040 $\pm$   0.0020 &      0.18 $\pm$    0.020 &      0.18 $\pm$  0.00100 \\
   9 &             ... &             ... &             ... &             ... &             ... &
      ... &             ... &             ... \\
  10 &             ... &      0.21 $\pm$    0.050 &             ... &       27 $\pm$      6.4 &       37 $\pm$      4.4 &      270 $\pm$      7.5 &             ... &             ... \\

\hline
\end{tabular}
\end{table}

\rotate
\tabletypesize{\scriptsize}
\begin{table}
\caption{\label{tabysolist3}$3.4-100$ \um\ Flux Densities, in milli-Jy, of YSOs in the c2d Clouds}
\begin{tabular}[htb]{rrrrrrrrr}\hline
\hline
   &  \emph{L} & \emph{M} & \emph{ISO} & \emph{IRAS}  & \emph{ISO} & \emph{IRAS} & \emph{IRAS}  
& \emph{IRAS}  \\
 Index  & 3.4 \um   &  5.0 \um   & 6.7 \um & 12 \um & 14.3 \um & 25 \um & 60 \um & 
100 \um  \\
\hline
   1 &     3500 $\pm$      32 &             ... &             ... &             ... &             ... &             ... &             ... &             ... \\
   2 &      540 $\pm$      5.0 &             ... &             ... &             ... &             ...
&             ... &             ... &             ... \\
   3 &    12000 $\pm$     230 &   21000 $\pm$ 590 &             ... &    44000 $\pm$    2200 &
  ... &   100000 $\pm$    5100 &   110000 $\pm$   11000 &   100000 $\pm$   10000 \\
   4 &             ... &             ... &             ... &      110 $\pm$      20 &             ... &      900 $\pm$     180 &     3600 $\pm$     730 &     5100 $\pm$    1100 \\
   5 &       58 $\pm$      6.5 &             ... &             ... &             ... &             ...
&       20 $\pm$      20 &             ... &             ... \\
   6 &       26 $\pm$      5.6 &             ... &             ... &             ... &             ...
&             ... &             ... &             ... \\
   7 &      290 $\pm$      2.7 &             ... &             ... &      520 $\pm$      50 &
   ... &      500 $\pm$      50 &      460 $\pm$      90 &     4300 $\pm$    1100 \\
   8 &             ... &             ... &       6.7 $\pm$     0.60 &             ... &       6.0 $\pm$      1.2 &             ... &             ... &             ... \\
   9 &             ... &             ... &      790 $\pm$      26 &      700 $\pm$      70 &     1200 $\pm$      25 &     4400 $\pm$     220 &    11000 $\pm$     530 &    18000 $\pm$    5400 \\
  10 &      840 $\pm$      7.7 &             ... &             ... &      380 $\pm$      40 &
   ... &      360 $\pm$      30 &             ... &             ... \\
\hline
\end{tabular}
\end{table}

\rotate
\tabletypesize{\scriptsize}
\begin{table}
\caption{\label{tabysolist4}$160-1300$ \um\ Flux Densities, in milli-Jy, of YSOs in the c2d Clouds}
\begin{tabular}[htb]{rrrrrrrr}\hline
\hline
   &  \emph{MIPS} & & & & & &   \\
 Index  & 160 \um   &  350 \um   & 450 \um & 850 \um & 1100 \um & 1200 \um & 1300 \um    \\
\hline
   1 &             ... &             ... &             ... &             ... &             ... &
      ... &             ... \\
   2 &             ... &             ... &             ... &             ... &             ... &
      ... &             ... \\
   3 &             ... &             ... &             ... &             ... &             ... &     1500 $\pm$     120 &      680 $\pm$      22 \\
   4 &    12000 $\pm$    2400 &             ... &             ... &             ... &             ... &             ... &       60 $\pm$      15 \\
   5 &             ... &             ... &             ... &             ... &             ... &
      ... &             ... \\
   6 &             ... &             ... &             ... &             ... &             ... &
      ... &             ... \\
   7 &             ... &             ... &             ... &             ... &             ... &
      ... &             ... \\
   8 &             ... &             ... &             ... &             ... &             ... &
      ... &             ... \\
   9 &    27000 $\pm$    2700 &             ... &             ... &             ... &             ... &     1900 $\pm$     130 &             ... \\
  10 &             ... &             ... &             ... &             ... &             ... &
      ... &             ... \\
\hline
\end{tabular}
\end{table}

\rotate
\tabletypesize{\scriptsize}
\begin{table}
\caption{\label{tabysoprops}Properties of YSOs in the c2d Clouds}
\begin{tabular}[htb]{rlclrrrrrrrcr}\hline
\hline
   &   & Spitzer &  & & & \multicolumn{3}{c}{Observed} & \multicolumn{3}{c}{Extinction Corrected} & 
  \\
               &                 & Source Name  & c2d            &         &
    & \tbol & \lbol    &          & \tbol$^{\prime}$ & {\lbol$^{\prime}$}      & &     \\
Index & Cloud & {(SSTc2d $+$)} & {Classification} & {A$_{\rm V}$\tablenotemark{a}} & {$\alpha$} &
 {(K)}    & {(\lsun)}   & {$\alpha^{\prime}$} & {(K)}    & {(\lsun)}   & {Envelope\tablenotemark{b}}
 & {Quality\tablenotemark{c}} \\
\hline
  1 & Cham II & J124825.74$-$770636.5 & YSOc\_star$+$dust(MP1) &  4.4 &   $-$2.19 &       2200 &
 11 &   $-$2.59 &       3000 &       28 & N &    G \\
   2 & Cham II & J125230.66$-$771513.0 & YSOc\_star$+$dust(MP1) &  4.3 &   $-$2.38 &       2100 &
 1.4 &   $-$2.71 &       2700 &       3.2 & N &    G \\
   3 & Cham II & J125317.23$-$770710.7 &              rising & 10.5 &   $-$0.72 &        550 &       38 &   $-$0.14 &       1400 &       72 & N &    G \\
   4 & Cham II & J125342.86$-$771511.5 &            YSOc\_red &  4.0 &    0.65 &        120 &       0.49 &    0.21 &        190 &       0.59 & Y &    G \\
   5 & Cham II & J125633.66$-$764545.3 & YSOc\_star$+$dust(IR2) &  2.4 &   $-$1.22 &       2200 &
 0.21 &   $-$1.35 &       2900 &       0.36 & N &    G \\
   6 & Cham II & J125658.68$-$764706.6 & YSOc\_star$+$dust(IR2) &  0.0 &   $-$1.84 & $\geq$4500 & $\geq$0.067 &   $-$1.84 & $\geq$4500 & $\geq$0.067 & N & L\_II \\
   7 & Cham II & J125711.77$-$764011.3 & YSOc\_star$+$dust(IR1) &  3.3 &   $-$0.85 &       1800 &
 0.86 &   $-$1.06 &       3100 &       1.8 & N &    G \\
   8 & Cham II & J125806.78$-$770909.4 & YSOc\_star$+$dust(IR1) &  5.0 &   $-$0.93 &       1400 &
 0.013 &   $-$1.09 &       1900 &       0.020 & N &    G \\
   9 & Cham II & J125906.58$-$770739.9 &            YSOc\_red &  4.0 &    0.68 &        220 &       1.9 &    0.49 &        270 &       2.4 & Y &    G \\
  10 & Cham II & J125926.45$-$774708.4 & YSOc\_star$+$dust(IR4) &  3.9 &   $-$1.78 &       2400 &
 3.7 &   $-$2.01 &       3300 &       9.8 & N &    G \\
\hline
\end{tabular}
\tablenotetext{a}{Value of A$_{\rm V}$ used for dereddening, as explained in the text.}
\tablenotetext{b}{Indicates whether or not the source is associated with an envelope as 
traced by extended millimeter emission (based primarily on the sample of Enoch et al. [2008]).}
\tablenotetext{c}{Indicates the quality of the calculated \lbol\ and \tbol\ values.\\
``G'': existing photometry provides sufficient spectral coverage for reliable calculations.  
Other flags mark sources without sufficient coverage:\\
``L\_I'': the source is associated with an envelope and thus likely an embedded source, but 
lacks spectral coverage between 24 \um\ and 1 mm.  The calculated values of \lbol\ and \tbol\ 
are thus considered lower and upper limits, respectively.\\
``L\_II'': the source is not associated with an envelope and has an observed $\alpha \leq$ $-$0.3, 
thus is not likely an embedded source, but no extinction correction could be derived based 
on available data, usually because of a lack of sufficient coverage in the near-infrared.  
The calculated values of \lbol\ and \tbol\ are thus considered lower limits.\\
``O'':  Other.  These three sources were added to the list of YSOs based on the searches for 
embedded protostars presented by J{\o}rgensen et al. (2007) (for Perseus) and J{\o}rgensen 
et al. (2008) (for Ophiuchus), but they are not associated with envelopes according to Enoch 
et al. (2008).  They are only detected at one wavelength (24 \um), thus calculations 
of $\alpha$, \tbol, and \lbol\ are not possible.
They are counted as Class I objects when classifying by $\alpha$ and $\alpha^{\prime}$; they are 
not counted at all when classifying by \tbol\ and \tbol$^{\prime}$.}
\end{table}

\end{document}